\documentclass[aps,prx,reprint,amsfonts,amsmath,amssymb,longbibliography,superscriptaddress]{revtex4-2}
\usepackage{xcolor}

\usepackage{hyperref}
\usepackage{graphicx}
\usepackage{bm}
\usepackage{newtxtext}
\usepackage[varg]{newtxmath}
\usepackage[normalem]{ulem}
\usepackage{lipsum}
\usepackage{bbold}
\usepackage{cases}
\usepackage{booktabs}
\usepackage{array}
\usepackage{tabularx}
\usepackage{multirow}

\hypersetup{colorlinks=true,linkcolor=blue,citecolor=blue,urlcolor=blue}

\begin{document}
\title{Quantum many-body scars with unconventional superconducting pairing symmetries \\via multibody interactions}
\author{Shohei Imai}
\affiliation{Department of Physics, University of Tokyo, Hongo, Tokyo 113-0033, Japan}
\author{Naoto Tsuji}
\affiliation{Department of Physics, University of Tokyo, Hongo, Tokyo 113-0033, Japan}
\affiliation{RIKEN Center for Emergent Matter Science (CEMS), Wako 351-0198, Japan}
\date{\today}

\begin{abstract}
We present a systematic framework to construct model Hamiltonians that have unconventional superconducting pairing states as exact energy eigenstates, by incorporating multibody interactions (i.e., interactions among more than two particles).
The multibody interactions are introduced in a form of the local density-density coupling in such a way that any pair configuration in real space has a constant interaction energy by canceling the two-body and multibody interactions.
Our approach is applicable to both spinless and spinful models in any spatial dimensions and on any bipartite lattices, facilitating an exhaustive extension of Yang's $s$-wave $\eta$-pairing state to various other unconventional pairing symmetries ($p$-wave, $d$-wave, $f$-wave, etc.).
Particularly, the constructed eigenstates have off-site pairs with finite center-of-mass momentum, which leads to superconducting states with either even-parity and spin-triplet or odd-parity and spin-singlet symmetry.
We verify that the two-dimensional spinful Hubbard model on a square lattice with the multibody interactions has the spin-triplet $d$-wave pairing state as an energy eigenstate, which can be regarded as a quantum many-body scar state as evidenced from the numerical analysis of the pair correlation function, entanglement entropy, and level statistics.
We also discuss other examples, including spin-triplet $f$-wave pairing states on a honeycomb lattice and spin-singlet $p$-wave pairing states in a one-dimensional chain.
These findings open up the possibility of realizing nonequilibrium unconventional superconductivity in a long-lived manner protected against thermalization.
\end{abstract}
\maketitle

\section{Introduction} \label{sec:introduction}
Out-of-equilibrium superconductivity offers a potential to realize exotic quantum many-body states that are otherwise challenging to observe in thermal equilibrium.
Notable candidates for those states include photoinduced superconducting-like states that have been experimentally observed even above the equilibrium transition temperature~\cite{Fausti2011, Kaiser2014, Hu2014, Mitrano2016a, Cremin2019} (see also~\cite{Katsumi2023, Zhang2024a}).
Another example is the Floquet engineering~\cite{Bukov2015a, Oka2019, Tsuji2023} (i.e., quantum states induced by a time-periodic drive), which has endowed superconductors with topological properties~\cite{Ezawa2015, Takasan2017b, Chono2020a, Wenk2022, Yanase2022, Kitamura2022a}.
A variety of transient drives have also been employed to induce superconductivity with unconventional pairing states---such as bulk odd-frequency pairings~\cite{Cayao2021, Cayao2022} and finite-momentum pairing states~\cite{Kaneko2019a, Malakhov2021a}---which are challenging to be realized in equilibrium.

Although numerous intriguing experimental and theoretical progress has been made, there still remains a problem of how such a nonthermal state can be made robust against thermalization.
To generate nonequilibrium states, one necessarily injects a finite amount of energy into the system, which typically leads to thermalization at elevated temperatures, thereby suppressing (or even erasing) superconducting correlations.
In order to overcome this difficulty, it is crucial to seek for a certain mechanism that protects nonequilibrium superconducting states from thermalization even in the presence of many-body interactions.

As a promising mechanism to maintain superconducting correlations for a sufficiently long time at highly excited states, there have been proposed quantum many-body scar (QMBS) states~\cite{Bernien2017, Shiraishi2017, Turner2018}, which are exceptional energy eigenstates that do not have thermal properties in a nonintegrable system.
According to the eigenstate thermalization hypothesis (ETH)~\cite{Deutsch1991, Srednicki1994, Rigol2008a, Deutsch2018a}, all excited energy eigenstates in a nonintegrable model are expected to be indistinguishable from thermal states as long as one refers to few-body observables.
QMBS states are considered to be an exception to the ETH, as characterized by long-lived nonthermal dynamics in nonintegrable systems~\cite{Serbyn2021, Papic2022, Moudgalya2022, Chandran2023, Moudgalya2018a, Schecter2019a, Pai2019, Ren2021a, Yu2018e, Scherg2021, Chattopadhyay2020a, Lin2020a, Kuno2020, Sugiura2021, Zhang2023h, Su2023, Omiya2023, Omiya2023a, Desaules2021, Kaneko2024, Matsui2024}.
Despite being exact energy eigenstates in a static Hamiltonian, QMBS states differ from equilibrium states: In equilibrium, if one takes the microcanonical ensemble, most of the eigenstates in the energy shell are thermal (according to the ETH).
Even if the QMBS states are included in the shell, they are overwhelmed by those thermal states.
Several methods have been proposed to construct a model Hamiltonian that accommodates a target state as a QMBS state, such as the (restricted) spectrum-generating algebra~\cite{Moudgalya2022}, the embedding formalism~\cite{Shiraishi2017}, and the symmetry-based formalism~\cite{Pakrouski2020a}.

A primary example of a nonthermal energy eigenstate in a nonintegrable system that supports superconducting correlations is the so-called $\eta$-pairing state, which is an exact energy eigenstate of the Hubbard model on a $d$-dimensional square lattice, as shown by C. N. Yang~\cite{Yang1989}.
Yang's $\eta$-pairing state corresponds to a condensate of doublons carrying finite center-of-mass momentum, and exhibits long-range superconducting correlations with the $s$-wave and spin-singlet pairing symmetry.
If one classifies the Hilbert space in terms of symmetry sectors, the $\eta$-pairing state is the only eigenstate in the corresponding sector, characterized by the so-called $\eta$-$\mathrm{SU}(2)$ symmetry~\cite{Yang1990, Vafek2017}.
In this respect, several proposals have been made to modify the Hamiltonian to break the $\eta$-$\mathrm{SU}(2)$ symmetry, thereby transforming the Yang's $\eta$-pairing state into a QMBS state~\cite{Pakrouski2020a, Li2020l, Mark2020, Moudgalya2020, Pakrouski2021, Wildeboer2022a, Sun2023b, Kolb2023a}.
Similar $\eta$-pairing states have recently been proposed to be realized by photoirradiation~\cite{Kaneko2019a}, and have attracted interest as a mechanism to induce superconductivity by light.
Once the system is excited to the $\eta$-pairing states, the system in principle stays there forever (since they are the exact eigenstates).
Moreover, in nonequilibrium situations one might expect to see rich superconducting states with unconventional pairing symmetries that are not typically found in equilibrium.

In this paper, we generalize Yang's $\eta$-pairing state to those with various unconventional pairing symmetries ($p$-wave, $d$-wave, $f$-wave, etc.), and establish a systematic framework to construct model Hamiltonians that have the unconventional pairing states as QMBS states.
Given the known instability of Yang's $\eta$-pairing state to perturbations, such as due to the long-range Coulomb interaction and the coupling to electromagnetic fields~\cite{Hoshino2014, Tsuji2021}, it will be important to explore possible pairing states with different symmetries, which may offer an opportunity to stabilize those states against perturbations.
Recent studies have extended Yang's $\eta$-pairing state to encompass spinless fermions~\cite{Shibata2020, Tamura2022, Gotta2022} and multi-component systems~\cite{Zhai2005, Nakagawa2022a, Yoshida2022a, Sun2023b, Ray2023a}.

\begin{figure}[t]
\centering
\includegraphics[width=1.0\columnwidth]{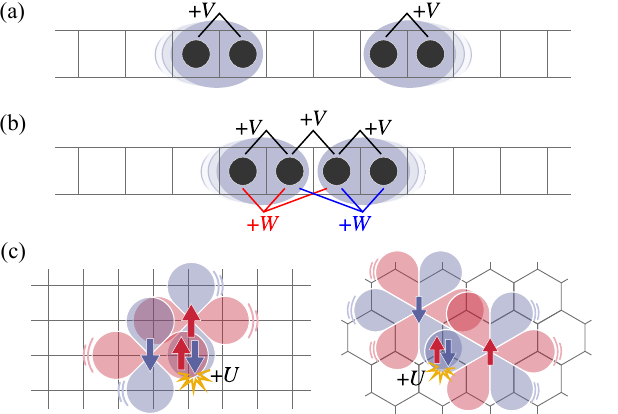}
\caption{
(a),~(b)~Examples of spatial configurations of the spinless nearest-neighbor pairing state in a one-dimensional lattice.
(a)~When two pairs (blue clouds) are separated to each other, the two-body interaction $V$ only acts within each pair.
(b)~When two pairs come next to each other, there is an additional energy increment $V$, which can be canceled by the three-body interaction $W$ (if one chooses $V+2W=0$).
(c)~Schematic picture of unconventional superconducting pairs of spinful fermions with $d$-wave (left panel) and $f$-wave (right) pairing symmetries (indicated by blue and red clouds) on the square and honeycomb lattices, respectively.
Two pairs feel an energy cost $U$ due to the two-body interaction when they overlap with each other.
}
\label{fig:schematic}
\end{figure}
If one naively extends the $\eta$-pairing states to those with other pairing symmetries, e.g., for spinless (or spin-triplet) nearest-neighbor pairing states in one dimension [see Fig.~\ref{fig:schematic}(a)], one can still see that they are eigenstates of the kinetic term in a Hubbard model (see below for more details), in much the same way as in the case of Yang's $\eta$-pairing state~\cite{Yang1989}.
The real challenge is in the interaction term: When two off-site pairs come close to each other, they feel an additional energy cost due to the two-body interaction.
This prevents the unconventional pairing states from being eigenstates of the interaction term (and hence the total Hamiltonian) in the Hubbard model.

To overcome this difficulty, we engineer multibody interactions (i.e., interactions among more than two particles) to cancel the energy increment of pairs due to the two-body interaction.
For example, in a one-dimensional spinless fermion model we can introduce a three-body interaction to cancel the nearest-neighbor two-body interaction acting on the pairs [Fig.~\ref{fig:schematic}(b)].
The cancellation works for arbitrary pair configurations, no matter how many pairs are distributed on the lattice.
In this way, the spinless pairing state can be made an exact eigenstate of the one-dimensional Hubbard model with the three-body interaction.
Previously, it has been proposed that the spinless pairing state becomes an exact eigenstate in a model with the three-body interaction and/or a density-dependent hopping~\cite{Mark2020, Shibata2020, Gotta2022, Tamura2022}, and a related model has been studied in the context of disordered quantum many-body scarred spin systems~\cite{Shibata2020}.

The advantage of our approach is that it can be systematically generalized to other pairing symmetries including spin degrees of freedom in higher dimensions [see Fig.~\ref{fig:schematic}(c)].
To this end, we introduce an extended number operator [$\mathbb{n}_{i\sigma}$ in Eq.~\eqref{eq:bold_number_operator}] that signals whether spin-$\sigma$ particles exist or not on lattice sites next to $i$ site.
Using this operator, we can systematically cancel the energy increment induced by the formation of doublons when off-site pairs overlap with each other, and obtain a model Hamiltonian with multibody interactions that has unconventional pairing states as exact eigenstates.
We note that the derived interaction is strictly local, and takes a form of the density-density coupling between $M$ particles ($M\le M_{\mathrm{max}}$ with $M_{\mathrm{max}}$ being finite and system-size independent).

Let us remark that the obtained eigenstates have off-site pairs with finite center-of-mass momentum.
When two fermions in a pair are exchanged with each other, there arises an additional minus sign coming from the sublattice degrees of freedom (on top of the orbital and spin degrees of freedom).
This leads to an unusual combination of the even-parity and spin-triplet (or odd-parity and spin-singlet) pairing symmetry.
A similar situation happens in the case of odd-frequency superconductors~\cite{Linder2019} and pair density wave (PDW)~\cite{Lu2014b, Chen2020d, Georgiou2020, Zhu2024}.
The generalized $\eta$-pairing states will open up a way to realize a family of unconventional superconducting nonthermal states, which are difficult to be accessed in equilibrium conditions.

We apply our construction to the spinful Hubbard model with multibody interactions on the two-dimensional square lattice, and show that the model has an exact eigenstate with the spin-triplet $d$-wave pairing symmetry [the left panel of Fig.~\ref{fig:schematic}(c)].
We numerically evidence that the model is nonintegrable from the analysis of energy level statistics.
We also find that the $d$-wave pairing state shows the subvolume law of entanglement entropy and the off-diagonal long-range order of superconductivity, which indicates that the $d$-wave pairing state is indeed a QMBS state.
Other examples are also discussed, including spin-triplet $f$-wave pairing states on a honeycomb lattice [the right panel of Fig.~\ref{fig:schematic}(c)] and spin-singlet $p$-wave pairing states in a one dimensional chain.

The rest of this paper is organized as follows.
In Sec.~\ref{sec:symmetry}, we introduce a generalized $\eta$-pairing operator that creates off-site pairs with finite center-of-mass momentum, and classify its symmetry in terms of the parity, spin exchange, and sublattice exchange.
In Sec.~\ref{sec:model}, we demonstrate the idea of cancelling two-body interactions with multibody ones to make the off-site pairing states exact eigenstates of a Hamiltonian.
We will mainly focus on the case of the spin-triplet $d$-wave pairing symmetry on the two-dimensional square lattice as an example for explanation.
We show that the unconventional pairing states become energy eigenstates of the multibody interacting model.
We numerically confirm that the unconventional pairing states are quantum many-body scar states based on the analysis of the pair correlation function, entanglement entropy (Sec.~\ref{sec:eta}), and the level-spacing statistics (Sec.~\ref{sec:scar}).
We will discuss generalizations to other pairing symmetries and other lattice structures, as well as extensions of the scarred models and the degeneracy structure of those pairing states, in Sec.~\ref{sec:generalization}.
Section~\ref{sec:summary} contains a summary of the paper and an outlook for possible experimental realizations.

Throughout the paper, we set the Dirac constant $\hbar = 1$ and the lattice constant $a = 1$.

\begin{table*}[t]
\caption{
Symmetry classification of conventional and unconventional $\eta$-pairing states on a bipartite lattice, which are characterized by inversion [$\mathcal{P}=\pm 1$, Eq.~\eqref{eq:parity_symmetry}], spin exchange [$\mathcal{S}=\pm 1$, Eq.~\eqref{eq:spin_symmetry}], and sublattice exchange [$\varLambda=\pm 1$, Eq.~\eqref{eq:sublattice_sign}] symmetry.
The product $\mathcal{P} \mathcal{S} \varLambda$ must satisfy $\mathcal{P} \mathcal{S} \varLambda=-1$.
The sublattice symmetry $\varLambda$ distinguishes pairs on the same sublattice ($\varLambda=+1$) from those on the different sublattices ($\varLambda=-1$).
In the case of the odd sublattice symmetry ($\varLambda=-1$), such as for nearest-neighbor (NN) pairings, a combination of the even-parity ($\mathcal{P}=+1$) and spin-triplet ($\mathcal{S}=+1$), or odd-parity ($\mathcal{P}=-1$) and spin-singlet ($\mathcal{S}=-1$) symmetry is allowed.
In the case of the even sublattice symmetry ($\varLambda=+1$), such as for on-site and next-nearest-neighbor (NNN) pairings, a combination of the even-parity ($\mathcal{P}=+1$) and spin-singlet ($\mathcal{S}=-1$), or odd-parity ($\mathcal{P}=-1$) and spin-triplet ($\mathcal{S}=+1$) symmetry is allowed.
* indicates that the mirror transformation is considered instead of the parity transformation.} \label{table:symmetry}
\centering
\begin{tabularx}{1.0\textwidth}{p{0.11\textwidth}p{0.17\textwidth}p{0.1\textwidth}p{0.09\textwidth}p{0.08\textwidth}p{0.13\textwidth}p{0.4\textwidth}}
\hline \hline 
\addlinespace[1.0mm]
Type of pairs & Lattice structure & Orbital & Parity ($\mathcal{P}$) & Spin ($\mathcal{S}$) & Sublattice ($\varLambda$) & Remarks \\
\addlinespace[1.0mm]
\hline
\addlinespace[1.0mm]
On-site pairs & 1D chain, square, cubic & $s$ wave & Even & Singlet & Even & Sec.~\ref{sec:s-wave}, Yang's $\eta$-pairing state~\cite{Yang1989} \\
\addlinespace[3.0mm]
NN pairs & 1D chain & $s$ wave & Even & Triplet & Odd & Secs.~\ref{sec:p-wave_eigenstate},~\ref{sec:p-wave_scar}, Refs.~\cite{Shibata2020, Gotta2022, Tamura2022}\\
 &  & $p$ wave & Odd & Singlet & Odd & Sec.~\ref{sec:pairing_symmetry}\\
 & Square  & $s$ wave & Even & Triplet & Odd & Sec.~\ref{sec:pairing_symmetry} \\
 & & $p_x$,$p_y$ wave & Odd & Singlet & Odd & Secs.~\ref{sec:pairing_symmetry},~\ref{sec:p-wave_scar} \\
 & & $d_{x^2-y^2}$ wave & Even & Triplet & Odd & Secs.~\ref{sec:eta_op}--\ref{sec:scar}\\
\addlinespace[3.0mm]
NNN pairs & Square & $d_{xy}$ wave & Even & Singlet & Even & Sec.~\ref{sec:pairing_symmetry}\\
 & Honeycomb & $f$ wave & Odd${}^{*}$ & Triplet & Even & Sec.~\ref{sec:honeycomb}, ${}^{*}\!$Mirror symmetry\\
\addlinespace[1.0mm]
\hline \hline
\end{tabularx}
\end{table*}
\section{Symmetry classification} \label{sec:symmetry}
We first classify symmetries of pairing states with pairs having finite momentum, which will be the target of our study.
As we mentioned in the introduction, finite-momentum pairs can exhibit unusual symmetries such as the spin-triplet and parity-odd symmetry.

We focus on off-site pairing states with staggered oscillating phases on a bipartite lattice, which can be viewed as an extension of Yang's on-site $\eta$-pairing state to those with unconventional pairing symmetries.
Let us define an off-site $\eta$-pairing operator as
\begin{align}
\eta^{+}_{\bm{\alpha},\sigma_1\sigma_2} &= \sum_{i}  f(\bm{r}_i) c_{\bm{r}_i,\sigma_1}^{\dagger} c_{\bm{r}_i+\bm{\alpha},\sigma_2}^{\dagger}, \label{eq:general_eta_operator} \\
f(\bm{r}_i) &= \begin{cases}
1 & (\bm{r}_i \in \mathrm{A}), \\
-1 & (\bm{r}_i \in \mathrm{B}),
\end{cases} \label{eq:staggered_phase}
\end{align}
where $c_{\bm{r}_i,\sigma}^{\dagger}$ ($c_{\bm{r}_i,\sigma}$) is the creation (annihilation) operator for fermions at site $\bm{r}_i$ with internal degrees of freedom (e.g., spin) denoted by $\sigma$, and $\bm{\alpha}$ is the relative displacement vector between fermions in a pair.
We partition the bipartite lattice into two sublattices labeled by $\mathrm{A}$ and $\mathrm{B}$, and $f(\bm{r}_i)$ represents an alternating phase between sites belonging to either the sublattice $\mathrm{A}$ or $\mathrm{B}$, as denoted by $\bm{r}_i \in \mathrm{A}$ or $\mathrm{B}$.
The generalized $\eta$-pairing states discussed in this paper are defined by applying these $\eta$-pairing operators to the vacuum state $|0\rangle$.
The $\eta$-pairing operators are a straightforward generalization of Yang's $\eta$-pairing operator $\eta^+=\sum_j \exp(\mathrm{i}\bm{\pi}\cdot\bm{r}_j)\, c_{\bm{r}_j\uparrow}^{\dagger} c_{\bm{r}_j\downarrow}^{\dagger}$~\cite{Yang1989} on a $d$-dimensional square lattice, which creates a doublon with the center-of-mass momentum $\bm{\pi}=(\pi,\pi,\cdots)$, corresponding to $\eta_{\bm\alpha, \sigma_1\sigma_2}^+$ with $\bm{\alpha}=(0,0,\cdots)$, $\sigma_1=\uparrow$, and $\sigma_2 = \downarrow$.

Any pairing state must be anti-symmetric against the fermion exchange due to fermion's anticommutation relation.
For on-site pairs, for example, the anti-symmetry requires a combination of the spin-singlet and parity-even symmetry.
Yang's $\eta$-pairing state belongs to this class, being classified to the $s$-wave (parity even) and spin-singlet state.
For off-site pairs spanned over different sublattices, on the other hand,
there appears an additional minus sign under the fermion exchange, which comes from the alternating phase between different sublattices in Eq.~(\ref{eq:general_eta_operator}).
Exchanging the fermion operators within Eq.~\eqref{eq:general_eta_operator} and relabeling the dammy position variables $\bm{r}_i + \bm{\alpha}$ as the new ones $\bm{r}'_i$, we obtain the following relation:
\begin{subnumcases}
{\label{eq:fermion_sign} \eta^{+}_{\bm{\alpha},\sigma_1\sigma_2} = }
-\eta_{-\bm{\alpha},\sigma_2\sigma_1}^+ &$ [f(\bm{r}_i-\bm{\alpha})=+f(\bm{r}_i)],$ \label{eq:same_sublattice} \\
+\eta_{-\bm{\alpha},\sigma_2\sigma_1}^+ &$ [f(\bm{r}_i-\bm{\alpha})=-f(\bm{r}_i)].$ \label{eq:diff_sublattice}
\end{subnumcases}
Here, Eq.~\eqref{eq:same_sublattice} corresponds to the case where paired fermions are sitting on the same sublattices, while Eq.~\eqref{eq:diff_sublattice} corresponds to the case where paired fermions are on the difference sublattices.
In the latter, the pair wavefunction changes its sign when the sublattices are exchanged ($\mathrm{A} \leftrightarrow \mathrm{B}$) due to the staggered phases (i.e., the pairs have finite momentum).

Based on this observation, we introduce a sublattice symmetry operation $\varLambda$ defined by
\begin{subnumcases}
{\label{eq:sublattice_sym} \varLambda^{-1} c_{\bm{r}_i\sigma} \varLambda =}
c_{\bm{r}_i\sigma} &$ (\bm{r}_i \in \mathrm{A}),$ \label{eq:A_sublattice_sym} \\
-c_{\bm{r}_i\sigma} &$ (\bm{r}_i \in \mathrm{B}).$ \label{eq:B_sublattice_sym}
\end{subnumcases}
Using the eigenvalues ($\pm 1$) of this sublattice symmetry, one can classify the symmetry of the $\eta$-pairing operators as follows:
\begin{subnumcases}
{\label{eq:sublattice_sign} \varLambda^{-1} \eta^{+}_{\bm{\alpha},\sigma_1\sigma_2} \varLambda = }
+\eta_{\bm{\alpha},\sigma_1\sigma_2}^+ &$ [f(\bm{r}_i-\bm{\alpha})=+f(\bm{r}_i)],$ \label{eq:sublattice_sign_same_sublattice} \\
-\eta_{\bm{\alpha},\sigma_1\sigma_2}^+ &$ [f(\bm{r}_i-\bm{\alpha})=-f(\bm{r}_i)].$ \label{eq:sublattice_sign_diff_sublattice}
\end{subnumcases}
As in Eqs.~\eqref{eq:fermion_sign}, the same-sublattice pairing operator has the $+1$ eigenvalue [Eq.~\eqref{eq:sublattice_sign_same_sublattice}], and the different-sublattice pairing operator has the $-1$ eigenvalue [Eq.~\eqref{eq:sublattice_sign_diff_sublattice}].

Furthermore, by combining the site-centered inversion symmetry $\mathcal{P}$ expressed as
\begin{equation}
\mathcal{P}^{-1} \eta_{\bm{\alpha},\sigma_1\sigma_2}^+ \mathcal{P} = \eta_{-\bm{\alpha},\sigma_1\sigma_2}^+, \label{eq:parity_symmetry}
\end{equation}
and the spin permutation symmetry $\mathcal{S}$ represented by
\begin{equation}
\mathcal{S}^{-1} \eta_{\bm{\alpha},\sigma_1\sigma_2}^+ \mathcal{S} = \eta_{\bm{\alpha},\sigma_2\sigma_1}^+,  \label{eq:spin_symmetry}
\end{equation}
one can symmetrize the extended $\eta$-pairing operators~\eqref{eq:general_eta_operator} via appropriate linear combinations, thereby fully specifying their symmetries under the restriction of $\mathcal{P}\mathcal{S}\varLambda = -1$ as imposed from the fermion anticommutation relation.

The symmetry classification of the on-site and off-site $\eta$-pairing states is summarized in Table~\ref{table:symmetry} for various pairing symmetries discussed in this paper.
For example, in a one-dimensional spinless fermion system, the nearest-neighbor $\eta$-pairing states, corresponding to the case of $\alpha=\pm 1$ with no internal degrees of freedom, are characterized by the $s$-wave ($\mathcal{P}=+1$), spin-triplet ($\mathcal{S}=+1$), and sublattice-odd ($\varLambda = -1$) pairing symmetry.
This is in contrast to the spinless $p$-wave topological superconductivity in the Kitaev chain~\cite{Kitaev2001}, which is even in the sublattice symmetry.
These unconventional pairing symmetries originate from the staggered phase oscillation of the nonlocal pairing between different sublattices.

We remark the case where the site-centered parity symmetry is absent.
A representative example is the honeycomb lattice, where the site-symmetry group (i.e., the point group that fixes one lattice site) is $C_{3v}$~\cite{Xu2022b}, which does not include the parity symmetry.
In those cases, one can consider the mirror symmetry instead of the parity symmetry (Table~\ref{table:symmetry}).
Actually, the irreducible representation corresponding to the $f$-wave symmetry belongs to $A_2$, which is odd under the mirror symmetry, being distinguished from other irreducible representations.

\section{Superconducting energy eigenstates with unconventional pairing symmetries} \label{sec:model}
In the previous section, we have established the symmetry classification of the off-site pairing states with finite center-of-mass momentum.
Here we present a systematic construction of model Hamiltonians that have unconventional off-site pairing states as exact eigenstates.

In Sec.~\ref{sec:p-wave_eigenstate}, we take a glance at the simplest case of the spinless (or spin-triplet) $s$-wave pairing state in a one-dimensional system, which contains the essence of our idea utilizing multibody interactions to construct the model Hamiltonians.
We then discuss how the model construction can be generalized to other pairing symmetries.
Here, we focus on a two-dimensional spinful model with multibody interactions on the square lattice.
In Sec.~\ref{sec:eta_op}, we will consider nearest-neighbor fermion pairs and introduce the $d$-wave spin-triplet $\eta$-pairing operator.
In Sec.~\ref{sec:two-body_interaction}, we see that the unconventional pairing states are not eigenstates of the two-body interacting model.
In Sec.~\ref{sec:multi-body_interaction}, we introduce an extended number operator, which measures whether neighboring sites are occupied by particles.
Using this operator, we can exactly and systematically cancel the energy increment between off-site pairs coming from the two-body interaction.
We show that the spin-triplet $d$-wave pairing state becomes an exact eigenstate of the multibody interacting model.

\subsection{Spinless $s$-wave pairing state in a one-dimensional system} \label{sec:p-wave_eigenstate}
Let us consider a one-dimensional spinless fermion model.
Based on the analogy of Kitaev's spinless fermion model of $p$-wave topological superconductors~\cite{Kitaev2001}, one can define a spinless $\eta$-pairing operator and a spinless $\eta$-pairing state~\cite{Shibata2020, Tamura2022, Gotta2022},
\begin{align}
\eta_{\mathrm{sl}}^{+} &= \sum_j (-1)^j c_j^{\dagger} c_{j+1}^{\dagger}, \label{eq:peta_operator} \\
|\Psi_{\mathrm{sl}}^N \rangle &= \frac{1}{\mathcal{N}_{\mathrm{sl}}^N} (\eta_{\mathrm{sl}}^{+})^{N/2} |0 \rangle, \label{eq:peta_state}
\end{align}
respectively.
Here, $c_i^{\dagger}$ ($c_i$) is the creation (annihilation) operator for spinless fermions at site $i$, $N$ is the number of fermions ($N$ is assumed to be even), and $\mathcal{N}_{\mathrm{sl}}^N$ is the normalization constant (such that $\langle \Psi_{\mathrm{sl}}^N|\Psi_{\mathrm{sl}}^N\rangle=1$).
The operator $\eta_{\mathrm{sl}}^{+}$ creates a nearest-neighbor pair with the center-of-mass momentum~$\pi$.
In contrast to Kitaev's $p$-wave superconductors, the pair created by $\eta_{\mathrm{sl}}^{+}$ has even parity ($\mathcal{P}^{-1} \eta_{\mathrm{sl}}^{+} \mathcal{P} = \eta_{\mathrm{sl}}^{+}$), so it should be classified to the $s$-wave pairing symmetry.
The spinless fermions can be regarded as spinful ones with fully polarized spins.
Hence the created pair has the spin-triplet symmetry, corresponding to the case of Eq.~\eqref{eq:diff_sublattice}.
The spinless (or spin-triplet) $\eta$-pairing state shows an off-diagonal long-range order with the $s$-wave pairing symmetry, which can be exactly evaluated in Ref.~\cite{Gotta2022}.
In Fig.~\ref{fig:schematic}(a) and (b), we illustrate examples of spatial configurations (Fock states in the coordinate basis) of the spinless $s$-wave $\eta$-pairing state for $N/2=2$.

The Hamiltonian of the spinless Hubbard model that we consider here is given by $\mathcal{H}=\mathcal{H}_t+\mathcal{H}_V$ with
\begin{align}
\mathcal{H}_t &= -t\sum_i \left( c_{i}^{\dagger} c_{i+1} + \mathrm{H.c.} \right), \label{eq:ham_t} \\
\mathcal{H}_V &= V \sum_i n_i n_{i+1},\label{eq:ham_V}
\end{align}
where $n_i=c_i^{\dagger} c_i$ is the number operator.
The first term $\mathcal{H}_t$ [Eq.~\eqref{eq:ham_t}] describes the nearest-neighbor particle hopping, and the second one $\mathcal{H}_V$ [Eq.~\eqref{eq:ham_V}] represents the nearest-neighbor two-body interaction with an interaction strength $V$.
For the kinetic term $\mathcal{H}_t$ [Eq.~\eqref{eq:ham_t}], the spinless $s$-wave $\eta$-pairing operator [Eq.~\eqref{eq:peta_operator}] satisfies the commutation relation $[\mathcal{H}_t,\ \eta_{\mathrm{sl}}^{+}]=0$ for both the periodic and open boundary conditions.
However, the spinless $s$-wave $\eta$-pairing state is not an eigenstate of the interaction term [Eq.~\eqref{eq:ham_V}], since adjacent pairs have an additional energy cost $V$, as displayed in Fig.~\ref{fig:schematic}(b).

To cancel the energy increment arising from the two-body interaction, we introduce the three-body interaction,
\begin{align}
\mathcal{H}_W = W \sum_i n_i n_{i+1} n_{i+2}, \label{eq:ham_W}
\end{align}
with the interaction strength $W$.
We observe that when two pairs come next to each other there always appear two combinations of neighboring three particles [Fig.~\ref{fig:schematic}(b)].
This motivates us to choose the three-body interaction parameter $W=-V/2$, which cancels the energy increase due to the two-body interaction.
The cancellation works for arbitrary number $N/2$ of pairs no matter how they distribute on the lattice.
It also works for both the periodic and open boundary conditions.
In the latter case, the interaction terms should be defined as $\mathcal{H}_V=V\sum_{i=1}^{L-1} n_i n_{i+1}$ and $\mathcal{H}_W=W\sum_{i=1}^{L-2} n_i n_{i+1} n_{i+2}$ with $L$ being the number of lattice sites.
Hence, the spinless $s$-wave $\eta$-pairing state becomes an exact eigenstate of the Hubbard model with the three-body interaction (with $W=-V/2$) under the periodic or open boundary condition,
\begin{align}
&\mathcal{H} = \mathcal{H}_{t} + \mathcal{H}_V + \mathcal{H}_{W}, \label{eq:spinless_hamiltonian} \\
&\mathcal{H} | \Psi_{\mathrm{sl}}^N \rangle = \frac{N}{2} V | \Psi_{\mathrm{sl}}^N \rangle,
\label{eq:p-wave_eigenstate}
\end{align}
with the eigenenergy $NV/2$ (coming from the nearest-neighbor two-body interaction within each pair).
Let us remark that the derived interaction part of the Hamiltonian takes a form of
\begin{align}
\mathcal{H}_V + \mathcal{H}_W =& V \sum_i n_i n_{i+1} \left(1 - \frac{1}{2}n_{i+2} \right), \label{eq:ham_modW} \\
=& \frac{V}{2} \sum_i n_i [1-(1-n_{i-1})(1 - n_{i+1})], \label{eq:ham_modW_bbn}
\end{align}
which suggests a hint for generalization to other pairing symmetries in higher dimensions.

The interaction term defined in Eq.~\eqref{eq:ham_modW} is a special case of Eq.~(13) in Ref.~\cite{Shibata2020} with parameters chosen as $c^{(1)}_j=-V/2$ and $c^{(2)}_j=c^{(3)}_j=0$, where the authors considered configurations that there is no isolated fermion in the $\eta$-pairing state.

\subsection{Off-site $\eta$-pairing operators and $d$-wave pairing states} \label{sec:eta_op}
Having established the spinless $s$-wave $\eta$-pairing state as an exact eigenstate of the three-body interacting system in the previous subsection, we discuss how the model construction can be generalized to other pairing symmetry cases including spin degrees of freedom in higher dimensions.
In the following subsections, we consider the system with the periodic boundary condition.

Let us define an off-site $\eta$-pairing creation operator in two dimensions as
\begin{equation}
\eta_{\bm{\alpha}}^{+} = \sum_i \mathrm{e}^{\mathrm{i}\bm{\pi} \cdot \bm{r}_i} c_{\bm{r}_i,\uparrow}^{\dagger} c_{\bm{r}_i+\bm{\alpha},\downarrow}^{\dagger}, \label{eq:eta_operator}
\end{equation}
where $\bm\alpha$ represents a separation between two particles in a pair, the site index $i$ runs over the entire lattice sites, $\bm{\pi}=(\pi,\pi)$ is the center-of-mass crystal momentum of pairs, and $c_{\bm{r}_i,\sigma}$ is the fermion annihilation operator with spin $\sigma$ ($= \uparrow,\downarrow$) at position $\bm{r}_{i}$.
This definition is the same as Eq.~(26) in Ref.~\cite{Yang1989}.
When we use the notation of Eq.~\eqref{eq:general_eta_operator}, $\eta_{\bm\alpha}^+$ in Eq.~\eqref{eq:eta_operator} can be written as $\eta_{\bm\alpha}^+=\eta_{\bm\alpha, \uparrow\downarrow}^+$.

Here we focus on the case of nearest-neighbor pairs on the square lattice, in which we can take four independent $\eta$-pairing operators $\eta_{\pm \bm{e}_x}^{+}$ and $\eta_{\pm \bm{e}_y}^{+}$ with $\bm{e}_x=(1,0)$ and $\bm{e}_y=(0,1)$ being the unit vectors.
Based on the point-group symmetry of the square lattice, we classify the $\eta$-pairing operators into those of the irreducible representations.
Among them, the $\eta$-pairing operator with the $d_{x^2-y^2}$-wave ($d$-wave) pairing symmetry corresponds to
\begin{equation}
\eta_d^{+} = \eta_{+\bm{e}_x}^+ - \eta_{+\bm{e}_y}^+ + \eta_{-\bm{e}_x}^+ - \eta_{-\bm{e}_y}^+. \label{eq:deta_operator}
\end{equation}
which creates a spin-triplet pair occupying different sublattices, corresponding to the case of Eq.~\eqref{eq:diff_sublattice}.
For other pairing states (such as those having $s$-, $p_x$-, and $p_y$-wave symmetries) which can also be created by the off-site $\eta$-pairing operators, we refer to Sec.~\ref{sec:pairing_symmetry}.
Using the $\eta$-pairing operator $\eta_d^+$, we can define the $d$-wave $\eta$-pairing state as
\begin{equation}
|\Psi_d^N \rangle = \frac{1}{\mathcal{N}_d^N} (\eta_{d}^{+})^{N/2} |0\rangle, \label{eq:deta_state}
\end{equation}
where $N$ is the number of fermions (which is even), and $\mathcal{N}_d^N$ is the normalization constant (such that $\langle\Psi_d^N|\Psi_d^N\rangle=1$).

\subsection{Hubbard model with the two-body interaction} \label{sec:two-body_interaction}
To seek for a lattice model in which the $d$-wave $\eta$-pairing state becomes an eigenstate, let us first consider the ordinary two-dimensional spinful Hubbard model with the two-body interaction.
The Hamiltonian is written as
\begin{align}
&\mathcal{H}_{\mathrm{H}} = \mathcal{H}_{\mathrm{kin}} + \mathcal{H}_{U}, \label{eq:Hubbard_Hamiltonian} \\
&\mathcal{H}_{\mathrm{kin}} = -t \sum_{\langle i,j \rangle \sigma} (c_{i\sigma}^{\dagger} c_{j\sigma} + \mathrm{H.c.}), \label{eq:kint_Hamiltonian} \\
&\mathcal{H}_{U} = U \sum_i n_{i\uparrow} n_{i\downarrow}. \label{eq:U_Hamiltonian}
\end{align}
Here, $\langle i, j \rangle$ represents the sum over a pair of nearest-neighbor sites, $c_{i\sigma}$ is a short-hand notation of $c_{\bm{r_i},\sigma}$, and $n_{i\sigma}=c_{i\sigma}^{\dagger} c_{i\sigma}$ is the number operator.
The term $\mathcal{H}_{\mathrm{kin}}$ [Eq.~\eqref{eq:kint_Hamiltonian}] describes the nearest-neighbor hopping with $t$ being the transfer integral, and the term $\mathcal{H}_{U}$ [Eq.~\eqref{eq:U_Hamiltonian}] represents the on-site two-body interaction with the strength $U$.

One can quickly see that $\eta$-pairing states created by the off-site $\eta$-pair operators $\eta_{\bm \alpha}^+$ in Eq.~\eqref{eq:eta_operator} are zero-energy eigenstates of $\mathcal{H}_{\mathrm{kin}}$ in Eq.~\eqref{eq:kint_Hamiltonian} under the periodic boundary condition: we can rewrite the kinetic term and the $\eta$-pairing operator in the momentum basis as $\mathcal{H}_{\mathrm{kin}} = \sum_{\bm{k}\sigma} \varepsilon(\bm{k}) c_{\bm{k}\sigma}^{\dagger} c_{\bm{k}\sigma}$ and $\eta_{\bm{\alpha}}^+ = \sum_{\bm{k}} \exp(-\mathrm{i} \bm{k}\cdot\bm{\alpha}) c_{\bm{\pi}-\bm{k},\uparrow}^{\dagger} c_{\bm{k},\downarrow}^{\dagger}$, where $\varepsilon(\bm{k})=-2t (\cos k_x + \cos k_y)$ is the dispersion relation for the square lattice, $c_{\bm{k}\sigma} = (L_x L_y)^{-1/2} \sum_{i} c_{\bm{r}_i\sigma} \exp(-\mathrm{i} \bm{k} \cdot \bm{r}_i)$ is the Fourier transformed form of the annihilation operator with crystal momentum $\bm{k}$, and $L_x$ ($L_y$) is the system length in the $x$ ($y$) direction.
One can immediately see that $[ \mathcal{H}_{\mathrm{kin}},\,\eta_{\bm{\alpha}}^+ ] = \sum_{\bm{k}} \exp(-\mathrm{i}\bm{k}\cdot \bm{\alpha})\, [\varepsilon(\bm{k}) + \varepsilon(\bm{\pi}-\bm{k})] c_{\bm{\pi}-\bm{k},\uparrow}^{\dagger} c_{\bm{k},\downarrow}^{\dagger} = 0$, since the dispersion relation satisfies $\varepsilon(\bm{k}) + \varepsilon(\bm{\pi}-\bm{k})=0$.
Hence, the $d$-wave $\eta$-pairing state $|\Psi_d^N\rangle$ in Eq.~\eqref{eq:deta_state} is a zero-energy eigenstate for the kinetic term,
\begin{equation}
\mathcal{H}_{\mathrm{kin}} | \Psi_d^N\rangle = 0, \label{eq:eigenstate kinetic term}
\end{equation}
on the square lattice with the periodic boundary condition.
We note that any pairing states created by $\eta_{\bm \alpha}^+$ (including $s$-, $p_x$-, and $p_y$-wave pairing symmetries) are also eigenstates of $\mathcal{H}_{\mathrm{kin}}$.

\begin{figure}[t]
\centering
\includegraphics[width=1.0\columnwidth]{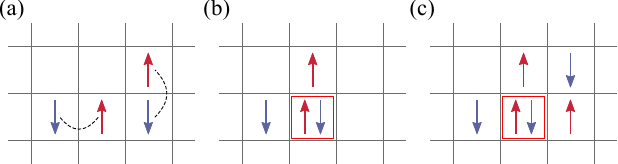}
\caption{
Examples of Fock states included in the $d$-wave $\eta$-pairing state [Eq.~\eqref{eq:deta_state}] on the square lattice: The case of (a) two pairs (connected by dashed curves) separated to each other, (b) two pairs overlapped to each other to form a doublon (enclosed by a red square), and (c) three pairs having a doublon surrounded by more than two particles.
Red (blue) arrows represent particles with spin up (down).
}
\label{fig:Fock basis}
\end{figure}
Next, we look at the interaction term $\mathcal{H}_U$ in Eq.~\eqref{eq:U_Hamiltonian}, for which we take the coordinate-space representation in the Fock basis.
Given a set of occupation numbers $\{\nu_{i\sigma}\}$, where $\nu_{i\sigma}$~($=0,1$) denotes the number of spin-$\sigma$ particles at site $i$, a Fock state is represented by
\begin{equation}
|\{ \nu_{i\sigma} \} \rangle = \prod_i \left(c^{\dagger}_{i\sigma}\right)^{\nu_{i\sigma}} |0\rangle. \label{eq:Fock basis}
\end{equation}
The $d$-wave $\eta$-pairing state is expanded as a linear combination of Fock states in which various oriented pairs are distributed on the square lattice.
Examples of those Fock states are shown in Fig.~\ref{fig:Fock basis}.
When two off-site pairs are present, they are either un-overlapped or overlapped, as shown in Fig.~\ref{fig:Fock basis}(a) and (b), respectively.
In the latter case, a doublon is formed, generating an energy cost due to the on-site two-body interaction.
This means that the latter Fock basis [Fig.~\ref{fig:Fock basis}(b)] has higher interaction energy (with the difference $U$) than the former basis [Fig.~\ref{fig:Fock basis}(a)].
Therefore the $d$-wave $\eta$-pairing state $|\Psi_d^N\rangle$ in Eq.~\eqref{eq:deta_state} is not an energy eigenstate of $\mathcal{H}_U$ [Eq.~\eqref{eq:U_Hamiltonian}].
When three (or more) pairs exist, more than two particles can occupy nearest-neighbor sites of a doublon, as shown in Fig.~\ref{fig:Fock basis}(c).
This example indicates that a naive application of the three-body interaction introduced in Sec.~\ref{sec:p-wave_eigenstate} is not sufficient to cancel the energy increment of the $d$-wave pairs due to the formation of doublons.
The question is how such an energy change can be canceled for all possible configurations of off-site pairs.

\subsection{Hubbard model with multibody interactions} \label{sec:multi-body_interaction}
Here we show that the unconventional $\eta$-pairing states can become exact energy eigenstates by introducing generalized multibody interactions to the Hubbard model.
As mentioned before, our motivation is to cancel the energy increment due to the formation of doublons by using the multibody interactions.
Since the number of particles that surround a doublon may vary depending on the configurations of pairs, but at least two are present (see Fig.~\ref{fig:Fock basis}), it will be convenient to introduce an operator that judges whether neighboring sites (next to a doublon) are occupied by particles or not.
This allows us to efficiently measure how much energy should be subtracted for each pair configuration to make the unconventional $\eta$-pairing states exact eigenstates.

\begin{figure}[t]
\centering
\includegraphics[width=1.0\columnwidth]{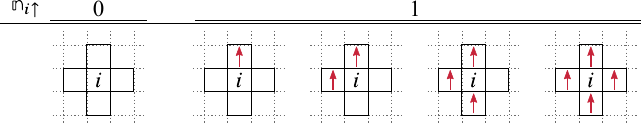}
\caption{
Values of the neighboring particle existence operator $\mathbb{n}_{i\sigma}$ in Eq.~\eqref{eq:bold_number_operator} with $\sigma=\uparrow$ for each configuration of spin-up particles (red arrows) at the nearest-neighbors sites of site $i$.
When there are no particles around site $i$, the operator takes a value of $0$.
Otherwise, if particles exist around site $i$, the operator takes a value of $1$.}
\label{fig:npe_operator}
\end{figure}

Such an operator, denoted by $\mathbb{n}_{i\sigma}$, is defined as follows.
In the coordinate basis spanned by Fock states in Eq.~\eqref{eq:Fock basis}, the operator $\mathbb{n}_{i\sigma}$ takes a value of 0 or 1 in such a way as
\begin{align}
\mathbb{n}_{i\sigma} = \begin{cases}
1 & [n_{\mathrm{nn}(i)\sigma}\neq 0] \\
0 & [n_{\mathrm{nn}(i)\sigma} = 0]
\end{cases}, \label{eq:bold_number_operator}
\end{align}
where $n_{\mathrm{nn}(i)\sigma}$ is the total number of spin-$\sigma$ particles occupying the nearest-neighbor sites of site $i$ [denoted by ${\mathrm{nn}}(i)$].
We call the operator $\mathbb{n}_{i\sigma}$ the neighboring particle existence operator.
In the case of the square lattice, we show the values of $\mathbb{n}_{i\sigma}$ (with $\sigma=\uparrow$) for each configuration of spin-up particles around site $i$ in Fig.~\ref{fig:npe_operator}.
Note that the values of $\mathbb{n}_{i\sigma}$ do not depend on the configuration of particles with the opposite spin $\overline{\!\sigma\!}$.
The operator in Eq.~\eqref{eq:bold_number_operator} tells us the presence (or absence) of a particle in the vicinity of a target site $i$, which can be used to avoid overcounting the number of neighboring particles, as described below.

The operator $\mathbb{n}_{i\sigma}$ in Eq.~\eqref{eq:bold_number_operator} is explicitly represented as a combination of multiples of the ordinary number operator $n_{i\sigma}$.
Let us label the nearest-neighbor sites of site $i$ by $i_1$, $i_2, \dots, i_z$, where $z$ is the coordination number of the site $i$.
Then $\mathbb{n}_{i\sigma}$ can be written as
\begin{align}
\mathbb{n}_{i\sigma}
&=
1-\prod_{m=1}^z (1-\hat n_{i_m \sigma})
\notag
\\
&=
-\sum_{m=1}^z (-1)^m \sideset{}{'}{\sum}_{j_1, \cdots, j_m \in \mathrm{nn}(i)} \hat{n}_{j_1\sigma} \cdots \hat{n}_{j_m\sigma},
\label{eq:definition_npe_operator}
\end{align}
where the prime summation means that $j_l$ ($l=1, \cdots, m$) runs over the nearest-neighbor sites of site $i$ with $j_l\neq j_{l'} (l<l')$, and different orders of $(j_1, j_2, \cdots, j_m)$ are not double-counted.
In general, $\mathbb{n}_{i\sigma}$ contains $m$-body operators with $1\le m\le z$.
The representation [Eq.~\eqref{eq:definition_npe_operator}] is applicable to arbitrary lattice structures.
If one performs the particle-hole transformation [$\tilde{c}_{j\sigma} = \exp(\mathrm{i}\bm{r}_j \cdot \bm{\pi})\, c_{j\sigma}^{\dagger}$], the expression for $\mathbb{n}_{i\sigma}$ is simplified to $\mathbb{n}_{i\sigma}=1-\prod_{j\in \mathrm{nn}(i)} \tilde{n}_{j\sigma}$ with $\tilde{n}_{j\sigma}=\tilde{c}_{j\sigma}^{\dagger} \tilde{c}_{j\sigma}$.

\begin{figure}[t]
\centering
\includegraphics[width=1.0\columnwidth]{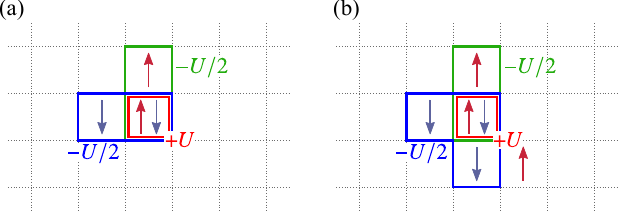}
\caption{
Interactions between off-site pairs in the $d$-wave $\eta$-pairing state in a spinful model with spin-$\uparrow$ (spin-$\downarrow$) particles indicated by the red (blue) arrows.
(a)~When two pairs are overlapped with each other to form a doublon, the on-site two-body interaction is canceled by the three-body interaction [Eq.~\eqref{eq:nd_ni}] acting between the doublon and a single particle.
(b)~When another pair comes next to the doublon, the on-site two-body interaction is not canceled by the three-body interaction [Eq.~\eqref{eq:nd_ni}], but is canceled by the multibody interaction in the form of Eq.~\eqref{eq:nd_bbn}.
Two particles forming a doublon are indicated by the red squares, while the multibody interacting doublon and spin-$\uparrow$ (spin-$\downarrow$) particles are enclosed by the blue (green) boxes.}
\label{fig:cancellation}
\end{figure}
In order to get an insight of how to cancel the energy increment of doublons with multibody interactions, we take two examples of Fock states relevant to the unconventional $\eta$-pairing states:
One is the case where two pairs are overlapped to each other to form a doublon, as shown in Fig.~\ref{fig:cancellation}(a).
The other is the case where there is another pair in the vicinity of the two pairs, as shown in Fig.~\ref{fig:cancellation}(b).
We label the site occupied by the doublon by $i$.
As shown in Fig.~\ref{fig:cancellation}(a), there are always two residual particles from the pairs that form the doublon, so that we can cancel the on-site two-body interaction $Un_{i\uparrow}n_{i\downarrow}$ by adding the following three-body interaction between the doublon and a particle,
\begin{equation}
-\frac{U}{2}\sum_{j\in \mathrm{nn}(i), \sigma} n_{i\uparrow}n_{i\downarrow} n_{j\sigma}, \label{eq:nd_ni}
\end{equation}
where $j$ runs over the nearest-neighbor sites of site $i$.
In Fig.~\ref{fig:cancellation}(b), however, the three-body interaction given by Eq.~\eqref{eq:nd_ni} amounts to $-3U/2$, which does not perfectly cancel the on-site two-body interaction $U$.
This is due to the overcounting of the three-body interaction energy from the neighboring pairs.
To avoid this overcounting, we replace $\sum_{j\in \mathrm{nn}(i)} n_{j\sigma}$ by $\mathbb{n}_{i\sigma}$ in Eq.~\eqref{eq:bold_number_operator} as follows:
\begin{equation}
-\frac{U}{2}\sum_{\sigma} n_{i\uparrow}n_{i\downarrow} \mathbb{n}_{i\sigma}. \label{eq:nd_bbn}
\end{equation}
The multibody interaction of Eq.~\eqref{eq:nd_bbn} contains $M$-body interactions with $3\le M\le z+2$.
With this form of the multibody interaction, it is possible to incorporate only the interaction between the doublon and a single spin-$\uparrow$ particle, and the interaction between the doublon and a single spin-$\downarrow$ particle.
One can see that the cancellation between the on-site two-body interaction $\mathcal{H}_U$ [Eq.~\eqref{eq:U_Hamiltonian}] and the multibody interaction [Eq.~\eqref{eq:nd_bbn}] works not only for the cases of Fig.~\ref{fig:cancellation}(a) and (b) but also for {\it all the possible configurations of pairs} in the unconventional $\eta$-pairing states, which can be easily checked since the cancellation occurs for each doublon one by one.

Based on the multibody interaction in Eq.~\eqref{eq:nd_bbn} introduced above, we define an extended Hubbard model having the unconventional $\eta$-pairing states as energy eigenstates.
The Hamiltonian of the extended Hubbard model, denoted by $\mathcal{H}_{\mathrm{extH}}$, is given as follows:
\begin{align}
&\mathcal{H}_{\mathrm{extH}} = \mathcal{H}_{\mathrm{kin}} + \mathcal{H}_{\mathrm{int}}, \label{eq:extended_Hubbard_Hamiltonian} \\
&\mathcal{H}_{\mathrm{int}} = U \sum_i n_{i\uparrow} n_{i\downarrow} \left( 1 - \frac{1}{2} \sum_{\sigma} \mathbb{n}_{i\sigma} \right). \label{eq:int_Hamiltonian}
\end{align}
As discussed in the previous paragraph, the interaction energy is always zero for the unconventional $\eta$-pairing states, due to the cancellation between the two-body interaction and the multibody interaction [Eq.~\eqref{eq:nd_bbn}].
In particular, the $d$-wave $\eta$-pairing state $|\Psi_d^N\rangle$ in Eq.~\eqref{eq:deta_state} is an exact eigenstate of $\mathcal{H}_{\mathrm{extH}}$ with a zero eigenenergy,
\begin{equation}
\mathcal{H}_{\mathrm{extH}} |\Psi_d^N \rangle = 0. \label{eq:energy_eigenstate}
\end{equation}
We remark that arbitrary $\eta$-pairing states created by $\eta_{\bm\alpha}^+$ [Eq.~\eqref{eq:eta_operator}] are also eigenstates of $\mathcal{H}_{\mathrm{extH}}$ (see Sec.~\ref{sec:pairing_symmetry}).
The same form of the Hamiltonian can be used for other bipartite lattices in arbitrary dimensions (e.g., the honeycomb lattice, the diamond lattice, etc.) to include unconventional $\eta$-pairing states as eigenstates.
The lattice structure must be bipartite, since the $\eta$-pairing states have staggered phases in real space in order to be eigenstates of the kinetic term $\mathcal{H}_{\mathrm{kin}}$.
Some of the examples are shown in Sec.~\ref{sec:generalization}.
So far, we have considered the case with the periodic boundary condition.
For the case of the open boundary condition, we refer to Sec.~\ref{sec:obc}.

We comment on the robustness against perturbations on the parameters of the multibody interactions.
When the two off-site pairs coexist, as shown in Fig.~\ref{fig:cancellation}(a), the three-body interaction in Eq.~\eqref{eq:nd_ni} is sufficient to cancel the on-site two-body interaction $U$ associated with the doublon formation.
However, when an additional pair comes next to this doublon, as shown in Fig.~\ref{fig:cancellation}(b), higher multibody interactions in Eq.~\eqref{eq:nd_bbn} become active.
These higher-order terms are generally required only when off-site pairs are densely distributed.
In cases with a low pair density, off-site pairs do not frequently come close to doublons, and higher multibody interactions become practically irrelevant.
Thus, we do not have to fine-tune the parameters of the higher multibody interactions.

Let us mention the relationship between our construction and previously proposed frameworks of QMBS states.
The obtained Hamiltonian $\mathcal{H}_{\mathrm{extH}}$ in Eq.~\eqref{eq:extended_Hubbard_Hamiltonian} satisfies the restricted spectrum-generating algebra~\cite{Moudgalya2022} in the Hilbert space spanned by a series of the $d$-wave $\eta$-pairing states $\{ |\Psi_d^N\rangle\, | \, N=2,4,\cdots \}$.
However, as shown in Eq.~\eqref{eq:definition_npe_operator}, the Hamiltonian $\mathcal{H}_{\mathrm{extH}}$ contains terms up to six-body density-density interactions in the case of the square lattice.
The conventional approach of assuming interaction forms and adjusting interaction coefficients to satisfy the restricted spectrum-generating algebra would be impractical due to the complexity of the commutation relations with the $d$-wave $\eta$-pairing operator in Eq.~\eqref{eq:deta_operator}.
The advantage of our approach is that we can systematically and directly construct model Hamiltonians that have off-site pair condensed states with unconventional pairing symmetry as their eigenstates in systems of higher dimensions and including spin degrees of freedom.

\section{Physical properties of the unconventional $\eta$-pairing states} \label{sec:eta}
In this section, we numerically confirm the nonthermal nature of the unconventional $\eta$-pairing states by computing two physical quantities: the superconducting correlation function and the entanglement entropy.
To obtain these quantities, we numerically calculated the entire exact wavefunction of the $d$-wave $\eta$-pairing state [Eq.~\eqref{eq:deta_state}] on a $L_x \times L_y$ square lattice with the periodic boundary condition.

\subsection{Off-diagonal long-range order} \label{sec:ODLRO}
We examine the nearest-neighbor pair correlation functions for the $d$-wave $\eta$-pairing state defined by
\begin{equation}
C_{\alpha\beta}(\bm{r}) = \langle \Psi_d^N| c_{\bm{r}+\bm{e}_{\alpha} \downarrow}^{\dagger} c_{\bm{r}\uparrow}^{\dagger} c_{\bm{0}\uparrow}  c_{\bm{e}_{\beta}\downarrow}  |\Psi_d^N \rangle, \label{eq:correlation_function}
\end{equation}
where the subscripts $\alpha$ and $\beta$ ($= x,y$) denote the orientation direction of the pairs, and $\bm{r}$ corresponds to the relative displacement between up-spin particles.
We consider a quarter-filled system with the lattice length $L_y=4$ and $L_x=4, 6, 8$ ($N=4, 6, 8$).
From the orientation dependence of the long-range pair correlations, one can discriminate different pairing symmetries.

\begin{figure}[t]
\centering
\includegraphics[width=\columnwidth]{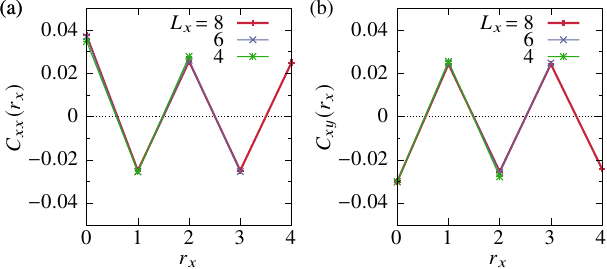}
\caption{
Pair correlation functions in Eq.~\eqref{eq:correlation_function} of the $d$-wave $\eta$-pairing state on the $L_x\times L_y$ lattice with $L_x=4,6,8$ and $L_y=4$.
(a)~The correlation between $x$-oriented pairs and (b) the correlation between $x$-oriented and $y$-oriented pairs
as a function of a distance $r_x$ between pairs along the $x$ axis.}
\label{fig:correlation_function}
\end{figure}
Figure~\ref{fig:correlation_function} shows the calculated pair correlation functions as a function of the distance along the $x$ axis (i.e., we set $\bm{r} = r_x \bm{e}_x$),
where Figs.~\ref{fig:correlation_function}(a) and (b) correspond to the parallel and orthogonal pair orientations, respectively.
We observe the staggered oscillations for both of the orientations, which do not quickly decay in a long distance (within the system size).
This is a direct consequence of the fact that the pairs carry a momentum $\bm{\pi}$ (and hence the staggered phase) in the $\eta$-pairing states in Eq.~\eqref{eq:eta_operator}.
We also see that the results of the pair correlations are well converged with respect to the system size with the particle density being fixed,
implying that the off-diagonal long-range order is present for the $d$-wave $\eta$-pairing state.
If one compares Fig.~\ref{fig:correlation_function}(a) with Fig.~\ref{fig:correlation_function}(b), one finds that the sign is reversed with respect to the pair orientation.
This reflects the $d$-wave pairing symmetry.

\subsection{Entanglement entropy} \label{sec:EE}
Next, we verify a nonthermal behavior of the $d$-wave $\eta$-pairing state through the entanglement entropy.
We consider the same $L_x \times L_y$ lattice as in the previous subsection with the periodic boundary condition.
The entire system (a torus) is divided into subsystems $A$ and $B$ (two tubes), where $A$ has the size of $r_x \times L_y$.
The entanglement entropy for the $d$-wave $\eta$-pairing state in Eq.~\eqref{eq:deta_state} is defined by 
$S_A(|\Psi_d^N\rangle)=-\mathrm{Tr}_{A}[ \rho_{A} \ln \rho_{A} ]$,
where $\rho_{A}$ is the reduced density matrix for the subsystem $A$, i.e., $\rho_A={\mathrm{Tr}}_{B} [|\Psi_d^N\rangle \langle \Psi_d^N|]$.
We compute $S_A(|\Psi_d^N\rangle)$ as a function of the volume fraction $f=V_A / V$, where $V=2L_x L_y$ ($V_A=2r_x L_y$) is the product of the system (subsystem) volume and the spin degrees of freedom.

The entanglement entropy of typical eigenstates of nonintegrable (chaotic) systems is expected to behave as the average entanglement entropy of a quantum pure-state ensemble, while that of integrable systems is to behave as that of a pure Gaussian state ensemble~\cite{Bianchi2022}.
In nonintegrable systems, the average entanglement entropy of a uniformly distributed pure state in a particle number conserving Hilbert space for a fermionic system is given by $\langle S_A \rangle_N = \sum_{N_A=0}^{\mathrm{min}(N,V_A)}  \left[ \langle S_A \rangle + \psi(d_N+1) - \psi(d_A d_B+1) \right] d_A d_B / d_N$, where $N_A$ is the particle number in the subspace $A$, $\langle S_A \rangle = \psi(d_A d_B + 1) - \psi(d_A+1) - (d_B-1)/2d_A$ is the Page formula in $d_A > d_B$~\cite{Page1993}, $\psi(x)$ is the digamma function, and $d_N=\binom{V}{N}$, $d_{A}=\binom{V_A}{N_A}$, and $d_B=\binom{V-V_A}{N-N_A}$ are the Hilbert-space dimensions of the entire system, subspace $A$, and complementary subspace $B$, respectively.
Similarly, in integrable systems, the average entanglement entropy over all pure fermionic Gaussian states in the thermodynamic limit is given by $\langle S_A \rangle_{G,N}= V \left\{ (n-1)\ln(1-n) + n[(f-1)\ln(1-f) - f\ln f -1] \right\}$~\cite{Bianchi2022}, where $n=N/V$ is the up-spin (or down-spin) particle density.

\begin{figure}[t]
\centering
\includegraphics[width=0.7\columnwidth]{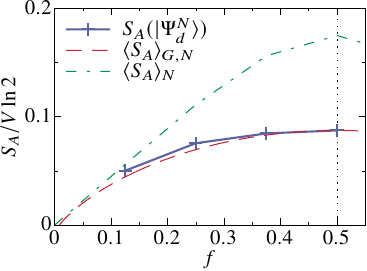}
\caption{
Entanglement entropy for fermionic systems on the $L_x\times L_y$ lattice as a function of the volume fraction $f=V_A/V$ with $L_x=8$, $L_y=4$, and the particle number $N=6$.
The blue cross points show the entanglement entropy of the $d$-wave $\eta$-pairing state.
The red dashed (green chained) curve represents the average entanglement entropy $\langle S_A \rangle_{G,N}$ ($\langle S_A \rangle_{N}$) typically obtained for a energy eigenstate in integrable (nonintegrable) systems.}
\label{fig:ee}
\end{figure}
Figure~\ref{fig:ee} shows the calculated entanglement entropy for the $d$-wave $\eta$-pairing state indicated by the blue cross points.
We also show the average entanglement entropy $\langle S_A \rangle_N$ and $\langle S_A \rangle_{G,N}$ with red dashed and green chained curves, respectively.
We observe that the entanglement entropy of the $d$-wave $\eta$-pairing state increases as the subsystem volume increases.
Particularly, we found that the subsystem size dependence of the entanglement entropy is in quantitative agreement with that typically observed in integrable systems, which is substantially smaller than the typical entanglement entropy in nonintegrable systems.
These numerical results support the nonthermal property of the $d$-wave $\eta$-pairing state.
In this study, due to the constraint on a computational cost, we have not examined the volume dependence of the entanglement entropy in the thermodynamic limit.
However, similarly to the Yang's $\eta$-pairing state~\cite{Vafek2017} and the spinless $\eta$-pairing state~\cite{Gotta2022}, we can expect that the entanglement entropy of the $d$-wave $\eta$-pairing state follows a logarithmic dependence on the subvolume.

\section{Nonintegrability of the multibody interacting model} \label{sec:scar}
In this section, we numerically confirm the nonintegrability of the extended Hubbard model $\mathcal{H}_{\mathrm{extH}}$ in Eq.~\eqref{eq:extended_Hubbard_Hamiltonian} from the energy level statistics.
Combining the results with those of Sec.~\ref{sec:eta}, we can judge whether the $d$-wave $\eta$-pairing state in the model is identified to be a QMBS state or not.
If the distribution of the difference between the nearest-neighbor eigenenergies follows the Wigner--Dyson distribution $P_{\mathrm{WD}}(s)=(\pi/2)s\exp(-\pi s^2/4)$ with $s$ being the level spacing, then the model is suggested to be nonintegrable and all the other energy eigenstates are thermal as expected from the argument of the ETH.

We consider the extended Hubbard model in Eq.~\eqref{eq:extended_Hubbard_Hamiltonian} on the square lattice under the periodic boundary condition with $L_x = L_y = 4$, $N=8$, and the total magnetization $S^z=0$.
We utilize the Lanczos method to numerically diagonalize the Hamiltonian.
Since the Hamiltonian has several global symmetries, we select a target subspace using symmetry projection operators.
As internal symmetries, we consider the time-reversal ($\mathcal{T}$) and spin-rotation symmetries.
It should be noted that our multibody interacting system does not have the particle-hole symmetry, unlike the two-body interacting Hubbard model in Eq.~\eqref{eq:Hubbard_Hamiltonian}.
Among the elements of the point group $C_{4v}$ (which is the symmetry of the square lattice), we take the reflections $\mathcal{P}_x$, $\mathcal{P}_y$, and $\mathcal{P}_d$ with respect to the $x$-, $y$-, and diagonal axes, respectively.
The fixed point of the point group operations is taken to be a site center, respecting the staggered phase oscillation of the $\eta$-pairing state.
Additionally, the system has the translation symmetries $\mathcal{X}$ and $\mathcal{Y}$ along the $x$ and $y$ directions, respectively.
We focus on the subspaces in which the $d$-wave $\eta$-pairing state exists, namely with $\mathcal{T}=+1$, the total spin $S=0$, $(\mathcal{P}_x, \mathcal{P}_y, \mathcal{P}_d) = (+1,+1,+1)$, and the total momentum $(0,0)$.
The last three are for the cases where $N/2$ is even.
The $d$-wave $\eta$-pairing state belongs to the total-spin eigenspace with $S=0,2,4$ when $N=8$.
Note that in a $4 \times 4$ periodic lattice system, there is an accidental hidden symmetry, i.e., the four-dimensional hypercubic symmetry~\cite{Bruus1997}.
If this higher symmetry is not taken into account to define the target space, the level-spacing distributions may behave differently from the Wigner--Dyson distribution (even though the system might be nonintegrable), which will be discussed later.

\begin{figure}[t]
\centering
\includegraphics[width=\columnwidth]{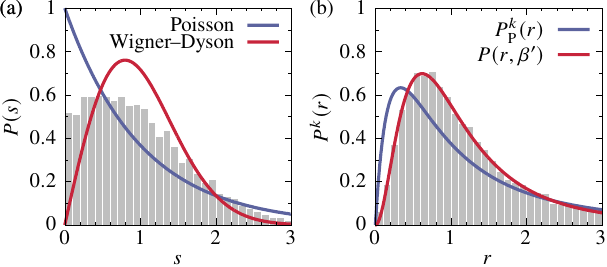}
\caption{
(a)~Nearest-neighbor level-spacing distribution $P(s)$ and (b)~higher-order spacing ratio distribution $P^{k=2}(r)$ (shown by the histograms) for the extended Hubbard model with the multibody interactions [Eq.~\eqref{eq:extended_Hubbard_Hamiltonian}] on the square lattice with the quantum numbers $(\mathcal{P}_x, \mathcal{P}_y, \mathcal{P}_d) = (+1,+1,+1)$.
The parameters are set to be $L_x = L_y = 4$, $N = 8$, and $U/t=2$.
}
\label{fig:level_statistics}
\end{figure}
Figure~\ref{fig:level_statistics}(a) shows the nearest-neighbor level-spacing distribution $P(s)$ in the subspace with the quantum numbers $(\mathcal{P}_x, \mathcal{P}_y, \mathcal{P}_d) = (+1,+1,+1)$, where we adopt the unfolding method with the Gaussian kernel density estimation~\cite{Gomez2002} with the smoothing parameter $\sigma_{\mathrm{Gauss}}=0.1$.
We find that the obtained level-spacing distribution has a shape in between the Poisson distribution $P_{\mathrm{P}}(s)=\exp(-s)$ and the Wigner--Dyson distribution $P_{\mathrm{WD}}(s)$.
A similar behavior is often observed when an extra discrete symmetry remains to be considered in a nonintegrable model.
For example, in the two-body interacting Hubbard model of Eq.~\eqref{eq:Hubbard_Hamiltonian} on the same $4 \times 4$ periodic lattice, the level-spacing distribution has been found to be neither the Poisson nor Wigner--Dyson distribution~\cite{Bruus1997}.
As we remarked before, this is due to the presence of the accidental hypercubic symmetry for the $4\times 4$ lattice geometry, which has not been taken into account in the level-spacing analysis.
Such an intermediate distribution between the Poisson and Wigner--Dyson ones has been reproduced by a spectrum of two independent GOE samples in the random matrix theory mixed with appropriate weights~\cite{Bruus1997}.

To extract the nonintegrability in systems with such a residual symmetry, we investigate higher-order spectral statistics~\cite{Tekur2020, Bhosale2021, Giraud2022}.
We define the distribution $P^k(r)$ of the $k$th order spacing ratio with $r = (E_{i+2k} - E_{i+k}) / (E_{i+k} - E_i)$ ($E_i$ is the $i$th energy eigenvalue).
Given superposed $m$ independent GOE samples, it has been numerically shown that $P^k(r) = P(r,\beta')$ holds with the relationship $\beta' = m = k$~\cite{Tekur2020}, where $P(r, \beta) = C_{\beta} \frac{(r + r^2)^{\beta}}{(1 + r + r^2)^{1 + 3\beta/2}}$ and $C_{\beta}$ is a normalization constant.
In the present case of $m = 2$, it is given by $C_2 = 81\sqrt{3}/4\pi$~\cite{Atas2013}.
In contrast, for integrable systems, the $k$th order spacing ratio is given by the distribution function $P^k_{\mathrm{P}}(r) = \frac{(2k-1)!}{[(k-1)!]^2} \frac{r^{k-1}}{(1+r)^{2k}}$~\cite{Tekur2020}.
Figure~\ref{fig:level_statistics}(b) presents a histogram of the numerically obtained distribution of the $k$th order spacing ratio ($k=2$) for the extended Hubbard model~\eqref{eq:extended_Hubbard_Hamiltonian}, under the same conditions as in Fig.~\ref{fig:level_statistics}(a).
The results confirm good agreement with $P(r, \beta')$, which shows the nonintegrability of the multibody interacting model of Eq.~\eqref{eq:extended_Hubbard_Hamiltonian}.

Given the above results with the nonthermal properties of the eigenstate seen in Sec.~\ref{sec:eta}, the $d$-wave $\eta$-pairing state [Eq.~\eqref{eq:deta_state}] can be regarded as a QMBS state in the nonintegrable system [Eq.~\eqref{eq:extended_Hubbard_Hamiltonian}].
The basic principle that makes the unconventional $\eta$-pairing states QMBSs is to use multibody interactions to protect the pairs from having the energy increase due to the two-body interaction.

\section{Generalizations} \label{sec:generalization}
In this section, we present several generalizations of our approach to construct superconducting scar states with the multibody interactions: generalized off-site $\eta$-pairing states (Sec.~\ref{sec:pairing_symmetry}), $f$-wave $\eta$-pairing states on the honeycomb lattice (Sec.~\ref{sec:honeycomb}), Yang's $\eta$-pairing states in a nearest-neighbor interacting system (Sec.~\ref{sec:s-wave}), and spinless $s$-wave $\eta$-pairing states in the one-dimensional system (Sec.~\ref{sec:p-wave_scar}).
Furthermore, we can also extend the form of the model Hamiltonian that has the unconventional $\eta$-pairing eigenstates.
In Sec.~\ref{sec:more multi-body}, we show that the $d$-wave $\eta$-pairing state also becomes an eigenstate in a model with higher multibody interaction.
In Sec.~\ref{sec:pair_number}, we argue that the eigenenergy of the unconventional $\eta$-pairing states can be controlled by an effective chemical potential in terms of pair numbers.
In Sec.~\ref{sec:obc}, we discuss the case of the open boundary condition, for which we determine the condition that the unconventional $\eta$-pairing states remain to be the eigenstates.

\subsection{General pairing symmetry and long-range pairing} \label{sec:pairing_symmetry}
The extended Hubbard model with the multibody interactions in Eq.~\eqref{eq:extended_Hubbard_Hamiltonian} has not only the $d$-wave $\eta$-pairing state but also several other $\eta$-pairing states with different pairing symmetries as energy eigenstates, since the energy cancellation holds for each Fock basis.
Let us consider $s$-, $p_x$-, and $p_y$-wave pairing symmetries, for which $\eta$-pairing operators are defined by
\begin{align}
&\eta_s^+ = \eta_{+\bm{e}_x}^+ + \eta_{+\bm{e}_y}^+ + \eta_{-\bm{e}_x}^+ + \eta_{-\bm{e}_y}^+, \label{eq:seta_operator} \\
&\eta_{p_x}^+= \eta_{+\bm{e}_x}^+ - \eta_{-\bm{e}_x}^+, \label{eq:pxeta_operator} \\
&\eta_{p_y}^+ = \eta_{+\bm{e}_y}^+ - \eta_{-\bm{e}_y}^+, \label{eq:pyete_operator}
\end{align}
respectively.
Acting each operator multiple times to the vacuum state $|0\rangle$ yields the $s$-, $p_x$-, and $p_y$-wave $\eta$-pairing states, as in the case of the $d$-wave symmetry.
Since these $\eta$-pairing states are represented by the same set of Fock bases in which the nearest-neighbor pairs are distributed, they are also energy eigenstates of $\mathcal{H}_{\mathrm{extH}}$ with zero eigenenergies, and hence are QMBS states.

Moreover, one can also combine these $\eta$-pairing operators with different pairing symmetries to define general $\eta$-pairing states,
\begin{equation}
| \nu_s \nu_{p_x} \nu_{p_y} \nu_{d} \rangle \propto \bigl( \eta_{s}^{+} \bigr)^{\nu_s} \bigl( \eta_{p_x}^{+} \bigr)^{\nu_{p_x}} \bigl( \eta_{p_y}^{+} \bigr)^{\nu_{p_y}} \bigl( \eta_{d}^{+} \bigr)^{\nu_{d}} |0\rangle, \label{eq:general_eta_states}
\end{equation}
where $\nu_{\gamma}$ ($\gamma=s, p_x, p_y,d$) denotes the number of pairs with the $\gamma$-wave pairing symmetry.
These $\eta$-pairing states are eigenstates of $\mathcal{H}_{\mathrm{extH}}$ in Eq.~\eqref{eq:extended_Hubbard_Hamiltonian}, which are $\binom{N/2+3}{3}$-fold degenerate with zero eigenenergy.
Note that an arbitrary linear combination of the general $\eta$-pairing states [Eq.~\eqref{eq:general_eta_states}] is also an eigenstate of $\mathcal{H}_{\mathrm{extH}}$ in Eq.~\eqref{eq:extended_Hubbard_Hamiltonian} (including, e.g., the $p_x+ip_y$-wave state and the $s+d$-wave state), since they are all degenerate at zero energy.
In order to induce the $d$-wave $\eta$-pairing state by a certain excitation protocol, it will be better if the $d$-wave pairing state is energetically separated from the other degenerate states.
In this paper, we leave it an open issue how to break the degeneracy, which would be effectively addressed by numerical methods for systematically constructing scarred Hamiltonians~\cite{Chertkov2018, Qi2019}.

While we have focused on the two-dimensional square lattice so far, most of the results can be straightforwardly extended to other dimensions and other bipartite lattice systems.
The bipartite lattice condition is necessary for the $\eta$-pairing states to be eigenstates of the kinetic term $\mathcal{H}_{\mathrm{kin}}$ [Eq.~\eqref{eq:kint_Hamiltonian}].
We can define the interaction term $\mathcal{H}_{\mathrm{int}}$ [Eq.~\eqref{eq:int_Hamiltonian}] and the neighboring particle existence operator $\mathbb{n}_{i\sigma}$ in Eqs.~\eqref{eq:bold_number_operator} and~\eqref{eq:definition_npe_operator} irrespective of dimensions and lattice structures.
Therefore one can use the Hamiltonian of Eq.~\eqref{eq:extended_Hubbard_Hamiltonian}, for example, on the one-dimensional chain, honeycomb, body-centered-cubic, and diamond lattices.

For the one-dimensional chain~\footnote{In the one-dimensional case, the spin-singlet and parity-odd $\eta$-pairing operator is mentioned in Ref.~\cite{Mark2020}.
Specifically, it is shown that the state obtained by acting with the $\eta_{p_x}^+$ operator [Eq.~\eqref{eq:pxeta_operator}] only once on Yang’s $\eta$-pairing state [Eq.~\eqref{eq:eta_state}] becomes an eigenstate of the Hirsch model.
To extend this result to states where the $\eta_{p_x}^+$ operator is applied multiple times or to higher-dimensional systems, the multibody interactions proposed in this study are required.}, we can obtain the spin-singlet $p$-wave pairing state $(\eta_{p_x}^+)^{N/2}|0\rangle$ with $\eta_{p_x}^+$ defined in Eq.~\eqref{eq:pxeta_operator}, as a QMBS state.
The scarred Hamiltonian consists of interaction terms involving up to a four-body interaction,
\begin{align}
\mathcal{H}_{\mathrm{int}} =& U\sum_i n_{i\uparrow}n_{i\downarrow} - \frac{U}{2} \sum_i n_{i\uparrow}n_{i\downarrow} (n_{i-1}+n_{i+1}) \nonumber \\
&+\frac{U}{2} \sum_i \sum_{\sigma} n_{i-1,\sigma} n_{i\uparrow}n_{i\downarrow} n_{i+1,\sigma}, \label{eq:1d_spinful_interaction_hamiltonian}
\end{align}
which is an explicit form of Eq.~\eqref{eq:int_Hamiltonian} in the case of the one-dimensional chain.
This state and model would be relevant for experimental demonstration, for example, with cold atoms in an optical lattice (see also Sec.~\ref{sec:summary}).

We can also generalize our construction to cases of long-range pairs.
Let $\bm{\alpha}$ refer to the $n$-th nearest-neighbor sites around site $i$, which is denoted by $\bm{\alpha} \in \mathrm{nn}(n,i)$.
The long-range $\eta$-pairing states are defined by $(\eta_{\bm\alpha}^+)^{N/2}|0\rangle$ with $\bm\alpha\in {\mathrm{nn}}(n,i)$.
The case of $n=1$ corresponds to the previous results.
On the square lattice, we can take $\bm{\alpha} = \pm \bm{e}_x \pm \bm{e}_y$ for $n=2$, and $\bm{\alpha} = \pm 2\bm{e}_x, \pm 2\bm{e}_y$ for $n=3$.
For example, in the case of $n=2$, we have the $d_{xy}$-wave spin-singlet $\eta$-pairing state $(\eta_{d_{xy}}^+)^{N/2}|0\rangle$ defined by
\begin{equation}
\eta_{d_{xy}}^+ = \eta_{+\bm{e}_x+\bm{e}_y}^+ - \eta_{-\bm{e}_x+\bm{e}_y}^+ + \eta_{-\bm{e}_x-\bm{e}_y}^+ - \eta_{+\bm{e}_x-\bm{e}_y}^+. \label{eq:dxy_eta_operator}
\end{equation}

To make the long-range $\eta$-pairing states exact eigenstates of the Hamiltonian, one should replace the form of the multibody interaction in Eq.~\eqref{eq:nd_bbn} by $-U\sum_{\sigma} n_{i\uparrow} n_{i\downarrow} \mathbb{n}_{i\sigma}^{(n)}/2$ with
\begin{align}
\mathbb{n}_{i\sigma}^{(n)}
&=
\begin{cases}
1 & [n_{{\mathrm{nn}}(n,i)\sigma}\neq 0] \\
0 & [n_{{\mathrm{nn}}(n,i)\sigma} = 0]
\end{cases},
\label{eq:bold n long-range}
\end{align}
where $n_{{\mathrm{nn}}(n,i)\sigma}$ is the number of spin-$\sigma$ particles on ${\mathrm{nn}}(n,i)$.
Using the ordinary density operator, one can represent $\mathbb{n}_{i\sigma}^{(n)}$ as $\mathbb{n}_{i\sigma}^{(n)}=1-\prod_{j\in \mathrm{nn}(n,i)} (1-\hat{n}_{j\sigma})$.

In this way, the $\eta$-pairing states, which consist of the long-range pairs and have different types of parity and spin symmetries, also become energy eigenstates due to the longer-range multibody interaction.
It should be noted that our method of realizing superconducting scar states using multibody interactions is only applicable to pairs of particles separated by a fixed distance.

\subsection{$f$-wave $\eta$-pairing state on the honeycomb lattice} \label{sec:honeycomb}
\begin{figure}[t]
\centering
\includegraphics[width=\columnwidth]{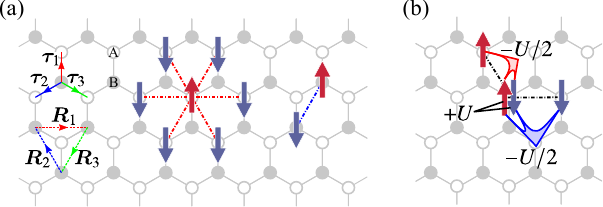}
\caption{
Generalized $\eta$-pairing states on the honeycomb lattice.
(a)~Six kinds of next-nearest-neighbor pairs (connected by the dash-dotted lines) in the $\eta$-pairing states.
The white (grey) circles represent $\mathrm{A}$ ($\mathrm{B}$) sublattice sites.
$\bm{\tau}_{\gamma}$ and $\bm{R}_{\gamma}$ ($\gamma=1,2,3$) denote the bond and lattice vectors, respectively.
A pair on the $\mathrm{B}$ sublattice (connected by the blue dash-dotted line) has the opposite sign of the wavefunction with respect to those on the $\mathrm{A}$ sublattice.
(b)~Energy cancellation between the two-body (black lines) and multibody (filled areas) interactions acting on overlapped pairs.}
\label{fig:honeycomb}
\end{figure}
Let us consider the multibody interacting model defined by Eq.~\eqref{eq:extended_Hubbard_Hamiltonian} on the honeycomb lattice, where an $\eta$-pairing state with the $f$-wave pairing symmetry becomes an eigenstate [see Fig.~\ref{fig:schematic}(c)].
The lattice structure is defined as follows.
We label sublattices by $\mathrm{A}$ and $\mathrm{B}$, as shown in Fig.~\ref{fig:honeycomb}(a).
Let $\bm{\tau}_{\gamma}$ ($\gamma=1,2,3$) be the bond vectors and $\bm{R}_{\gamma}$ the lattice vectors, which are explicitly given by
\begin{alignat}{3}
&\bm{\tau}_1 = 
\left(
\begin{array}{c}
0 \\
1/\sqrt{3}
\end{array}
\right),\ &
&\bm{\tau}_2 = 
\left(
\begin{array}{c}
-1/2 \\
-1/2\sqrt{3}
\end{array}
\right),\ & 
&\bm{\tau}_3 = 
\left(
\begin{array}{c}
1/2 \\
-1/2\sqrt{3}
\end{array}
\right), \label{eq:tau_honeycomb} \\
&\bm{R}_1 = \bm{\tau}_3 - \bm{\tau}_2,& &\bm{R}_2 = \bm{\tau}_1 - \bm{\tau}_3,& &\bm{R}_3 = \bm{\tau}_2 - \bm{\tau}_1. \label{eq:R_honeycomb}
\end{alignat}
The Fourier transform of the annihilation operator $c^{l}_{i\sigma}$ of fermions for each sublattice $l$ ($=\mathrm{A},\mathrm{B}$) is given by $c^{l}_{i\sigma} = N_{\mathrm{s}}^{-1/2}\sum_{\bm{k}} \exp(\mathrm{i}\bm{k}\cdot \bm{r}_i)\, c^{l}_{\bm{k}\sigma}$ with $N_{\mathrm{s}}$ being the total number of unit cells.
After the Fourier transformation, the kinetic Hamiltonian in Eq.~\eqref{eq:kint_Hamiltonian} is diagonalized as
\begin{align}
\mathcal{H}_{\mathrm{kin}} &= \sum_{\bm{k}\sigma} 
\left( {c^{\mathrm{A}}_{\bm{k}\sigma}}^{\dagger},\  {c^{\mathrm{B}}_{\bm{k}\sigma}}^{\dagger} \right)   
\left( \begin{array}{cc}
0 & g(\bm{k}) \\
g^*(\bm{k}) & 0 \\
\end{array} \right)
\left( \begin{array}{c}
{c^{\mathrm{A}}_{\bm{k}\sigma}} \\[5pt]
{c^{\mathrm{B}}_{\bm{k}\sigma}}
\end{array} \right),\label{eq:kint_honeycomb_AB} \\
&= \sum_{\bm{k}\sigma} |g(\bm{k})| \left[ {c^{(+)}_{\bm{k}\sigma}}^{\dagger} {c^{(+)}_{\bm{k}\sigma}} - {c^{(-)}_{\bm{k}\sigma}}^{\dagger} {c^{(-)}_{\bm{k}\sigma}} \right], \label{eq:kint_honeycomb_+-}
\end{align}
where the hopping amplitude $g(\bm{k})$ and the creation operators of the energy eigenstates labeled by $(+)$ and $(-)$ are given by
\begin{align}
g(\bm{k}) &= \sum_{\gamma=1,2,3} \mathrm{e}^{\mathrm{i}\bm{k}\cdot \bm{\tau}_{\gamma}}, \label{eq:g-factor} \\
\left( {c^{(+)}_{\bm{k}\sigma}}^{\dagger},\ {c^{(-)}_{\bm{k}\sigma}}^{\dagger} \right) &= \left( {c^{\mathrm{A}}_{\bm{k}\sigma}}^{\dagger},\  {c^{\mathrm{B}}_{\bm{k}\sigma}}^{\dagger} \right)   
\frac{1}{\sqrt{2}|g(\bm{k})|} \left( \begin{array}{cc}
g(\bm{k}) & g(\bm{k}) \\
|g(\bm{k})| & -|g(\bm{k})| \\
\end{array} \right), \label{eq:eigenvector_honeycomb}
\end{align}
respectively.

On the honeycomb lattice, the off-site $\eta$-pairing operators can be defined as [see Eq.~\eqref{eq:general_eta_operator}]
\begin{equation}
\eta^{+}_{\bm{\alpha}} = \sum_{i}  f(\bm{r}_i) c_{\bm{r}_i,\uparrow}^{\dagger} c_{\bm{r}_i+\bm{\alpha},\downarrow}^{\dagger}. \label{eq:honeycomb_eta_operator}
\end{equation}
Here, the direction vector of the pairs is given by $\bm{\alpha} = \pm \bm{R}_{\gamma}$ ($\gamma=1,2,3$), which point to the second nearest-neighbor sites, namely $\bm{\alpha} \in \mathrm{nn}(2,i)$.
These six $\eta$-pairing operators create the next-nearest neighbor pairing states with the staggered phase factors, as shown in Fig.~\ref{fig:honeycomb}(a).
Since the paired fermions are on the same sublattices [as in the case of Eq.~\eqref{eq:same_sublattice}], these $\eta$-pairing operators show the even sublattice symmetry.

The unconventional $\eta$-pairing states on the honeycomb lattice can be constructed from the $\eta$-pairing operators in Eq.~\eqref{eq:honeycomb_eta_operator}.
As an example, let us introduce an $\eta$-pairing state with the spin-triplet $f$-wave pairing symmetry (see Table~\ref{table:symmetry}) as follows:
\begin{align}
&\eta_{f}^+ = \eta_{+\bm{R}_1}^+ - \eta_{-\bm{R}_3}^+ + \eta_{+\bm{R}_2}^+
-\eta_{-\bm{R}_1}^+ + \eta_{+\bm{R}_3}^+ - \eta_{-\bm{R}_2}^+, \label{eq:f-wave_eta_operator} \\
&|\Psi_f^N \rangle \propto \left( \eta_{f}^+ \right)^{N/2} |0\rangle. \label{eq:f-wave_eta_state}
\end{align}
The following discussion also holds for other pairing symmetries as discussed in Sec.~\ref{sec:pairing_symmetry}.

The $\eta$-pairing operators on the honeycomb lattice commute with the kinetic Hamiltonian $\mathcal{H}_{\mathrm{kin}}$ in Eqs.~\eqref{eq:kint_honeycomb_AB} and~\eqref{eq:kint_honeycomb_+-}.
In the momentum representation, the $\eta$-pairing operators in Eq.~\eqref{eq:honeycomb_eta_operator} are expressed as
\begin{align}
\eta^{+}_{\bm{\alpha}} &= \sum_{\bm{k}} \mathrm{e}^{\mathrm{i}\bm{k}\cdot\bm{\alpha}} \left( {c_{\bm{k},\uparrow}^{\mathrm{A}}}^{\dagger} {c_{-\bm{k},\downarrow}^{\mathrm{A}}}^{\dagger}
- {c_{\bm{k},\uparrow}^{\mathrm{B}}}^{\dagger} {c_{-\bm{k},\downarrow}^{\mathrm{B}}}^{\dagger} \right), \label{eq:eta-operator_AB_k} \\
&= \sum_{\bm{k}} \mathrm{e}^{\mathrm{i}\bm{k}\cdot\bm{\alpha}} \left( {c_{\bm{k},\uparrow}^{(+)}}^{\dagger} {c_{-\bm{k},\downarrow}^{(-)}}^{\dagger}
+{c_{\bm{k},\uparrow}^{(-)}}^{\dagger} {c_{-\bm{k},\downarrow}^{(+)}}^{\dagger} \right). \label{eq:eta-operator_+-_k}
\end{align}
The commutation relation between $\mathcal{H}_{\mathrm{kin}}$ [Eq.~\eqref{eq:kint_honeycomb_AB}] and $\eta_{\bm\alpha}^+$ [Eq.~\eqref{eq:eta-operator_AB_k}] are evaluated as
\begin{align}
\left[ \mathcal{H}_{\mathrm{kin}}, \eta_{\bm{\alpha}}^+ \right] =& \sum_{\bm{k}} \mathrm{e}^{\mathrm{i}\bm{k}\cdot\bm{\alpha}} \Bigl( g^*(\bm{k}) {c^{\mathrm{B}}_{\bm{k}\uparrow}}^{\dagger} {c^{\mathrm{A}}_{-\bm{k}\downarrow}}^{\dagger}
+g^*(-\bm{k}) {c^{\mathrm{A}}_{\bm{k}\uparrow}}^{\dagger} {c^{\mathrm{B}}_{-\bm{k}\downarrow}}^{\dagger} \nonumber \\
&-g(\bm{k}) {c^{\mathrm{A}}_{\bm{k}\uparrow}}^{\dagger} {c^{\mathrm{B}}_{-\bm{k}\downarrow}}^{\dagger}
-g(-\bm{k}) {c^{\mathrm{B}}_{\bm{k}\uparrow}}^{\dagger} {c^{\mathrm{A}}_{-\bm{k}\downarrow}}^{\dagger} \Bigr). \label{eq:commutation_H_eta_AB}
\end{align}
Given $g(\bm{k}) = g^{*}(-\bm{k})$, the first and fourth terms and the second and third terms cancel each other, indicating that $[\mathcal{H}_{\mathrm{kin}}, \eta_{\bm\alpha}^+]=0$.
This is also confirmed from the commutation relation between Eqs.~\eqref{eq:kint_honeycomb_+-} and~\eqref{eq:eta-operator_+-_k} as $\left[ \mathcal{H}_{\mathrm{kin}}, \eta_{\bm{\alpha}}^+ \right]= \sum_{\bm{k}} \exp(\mathrm{i}\bm{k}\cdot\bm{\alpha})\, (|g(\bm{k})| - |g(-\bm{k})|) ( {c_{\bm{k},\uparrow}^{(+)}}^{\dagger} {c_{-\bm{k},\downarrow}^{(-)}}^{\dagger}
+{c_{\bm{k},\uparrow}^{(-)}}^{\dagger} {c_{-\bm{k},\downarrow}^{(+)}}^{\dagger} ) = 0$.
Therefore the $f$-wave $\eta$-pairing state is a zero-energy eigenstate of the kinetic term in Eq.~\eqref{eq:kint_Hamiltonian}.

By applying the neighboring particle existence operator $\mathbb{n}_{i\sigma}(2)$ defined in Eq.~\eqref{eq:bold n long-range} to the case of $\mathrm{nn}(2,i)$, the $f$-wave $\eta$-pairing state also becomes an energy eigenstate of the Hamiltonian in Eq.~\eqref{eq:extended_Hubbard_Hamiltonian}.
The energy increase due to the two-body interaction $U$ is canceled by the multibody interactions between a doublon and a next-nearest neighboring particle, as shown in Fig.~\ref{fig:honeycomb}(b).
Therefore, using the multibody interactions, we can realize the $\eta$-pairing states with unconventional pairing symmetries in various lattice systems.

\subsection{Yang's $s$-wave $\eta$-pairing state in a nearest-neighbor interacting system} \label{sec:s-wave}
Yang's $\eta$-pairing state is no longer an energy eigenstate when one adds nearest-neighbor two-body interactions to the Hubbard model in Eq.~\eqref{eq:Hubbard_Hamiltonian} due to the breaking of $\eta$-$\mathrm{SU}(2)$ symmetry~\cite{Yang1990, Vafek2017}.
However, one can make Yang's $\eta$-pairing state the energy eigenstate in nearest-neighbor interacting systems by utilizing appropriate multibody interactions like those discussed above.
In Ref.~\cite{Mark2020}, it has been studied how to make Yang's $\eta$-pairing state a QMBS state, where the multibody interaction Hamiltonian presented below can be reproduced.
Here, we include this example to show the role of the multibody interaction discussed above.
The following discussions in this subsection are applicable to systems in arbitrary dimensions and with both the periodic and open boundary conditions.

\begin{figure}[t]
\centering
\includegraphics[width=\columnwidth]{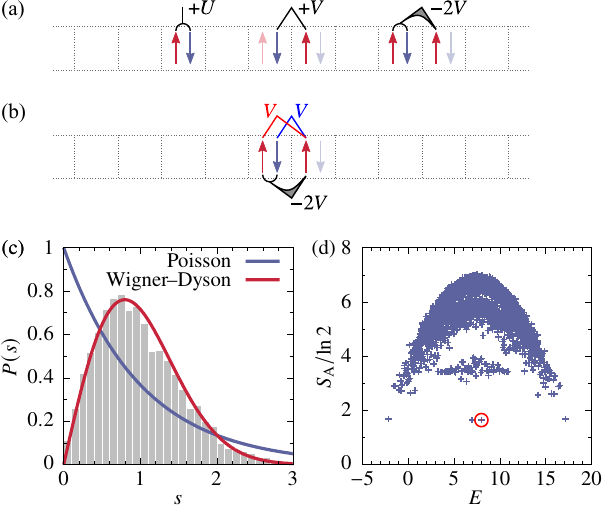}
\caption{
(a) Interactions between doublons in Yang's $s$-wave $\eta$-pairing state in the model with the on-site interaction $U$ and the nearest-neighbor two-body and three-body interactions in the form of Eq.~\eqref{eq:doublon_int_Hamiltonian}.
(b)~Cancellation between the two-body and three-body interactions.
(c)~Nearest-neighbor level-spacing distribution $P(s)$ in the model $\mathcal{H}_{\mathrm{H}}+\mathcal{H}_{\mathrm{mod}V}$ with $L=12$.
(d)~Entanglement entropy spectrum of the energy eigenstates in the corresponding model with $L=8$.
The red circle indicates Yang's $\eta$-pairing state.
The parameters are set to be $t=1$, $U=2$, $V=1$ and $N=L$.}
\label{fig:Yang}
\end{figure}
Let us consider a $d$-dimensional periodic bipartite lattice, in which Yang's $\eta$-pairing state is defined as
\begin{align}
&\eta^{+} = \sum_{i} \mathrm{e}^{\mathrm{i} \bm{\pi} \cdot \bm{r}_i} c_{i\uparrow}^{\dagger} c_{i\downarrow}^{\dagger}, \label{eq:Yang_eta_operator} \\
&|\Psi^{N} \rangle = \frac{1}{\mathcal{N}^N} (\eta^{+})^{N/2} |0\rangle,
\label{eq:eta_state}
\end{align}
with the normalization constant $\mathcal{N}^N$ (such that $\langle \Psi^N|\Psi^N\rangle=1$).
Yang's $\eta$-pairing state is a linear combination of Fock states with various configurations of doublons.
It is known that this $\eta$-pairing state is the energy eigenstate of the Hubbard model in Eq.~\eqref{eq:Hubbard_Hamiltonian}, i.e., $\mathcal{H}_{\mathrm{H}} | \Psi^{N} \rangle = NU/2 |\Psi^{N} \rangle$~\cite{Yang1989}.
However, Yang's $\eta$-pairing state is no longer an eigenstate if one includes a nearest-neighbor interaction defined by
\begin{equation}
\mathcal{H}_V = \frac{V}{2} \sum_{\langle i,j\rangle} n_i n_j, \label{eq:V_Hamiltonian}
\end{equation}
with $n_i=\sum_\sigma n_{i\sigma}$.
As shown in Fig.~\ref{fig:Yang}(a), two doublons placed at adjacent sites gain additional energy of $4V$.
Due to this, the interaction energy for each Fock state may change depending on the doublon configuration, which prevents Yang's $\eta$-pairing state from being an eigenstate of the Hubbard model with the nearest-neighbor interaction of Eq.~\eqref{eq:V_Hamiltonian}.

If we consider three-body interactions, on the other hand, we notice that an interaction between a doublon and a nearby particle can be used to cancel the nearest-neighbor two-body interaction of Eq.~\eqref{eq:V_Hamiltonian}, as shown in Fig.~\ref{fig:Yang}(b).
In Yang's $\eta$-pairing state, the number of particles occupying each site is equal to twice the number of doublons at the same site, since the particles always exist as doublons.
From this fact, one can see that Yang's $\eta$-pairing state satisfies
\begin{align}
&\left( \mathcal{H}_{\mathrm{H}} + \mathcal{H}_{\mathrm{mod}V} \right) |\Psi^{N} \rangle = \frac{N}{2}U |\Psi^{N} \rangle,\label{eq:doublon_eigenstate}\\
&\mathcal{H}_{\mathrm{mod}V} = \frac{V}{2} \sum_{\langle i,j \rangle} n_{i} \left( n_j - 2 n_{j\uparrow}n_{j\downarrow} \right).\label{eq:doublon_int_Hamiltonian}
\end{align}
Therefore, utilizing the multibody interaction, one can make Yang's $\eta$-pairing state an energy eigenstate even in the presence of the nearest-neighbor interaction of Eq.~\eqref{eq:V_Hamiltonian}.

We numerically confirm that Yang's $\eta$-pairing state can be regarded as a QMBS state
in the model with the Hamiltonian $\mathcal{H}_{\mathrm{H}} + \mathcal{H}_{\mathrm{mod}V}$ in Eq.~\eqref{eq:doublon_eigenstate}.
For the half-filled one-dimensional system with the periodic boundary condition of length $L$, we numerically obtain the level-spacing distribution $P(s)$ [Fig.~\ref{fig:Yang}(c)] in the subspace with even parity, even time-reversal symmetry, and zero total momentum.
The level-spacing distribution agrees well with the Winger--Dyson distribution $P_{\mathrm{WD}}(s)$, implying that the model ($\mathcal{H}_{\mathrm{H}} + \mathcal{H}_{\mathrm{mod}V}$) is nonintegrable.
We also consider the corresponding model with the open boundary condition, for which the bipartite entanglement entropy for each energy eigenstate is shown in Fig.~\ref{fig:Yang}(d).
We find that Yang's $\eta$-pairing state, indicated by the red circle, has much smaller entanglement entropy than the other thermal states (with $S_{\mathrm{A}} \gtrsim 3 \ln 2$).
Therefore we conclude that Yang's $\eta$-pairing state is a nonthermal energy eigenstate (i.e., a QMBS state) in the multibody interacting system with the Hamiltonian $\mathcal{H}_{\mathrm{H}} + \mathcal{H}_{\mathrm{mod}V}$.

\subsection{The spinless $s$-wave $\eta$-pairing state in the one-dimensional system} \label{sec:p-wave_scar}
\begin{figure}[t]
\centering
\includegraphics[width=\columnwidth]{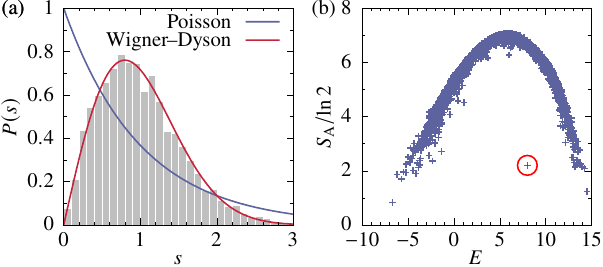}
\caption{
(a)~Nearest-neighbor level-spacing distribution $P(s)$ for the one-dimensional spinless Hubbard model with the three-body interaction [Eq.~\eqref{eq:spinless_hamiltonian}] with $L=20$.
The blue and red curves correspond to the Poisson and Wigner--Dyson distribution, respectively.
(b)~Entanglement entropy spectrum of the energy eigenstates in the corresponding model with $L=16$.
The red circle indicates the spinless $s$-wave $\eta$-pairing state [Eq.~\eqref{eq:peta_state}].
The parameters are set to be $t=1$, $V=2$, $W=-V/2$, $N=L/2$, and $\sigma_{\mathrm{Gauss}}=0.5$.}
\label{fig:p-wave}
\end{figure}
The spinless $s$-wave $\eta$-pairing state in Eq.~\eqref{eq:peta_state} discussed in Sec.~\ref{sec:p-wave_eigenstate} is one of the simplest examples that show unconventional pairing symmetries, since the conventional pairings have the spin-singlet and $s$-wave symmetry.
Let us first check that the spinless $s$-wave $\eta$-pairing state satisfies the conditions for a QMBS state in the extended spinless Hubbard model with the three-body interaction in Eq.~\eqref{eq:spinless_hamiltonian}.
Figure~\ref{fig:p-wave}(a) shows the level-spacing distribution $P(s)$ of the energy eigenstates on the half-filled periodic lattice in the subspace with even parity and total momentum $\pi$.
We observe good agreement between $P(s)$ and the Winger--Dyson distribution.
Figure~\ref{fig:p-wave}(b) shows the bipartite entanglement entropy spectrum in the open boundary chain at half filling.
We find that the spinless $s$-wave $\eta$-pairing state, indicated by the red circle, has much smaller entanglement entropy than those of the other eigenstates that form a convex upward curve.
Therefore the spinless $s$-wave $\eta$-pairing state can be regarded as a QMBS state in the multibody interacting system.

Next, we clarify the relationship between the spinless Hamiltonian in Eq.~\eqref{eq:spinless_hamiltonian} and our construction of scarred spinful Hamiltonians using the operator $\mathbb{n}_{i\sigma}$ in Eq.~\eqref{eq:bold_number_operator} introduced in Sec.~\ref{sec:multi-body_interaction}.
The spinless system can be described by a spinful model in which all the spins of the particles are polarized to the same direction (say, $\sigma$).
The spinless $\eta$-pairing operator $\eta_{\mathrm{sl}}^+$ in Eq.~\eqref{eq:peta_operator} corresponds to the general $\eta$-pairing operator in Eq.~\eqref{eq:general_eta_operator} where two fermions have the same $\sigma$-spin, i.e., $\eta_{\mathrm{sl}}^+=\eta_{+1,\sigma\sigma}^+$.
The two-body interaction term $\mathcal{H}_V$ in Eq.~\eqref{eq:ham_V} corresponds to a nearest-neighbor interaction in a spinful model, $\sum_i n_{i\sigma} n_{i+1,\sigma}$.
The scarred Hamiltonian including the three-body interaction in Eq.~\eqref{eq:ham_modW} can be rewritten as $V \sum_i n_{i\sigma} \mathbb{n}_{i\sigma}/2$ using the neighboring particle existence operator $\mathbb{n}_{i\sigma}$ in the one-dimensional chain, which is explicitly represented by $\mathbb{n}_{i\sigma} = 1-(1-n_{i-1,\sigma})(1-n_{i+1,\sigma}) = n_{i-1,\sigma} + n_{i+1,\sigma} - n_{i-1,\sigma}n_{i+1,\sigma}$.
The term $V \sum_i n_{i\sigma} \mathbb{n}_{i\sigma}/2$ means that each fermion pair always has the constant energy $V$ in any pair configuration, no matter how close the pairs are coming to each other.
We will discuss the spinful version of this term in more detail in Sec.~\ref{sec:pair_number}.

\subsection{Higher multibody interactions} \label{sec:more multi-body}
\begin{figure}[t]
\centering
\includegraphics[width=0.5\columnwidth]{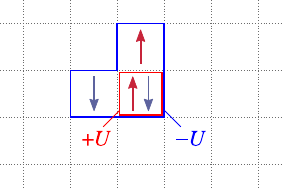}
\caption{
Energy cancellation between the two-body on-site interaction (red box) and the multibody interaction acting among a doublon and two neighboring particles (blue box).}
\label{fig:four-body}
\end{figure}
To achieve superconducting scar states, it is also possible to exploit higher multibody interactions (i.e., $M$-body interactions with larger $M$).
In the case of the $d$-wave $\eta$-pairing state on the square lattice (discussed in Sec.~\ref{sec:model}), we find that a doublon created by an overlap between two pairs is always accompanied by two particles around the doublon.
We can cancel the on-site two-body interaction with a multibody interaction acting among a doublon and two associated particles, as shown by the blue box in Fig.~\ref{fig:four-body}.
Motivated by this observation, we consider the following interaction term in the Hamiltonian, 
\begin{equation}
\mathcal{H}_{\mathrm{int}}' = U\sum_i n_{i\uparrow} n_{i\downarrow} \left[ 1 - \mathbb{n}_{i\uparrow} \mathbb{n}_{i\downarrow} \right],\label{eq:4body-interaction}
\end{equation}
which contains up to $(2z+2)$-body interactions.
The $d$-wave $\eta$-pairing state is also an exact energy eigenstate of the model with the Hamiltonian $\mathcal{H}_{\mathrm{kin}}+\mathcal{H}_{\mathrm{int}}'$.
Here we need to use the neighboring particle existence operator $\mathbb{n}_{i\sigma}$ [Eq.~\eqref{eq:bold_number_operator}] to correctly cancel the energy increment of overlapped pairs, as discussed in Sec.~\ref{sec:multi-body_interaction}.

\subsection{Pair number operator} \label{sec:pair_number}
\begin{figure}[t]
\centering
\includegraphics[width=1.0\columnwidth]{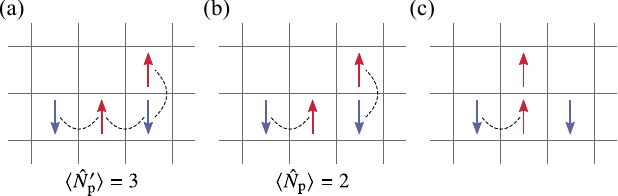}
\caption{
Number of off-site pairs in examples of Fock states.
(a)~Expectation value of $\hat{N}'_{\mathrm{p}}$ in Eq.~\eqref{eq:pre_pair-number-operator} for the case of two neighboring off-site pairs.
(b)~Expectation value of $\hat{N}_{\mathrm{p}}$ in Eq.~\eqref{eq:pair_number_operator} for the same state as in (a).
(c)~The case in which one pair is broken.
The number of particles is given by $\langle \hat{N} \rangle/2 = 2$.}
\label{fig:pair number}
\end{figure}
The unconventional $\eta$-pairing states studied in this paper have high eigenenergies as compared to the ground-state energy.
From a practical point of view, it will be convenient if we could control the eigenenergies of those pairing states.
To this end, we introduce an operator that counts the number of pairs, extending the concept of the number operator for particles ($\hat N=\sum_{i\sigma} n_{i\sigma}$).
Since off-site pairs considered in this paper are composed of two nearest-neighbor particles, one might think that the following operator could be used:
\begin{equation}
\hat{N}'_{\mathrm{p}} = \frac{1}{2} \sum_{\langle i,j \rangle\sigma} n_{i\sigma} n_{j\overline{\!\sigma\!}}, \label{eq:pre_pair-number-operator}
\end{equation}
with $\overline{\!\sigma\!}=\downarrow$ ($\uparrow$) for $\sigma=\uparrow$ ($\downarrow$).
This operator is a part of the nearest-neighbor interaction in Eq.~\eqref{eq:V_Hamiltonian}.
However, as discussed in Sec.~\ref{sec:two-body_interaction}, the overcounting occurs when pairs are close to each other.
Figure~\ref{fig:pair number}(a) shows one example of Fock states representing the $d$-wave $\eta$-pairing state.
Although there are two pairs, the expectation value of $\hat{N}'_{\mathrm{p}}$ is $3$.
Indeed, the appropriate operator to count the number of pairs is given by
\begin{equation}
\hat{N}_{\mathrm{p}} = \frac{1}{2} \sum_{i\sigma} n_{i\sigma} \mathbb{n}_{i\overline{\!\sigma\!}}. \label{eq:pair_number_operator}
\end{equation}
Using the neighboring particle existence operator $\mathbb{n}_{i\sigma}$ [Eq.~\eqref{eq:bold_number_operator}], we can measure the number of pairs correctly, as shown in Fig.~\ref{fig:pair number}(b).
The $\eta$-pairing states become an eigenstate of $\hat{N}_{\mathrm{p}}$:
\begin{equation}
\hat{N}_{\mathrm{p}} |\Psi_{\bm{\alpha}}^N \rangle = \frac{N}{2} | \Psi_{\bm{\alpha}}^N \rangle, \label{eq:pair_number_eigenstate}
\end{equation}
where $|\Psi_{\bm\alpha}^N\rangle\propto (\eta_{\bm\alpha}^+)^{N/2}|0\rangle$.

The operator $\hat{N}_{\mathrm{p}}$ is used to stabilize the unconventional $\eta$-pairing states over unpaired states.
Figure~\ref{fig:pair number}(c) shows an example of a Fock state with one broken pair.
The expectation value of $\hat{N}_{\mathrm{p}}$ for this state is $3/2$.
If we introduce a pair chemical potential $\mu_{\mathrm{p}}$ and replace the Hamiltonian $\mathcal{H}_{\mathrm{extH}}$ with $\mathcal{H}_{\mathrm{extH}} - \mu_{\mathrm{p}} \hat{N}_{\mathrm{p}}$, we can decrease the eigenenergy of the unconventional pairing states than other unpaired eigenstates.
Since the pair chemical potential $\mu_{\mathrm{p}}$ lowers the energy of the entire subspace where all particles form pairs (not limited to the unconventional $\eta$-pairing states), an important remaining question is whether superconductivity may emerge in the ground state within this subspace.

\subsection{Open boundary condition} \label{sec:obc}
In the open boundary condition, the off-site $\eta$-pairing operator $\eta_{\bm\alpha, \sigma_1\sigma_2}^+$ is defined in a similar way as in Eq.~\eqref{eq:general_eta_operator} but the summation over the site index $i$ should be taken as long as both $\bm r_i$ and $\bm r_i+\bm\alpha$ belong to the lattice with the open boundary condition.
The off-site $\eta$-pairing states created by this $\eta_{\bm\alpha, \sigma_1\sigma_2}^+$ are generally not eigenstates of the kinetic Hamiltonian in Eq.~\eqref{eq:kint_Hamiltonian} with the open boundary condition.
To see this, let us consider an open one-dimensional chain of length $L$ as an example.
The commutation relation between the kinetic term of the Hamiltonian and the off-site $\eta$-pairing operator is evaluated as $[\mathcal{H}_{\mathrm{kin}},\, \eta^{+}_{\alpha,\sigma_1\sigma_2}] = [\mathcal{H}_{\mathrm{kin}},\, \sum_{i=1}^{L-\alpha} f(i) c_{i\sigma_1}^{\dagger} c_{i+\alpha,\sigma_2}^{\dagger}] = f(1) c_{1\sigma_1}^{\dagger} c_{\alpha,\sigma_2}^{\dagger} + f(L-\alpha) c_{L-\alpha+1,\sigma_1}^{\dagger} c_{L\sigma_2}^{\dagger}$, indicating that the $\eta$-pairing states are generally not eigenstates of $\mathcal{H}_{\mathrm{kin}}$ due to the presence of the boundary terms.

In the case of $\alpha = 1$ (nearest-neighbor pairing) and the spin-triplet pairing (i.e., $\eta_{1,\sigma\sigma}^+$ or $\eta_{1,\sigma_1\sigma_2}^+ + \eta_{1, \sigma_2\sigma_1}^+$), however, the remaining boundary terms are canceled due to the fermion anticommutation relation.
This also applies to the case of the spinless fermions, discussed in Secs.~\ref{sec:p-wave_eigenstate} and \ref{sec:p-wave_scar}, since they can be mapped to the spin-triplet states.
In higher-dimensional systems, a similar argument can be applied by regarding the vertical hopping relative to the boundary as the one in the above one-dimensional case.

The interaction term $\mathcal{H}_{\mathrm{int}}$ of the Hamiltonian in the open boundary condition is defined as in Eq.~\eqref{eq:int_Hamiltonian} with the neighboring particle existence operator $\mathbb{n}_{i\sigma}$ given by Eq.~\eqref{eq:definition_npe_operator}, where the coordination number $z$ changes at the boundary.
One can confirm that the off-site $\eta$-pairing states are eigenstates of the interaction term $\mathcal{H}_{\mathrm{int}}$ with the appropriately defined $\mathbb n_{i\sigma}$ in the open boundary condition.
As a result, the spin-triplet $d$-wave $\eta$-pairing states given in Eq.~\eqref{eq:deta_state}, for instance, are the exact energy eigenstates of $\mathcal{H}_{\mathrm{extH}}$ [Eq.~\eqref{eq:extended_Hubbard_Hamiltonian}] even in the open boundary condition.

Other systems with complex boundary conditions will be the subject of a future work.

\section{Summary and outlook} \label{sec:summary}
In this paper, we have presented a systematic framework to construct model Hamiltonians that have superconducting quantum many-body scar states with unconventional pairing symmetries, e.g., the $s$-wave spin-triplet (or spinless), $p$-wave spin-singlet, $d$-wave spin-triplet, and $f$-wave spin-triplet symmetries, by engineering multibody interactions.
The key idea was to cancel the energy increment due to the formation of doublons by multibody interactions.
To this end, we have introduced the neighboring particle existence operator $\mathbb{n}_{i\sigma}$ in Eq.~\eqref{eq:bold_number_operator}, which judges whether there exist spin-$\sigma$ particles on the nearest-neighbor sites of site $i$.
Using this operator, we have defined the multibody interaction terms in the form of $\sum_{\sigma} n_{i\uparrow}n_{i\downarrow}\mathbb{n}_{i\sigma}$ in Eq.~\eqref{eq:nd_bbn}, which precisely cancel the two-body interactions acting among off-site pairs (while working non-trivially on unpaired particles).
We have applied our approach to the two-dimensional extended Hubbard model with the multibody interactions, and numerically confirmed that the spin-triplet $d$-wave $\eta$-pairing state is indeed the QMBS eigenstate.
Our construction is flexible enough that it can be applied to various pairing symmetries and arbitrary bipartite lattice models, some of which are discussed in the paper (see Table~\ref{table:symmetry}).
The derived unconventional pairing states are natural extensions of Yang's $s$-wave spin-single $\eta$-pairing states, and will be relevant for applications to nonthermal and long-lived nonequilibrium superconductivity with unconventional pairing symmetries that might be difficult to achieve in thermal equilibrium.

There are various future problems.
From a practical point of view, it will be important to find a protocol to reach the unconventional superconducting QMBS states from experimentally accessible initial states.
For the conventional $s$-wave $\eta$-pairing states, there have been several proposals including the periodic drive~\cite{Kitamura2016, Peronaci2020, Cook2020, Tindall2021}, photo-doping~\cite{Werner2018a, Kaneko2019a, Werner2019, Li2020j, Murakami2022f, Murakami2023b, Ray2023a}, dissipation engineering~\cite{Diehl2008, Kraus2008, Nakagawa2021, Yang2022e}, and adiabatic driving~\cite{Kantian2010}, some of which could be extended to the unconventional cases.
Another issue is the stability of the unconventional superconducting scar states against various perturbations~\cite{Kolb2023a, Gotta2023a}.
It is necessary to understand how the scar states can persist (or not) when the parameters of the multibody interactions deviate from the ones we obtained.
Furthermore, it is known that Yang’s $\eta$-pairing state exhibits electromagnetic instability in the presence of external electromagnetic fields~\cite{Hoshino2014, Tsuji2021}.
However, in the case of unconventional superconducting scar states, $\eta$-pairing states with different (odd) parity can potentially lead to qualitatively distinct electromagnetic stability, which warrants further investigation.
Topological aspects of the scar states (such as the $p_x+ip_y$-wave $\eta$-pairing state discussed in Sec.~\ref{sec:pairing_symmetry}) will be another interesting topic, where the question is whether a bulk-boundary correspondence also holds for quantum many-body scar states.
It will also be interesting to explore the relation between the present unconventional superconducting scar states and the fracton physics~\cite{Nandkishore2019, Pretko2020a, Xavier2021, Ren2021a, Gotta2022}, since our model exhibits a high degree of degeneracy of the eigenstates which grow as a function of the system size.

Finally, we briefly comment on possible experimental realizations of the superconducting QMBS states with unconventional pairing symmetries.
It is required that the scarred models have a controllable multibody interaction.
The spinless/spin-polarized $s$-wave pairing states discussed in Secs.~\ref{sec:p-wave_eigenstate} and~\ref{sec:p-wave_scar} can be implemented in the one-dimensional system with the density-density-type three-body interaction in Eq.~\eqref{eq:ham_W}.
The three-body interaction has been realized in a controllable manner in cold polar-molecular gases trapped in an optical lattice by means of effective interactions mediated by dipole-dipole interactions~\cite{Buchler2007, Han2010, Will2010, Hammer2013, Ren2015, Valiente2019a}.
Thus, the unconventional superconducting scar state in Eq.~\eqref{eq:peta_state} in the system described by Eq.~\eqref{eq:spinless_hamiltonian} is a primary example of what can be achieved with the state-of-the-art experimental techniques for ultracold atomic systems.

Another example is the spin-singlet $p$-wave pairing state $(\eta_{p_x}^+)^{N/2}|0\rangle$ discussed in Sec.~\ref{sec:pairing_symmetry}.
In a one-dimensional chain, it becomes a superconducting scar state in the model with the interaction terms containing up to the four-body interaction in Eq.~\eqref{eq:1d_spinful_interaction_hamiltonian}.
As in the previous case, similar four-body interactions have been realized in recent experiments on cold atomic systems~\cite{Gurian2012, Honda2024}.
So far, the four-body interactions have been implemented in bosonic systems, but in principle they could be extended to fermionic systems, allowing one to explore the rich physics, as seen in odd-frequency superconductivity~\cite{Linder2019}, through the nonequilibrium dynamics of cold atoms protected against thermalization.

\begin{acknowledgments}
The authors thank Hosho Katsura for fruitful discussions.
This work was supported by JST FOREST (Grant No.~JPMJFR2131) and JSPS KAKENHI (Grant Nos.~JP20K03811, JP23K19030, and JP24H00191).
\end{acknowledgments}

\bibliography{ref}

\begin{thebibliography}{112}%
\makeatletter
\providecommand \@ifxundefined [1]{%
 \@ifx{#1\undefined}
}%
\providecommand \@ifnum [1]{%
 \ifnum #1\expandafter \@firstoftwo
 \else \expandafter \@secondoftwo
 \fi
}%
\providecommand \@ifx [1]{%
 \ifx #1\expandafter \@firstoftwo
 \else \expandafter \@secondoftwo
 \fi
}%
\providecommand \natexlab [1]{#1}%
\providecommand \enquote  [1]{``#1''}%
\providecommand \bibnamefont  [1]{#1}%
\providecommand \bibfnamefont [1]{#1}%
\providecommand \citenamefont [1]{#1}%
\providecommand \href@noop [0]{\@secondoftwo}%
\providecommand \href [0]{\begingroup \@sanitize@url \@href}%
\providecommand \@href[1]{\@@startlink{#1}\@@href}%
\providecommand \@@href[1]{\endgroup#1\@@endlink}%
\providecommand \@sanitize@url [0]{\catcode `\\12\catcode `\$12\catcode `\&12\catcode `\#12\catcode `\^12\catcode `\_12\catcode `\%12\relax}%
\providecommand \@@startlink[1]{}%
\providecommand \@@endlink[0]{}%
\providecommand \url  [0]{\begingroup\@sanitize@url \@url }%
\providecommand \@url [1]{\endgroup\@href {#1}{\urlprefix }}%
\providecommand \urlprefix  [0]{URL }%
\providecommand \Eprint [0]{\href }%
\providecommand \doibase [0]{https://doi.org/}%
\providecommand \selectlanguage [0]{\@gobble}%
\providecommand \bibinfo  [0]{\@secondoftwo}%
\providecommand \bibfield  [0]{\@secondoftwo}%
\providecommand \translation [1]{[#1]}%
\providecommand \BibitemOpen [0]{}%
\providecommand \bibitemStop [0]{}%
\providecommand \bibitemNoStop [0]{.\EOS\space}%
\providecommand \EOS [0]{\spacefactor3000\relax}%
\providecommand \BibitemShut  [1]{\csname bibitem#1\endcsname}%
\let\auto@bib@innerbib\@empty
\bibitem [{\citenamefont {Fausti}\ \emph {et~al.}(2011)\citenamefont {Fausti}, \citenamefont {Tobey}, \citenamefont {Dean}, \citenamefont {Kaiser}, \citenamefont {Dienst}, \citenamefont {Hoffmann}, \citenamefont {Pyon}, \citenamefont {Takayama}, \citenamefont {Takagi},\ and\ \citenamefont {Cavalleri}}]{Fausti2011}%
  \BibitemOpen
  \bibfield  {author} {\bibinfo {author} {\bibfnamefont {D.}~\bibnamefont {Fausti}}, \bibinfo {author} {\bibfnamefont {R.~I.}\ \bibnamefont {Tobey}}, \bibinfo {author} {\bibfnamefont {N.}~\bibnamefont {Dean}}, \bibinfo {author} {\bibfnamefont {S.}~\bibnamefont {Kaiser}}, \bibinfo {author} {\bibfnamefont {A.}~\bibnamefont {Dienst}}, \bibinfo {author} {\bibfnamefont {M.~C.}\ \bibnamefont {Hoffmann}}, \bibinfo {author} {\bibfnamefont {S.}~\bibnamefont {Pyon}}, \bibinfo {author} {\bibfnamefont {T.}~\bibnamefont {Takayama}}, \bibinfo {author} {\bibfnamefont {H.}~\bibnamefont {Takagi}},\ and\ \bibinfo {author} {\bibfnamefont {A.}~\bibnamefont {Cavalleri}},\ }\bibfield  {title} {\bibinfo {title} {Light-induced superconductivity in a stripe-ordered cuprate},\ }\href {https://doi.org/10.1126/science.1197294} {\bibfield  {journal} {\bibinfo  {journal} {Science}\ }\textbf {\bibinfo {volume} {331}},\ \bibinfo {pages} {189} (\bibinfo {year} {2011})}\BibitemShut {NoStop}%
\bibitem [{\citenamefont {Kaiser}\ \emph {et~al.}(2014)\citenamefont {Kaiser}, \citenamefont {Hunt}, \citenamefont {Nicoletti}, \citenamefont {Hu}, \citenamefont {Gierz}, \citenamefont {Liu}, \citenamefont {Le~Tacon}, \citenamefont {Loew}, \citenamefont {Haug}, \citenamefont {Keimer},\ and\ \citenamefont {Cavalleri}}]{Kaiser2014}%
  \BibitemOpen
  \bibfield  {author} {\bibinfo {author} {\bibfnamefont {S.}~\bibnamefont {Kaiser}}, \bibinfo {author} {\bibfnamefont {C.~R.}\ \bibnamefont {Hunt}}, \bibinfo {author} {\bibfnamefont {D.}~\bibnamefont {Nicoletti}}, \bibinfo {author} {\bibfnamefont {W.}~\bibnamefont {Hu}}, \bibinfo {author} {\bibfnamefont {I.}~\bibnamefont {Gierz}}, \bibinfo {author} {\bibfnamefont {H.~Y.}\ \bibnamefont {Liu}}, \bibinfo {author} {\bibfnamefont {M.}~\bibnamefont {Le~Tacon}}, \bibinfo {author} {\bibfnamefont {T.}~\bibnamefont {Loew}}, \bibinfo {author} {\bibfnamefont {D.}~\bibnamefont {Haug}}, \bibinfo {author} {\bibfnamefont {B.}~\bibnamefont {Keimer}},\ and\ \bibinfo {author} {\bibfnamefont {A.}~\bibnamefont {Cavalleri}},\ }\bibfield  {title} {\bibinfo {title} {Optically induced coherent transport far above ${T}_{c}$ in underdoped ${{\mathrm{YBa}}}_{2}{{\mathrm{Cu}}}_{3}{{\mathrm{O}}}_{6+\ensuremath{\delta}}$},\ }\href {https://doi.org/10.1103/PhysRevB.89.184516} {\bibfield  {journal} {\bibinfo  {journal} {Phys. Rev. B}\
  }\textbf {\bibinfo {volume} {89}},\ \bibinfo {pages} {184516} (\bibinfo {year} {2014})}\BibitemShut {NoStop}%
\bibitem [{\citenamefont {Hu}\ \emph {et~al.}(2014)\citenamefont {Hu}, \citenamefont {Kaiser}, \citenamefont {Nicoletti}, \citenamefont {Hunt}, \citenamefont {Gierz}, \citenamefont {Hoffmann}, \citenamefont {{Le Tacon}}, \citenamefont {Loew}, \citenamefont {Keimer},\ and\ \citenamefont {Cavalleri}}]{Hu2014}%
  \BibitemOpen
  \bibfield  {author} {\bibinfo {author} {\bibfnamefont {W.}~\bibnamefont {Hu}}, \bibinfo {author} {\bibfnamefont {S.}~\bibnamefont {Kaiser}}, \bibinfo {author} {\bibfnamefont {D.}~\bibnamefont {Nicoletti}}, \bibinfo {author} {\bibfnamefont {C.~R.}\ \bibnamefont {Hunt}}, \bibinfo {author} {\bibfnamefont {I.}~\bibnamefont {Gierz}}, \bibinfo {author} {\bibfnamefont {M.~C.}\ \bibnamefont {Hoffmann}}, \bibinfo {author} {\bibfnamefont {M.}~\bibnamefont {{Le Tacon}}}, \bibinfo {author} {\bibfnamefont {T.}~\bibnamefont {Loew}}, \bibinfo {author} {\bibfnamefont {B.}~\bibnamefont {Keimer}},\ and\ \bibinfo {author} {\bibfnamefont {A.}~\bibnamefont {Cavalleri}},\ }\bibfield  {title} {\bibinfo {title} {{Optically enhanced coherent transport in ${{\mathrm{YBa}}_2{\mathrm{Cu}}_3{\mathrm{O}}_{6.5}}$ by ultrafast redistribution of interlayer coupling}},\ }\href {https://doi.org/10.1038/nmat3963} {\bibfield  {journal} {\bibinfo  {journal} {Nat. Mater.}\ }\textbf {\bibinfo {volume} {13}},\ \bibinfo {pages} {705} (\bibinfo
  {year} {2014})}\BibitemShut {NoStop}%
\bibitem [{\citenamefont {Mitrano}\ \emph {et~al.}(2016)\citenamefont {Mitrano}, \citenamefont {Cantaluppi}, \citenamefont {Nicoletti}, \citenamefont {Kaiser}, \citenamefont {Perucchi}, \citenamefont {Lupi}, \citenamefont {{Di Pietro}}, \citenamefont {Pontiroli}, \citenamefont {Ricc{\`{o}}}, \citenamefont {Clark}, \citenamefont {Jaksch},\ and\ \citenamefont {Cavalleri}}]{Mitrano2016a}%
  \BibitemOpen
  \bibfield  {author} {\bibinfo {author} {\bibfnamefont {M.}~\bibnamefont {Mitrano}}, \bibinfo {author} {\bibfnamefont {A.}~\bibnamefont {Cantaluppi}}, \bibinfo {author} {\bibfnamefont {D.}~\bibnamefont {Nicoletti}}, \bibinfo {author} {\bibfnamefont {S.}~\bibnamefont {Kaiser}}, \bibinfo {author} {\bibfnamefont {A.}~\bibnamefont {Perucchi}}, \bibinfo {author} {\bibfnamefont {S.}~\bibnamefont {Lupi}}, \bibinfo {author} {\bibfnamefont {P.}~\bibnamefont {{Di Pietro}}}, \bibinfo {author} {\bibfnamefont {D.}~\bibnamefont {Pontiroli}}, \bibinfo {author} {\bibfnamefont {M.}~\bibnamefont {Ricc{\`{o}}}}, \bibinfo {author} {\bibfnamefont {S.~R.}\ \bibnamefont {Clark}}, \bibinfo {author} {\bibfnamefont {D.}~\bibnamefont {Jaksch}},\ and\ \bibinfo {author} {\bibfnamefont {A.}~\bibnamefont {Cavalleri}},\ }\bibfield  {title} {\bibinfo {title} {{Possible light-induced superconductivity in $\mathrm{K}_3\mathrm{C}_{60}$ at high temperature}},\ }\href {https://doi.org/10.1038/nature16522} {\bibfield  {journal} {\bibinfo
  {journal} {Nature}\ }\textbf {\bibinfo {volume} {530}},\ \bibinfo {pages} {461} (\bibinfo {year} {2016})}\BibitemShut {NoStop}%
\bibitem [{\citenamefont {Cremin}\ \emph {et~al.}(2019)\citenamefont {Cremin}, \citenamefont {Zhang}, \citenamefont {Homes}, \citenamefont {Gu}, \citenamefont {Sun}, \citenamefont {Fogler}, \citenamefont {Millis}, \citenamefont {Basov},\ and\ \citenamefont {Averitt}}]{Cremin2019}%
  \BibitemOpen
  \bibfield  {author} {\bibinfo {author} {\bibfnamefont {K.~A.}\ \bibnamefont {Cremin}}, \bibinfo {author} {\bibfnamefont {J.}~\bibnamefont {Zhang}}, \bibinfo {author} {\bibfnamefont {C.~C.}\ \bibnamefont {Homes}}, \bibinfo {author} {\bibfnamefont {G.~D.}\ \bibnamefont {Gu}}, \bibinfo {author} {\bibfnamefont {Z.}~\bibnamefont {Sun}}, \bibinfo {author} {\bibfnamefont {M.~M.}\ \bibnamefont {Fogler}}, \bibinfo {author} {\bibfnamefont {A.~J.}\ \bibnamefont {Millis}}, \bibinfo {author} {\bibfnamefont {D.~N.}\ \bibnamefont {Basov}},\ and\ \bibinfo {author} {\bibfnamefont {R.~D.}\ \bibnamefont {Averitt}},\ }\bibfield  {title} {\bibinfo {title} {Photoenhanced metastable c-axis electrodynamics in stripe-ordered cuprate ${{\mathrm{La}}_{1.885}{\mathrm{Ba}}_{0.115}{\mathrm{CuO}}_4}$},\ }\href {https://doi.org/10.1073/pnas.1908368116} {\bibfield  {journal} {\bibinfo  {journal} {Proc. Natl. Acad. Sci.}\ }\textbf {\bibinfo {volume} {116}},\ \bibinfo {pages} {19875} (\bibinfo {year} {2019})}\BibitemShut {NoStop}%
\bibitem [{\citenamefont {Katsumi}\ \emph {et~al.}(2023)\citenamefont {Katsumi}, \citenamefont {Nishida}, \citenamefont {Kaiser}, \citenamefont {Miyasaka}, \citenamefont {Tajima},\ and\ \citenamefont {Shimano}}]{Katsumi2023}%
  \BibitemOpen
  \bibfield  {author} {\bibinfo {author} {\bibfnamefont {K.}~\bibnamefont {Katsumi}}, \bibinfo {author} {\bibfnamefont {M.}~\bibnamefont {Nishida}}, \bibinfo {author} {\bibfnamefont {S.}~\bibnamefont {Kaiser}}, \bibinfo {author} {\bibfnamefont {S.}~\bibnamefont {Miyasaka}}, \bibinfo {author} {\bibfnamefont {S.}~\bibnamefont {Tajima}},\ and\ \bibinfo {author} {\bibfnamefont {R.}~\bibnamefont {Shimano}},\ }\bibfield  {title} {\bibinfo {title} {{Near-infrared light-induced superconducting-like state in underdoped $\mathrm{Y}{\mathrm{Ba}}_{2}{\mathrm{Cu}}_{3}{\mathrm{O}}_{y}$ studied by $c$-axis terahertz third-harmonic generation}},\ }\href {https://doi.org/10.1103/PhysRevB.107.214506} {\bibfield  {journal} {\bibinfo  {journal} {Phys. Rev. B}\ }\textbf {\bibinfo {volume} {107}},\ \bibinfo {pages} {214506} (\bibinfo {year} {2023})}\BibitemShut {NoStop}%
\bibitem [{\citenamefont {Zhang}\ \emph {et~al.}(2024)\citenamefont {Zhang}, \citenamefont {Zhou}, \citenamefont {Xu}, \citenamefont {Wu}, \citenamefont {Yue}, \citenamefont {Liu}, \citenamefont {Hu}, \citenamefont {Li}, \citenamefont {Yuan}, \citenamefont {Homes}, \citenamefont {Gu}, \citenamefont {Dong},\ and\ \citenamefont {Wang}}]{Zhang2024a}%
  \BibitemOpen
  \bibfield  {author} {\bibinfo {author} {\bibfnamefont {S.~J.}\ \bibnamefont {Zhang}}, \bibinfo {author} {\bibfnamefont {X.~Y.}\ \bibnamefont {Zhou}}, \bibinfo {author} {\bibfnamefont {S.~X.}\ \bibnamefont {Xu}}, \bibinfo {author} {\bibfnamefont {Q.}~\bibnamefont {Wu}}, \bibinfo {author} {\bibfnamefont {L.}~\bibnamefont {Yue}}, \bibinfo {author} {\bibfnamefont {Q.~M.}\ \bibnamefont {Liu}}, \bibinfo {author} {\bibfnamefont {T.~C.}\ \bibnamefont {Hu}}, \bibinfo {author} {\bibfnamefont {R.~S.}\ \bibnamefont {Li}}, \bibinfo {author} {\bibfnamefont {J.~Y.}\ \bibnamefont {Yuan}}, \bibinfo {author} {\bibfnamefont {C.~C.}\ \bibnamefont {Homes}}, \bibinfo {author} {\bibfnamefont {G.~D.}\ \bibnamefont {Gu}}, \bibinfo {author} {\bibfnamefont {T.}~\bibnamefont {Dong}},\ and\ \bibinfo {author} {\bibfnamefont {N.~L.}\ \bibnamefont {Wang}},\ }\bibfield  {title} {\bibinfo {title} {{Light-Induced Melting of Competing Stripe Orders without Introducing Superconductivity in
  ${\mathrm{La}}_{2\ensuremath{-}x}{\mathrm{Ba}}_{x}{\mathrm{CuO}}_{4}$}},\ }\href {https://doi.org/10.1103/PhysRevX.14.011036} {\bibfield  {journal} {\bibinfo  {journal} {Phys. Rev. X}\ }\textbf {\bibinfo {volume} {14}},\ \bibinfo {pages} {011036} (\bibinfo {year} {2024})}\BibitemShut {NoStop}%
\bibitem [{\citenamefont {Marin~Bukov}\ and\ \citenamefont {Polkovnikov}(2015)}]{Bukov2015a}%
  \BibitemOpen
  \bibfield  {author} {\bibinfo {author} {\bibfnamefont {L.~D.}\ \bibnamefont {Marin~Bukov}}\ and\ \bibinfo {author} {\bibfnamefont {A.}~\bibnamefont {Polkovnikov}},\ }\bibfield  {title} {\bibinfo {title} {Universal high-frequency behavior of periodically driven systems: from dynamical stabilization to {Floquet} engineering},\ }\href {https://doi.org/10.1080/00018732.2015.1055918} {\bibfield  {journal} {\bibinfo  {journal} {Adv. Phys.}\ }\textbf {\bibinfo {volume} {64}},\ \bibinfo {pages} {139} (\bibinfo {year} {2015})}\BibitemShut {NoStop}%
\bibitem [{\citenamefont {Oka}\ and\ \citenamefont {Kitamura}(2019)}]{Oka2019}%
  \BibitemOpen
  \bibfield  {author} {\bibinfo {author} {\bibfnamefont {T.}~\bibnamefont {Oka}}\ and\ \bibinfo {author} {\bibfnamefont {S.}~\bibnamefont {Kitamura}},\ }\bibfield  {title} {\bibinfo {title} {{Floquet Engineering of Quantum Materials}},\ }\href {https://doi.org/10.1146/annurev-conmatphys-031218-013423} {\bibfield  {journal} {\bibinfo  {journal} {Annu. Rev. Condens. Matter Phys.}\ }\textbf {\bibinfo {volume} {10}},\ \bibinfo {pages} {387} (\bibinfo {year} {2019})}\BibitemShut {NoStop}%
\bibitem [{\citenamefont {Tsuji}(2024)}]{Tsuji2023}%
  \BibitemOpen
  \bibfield  {author} {\bibinfo {author} {\bibfnamefont {N.}~\bibnamefont {Tsuji}},\ }\bibfield  {title} {\bibinfo {title} {Floquet states},\ }in\ \href {https://doi.org/https://doi.org/10.1016/B978-0-323-90800-9.00241-9} {\emph {\bibinfo {booktitle} {Encyclopedia of Condensed Matter Physics (Second Edition)}}},\ \bibinfo {editor} {edited by\ \bibinfo {editor} {\bibfnamefont {T.}~\bibnamefont {Chakraborty}}}\ (\bibinfo  {publisher} {Academic Press},\ \bibinfo {address} {Oxford},\ \bibinfo {year} {2024})\ pp.\ \bibinfo {pages} {967--980}\BibitemShut {NoStop}%
\bibitem [{\citenamefont {Ezawa}(2015)}]{Ezawa2015}%
  \BibitemOpen
  \bibfield  {author} {\bibinfo {author} {\bibfnamefont {M.}~\bibnamefont {Ezawa}},\ }\bibfield  {title} {\bibinfo {title} {{Photo-Induced Topological Superconductor in Silicene, Germanene, and Stanene}},\ }\href {https://doi.org/10.1007/s10948-014-2900-x} {\bibfield  {journal} {\bibinfo  {journal} {J. Supercond. Nov. Magn.}\ }\textbf {\bibinfo {volume} {28}},\ \bibinfo {pages} {1249} (\bibinfo {year} {2015})}\BibitemShut {NoStop}%
\bibitem [{\citenamefont {Takasan}\ \emph {et~al.}(2017)\citenamefont {Takasan}, \citenamefont {Daido}, \citenamefont {Kawakami},\ and\ \citenamefont {Yanase}}]{Takasan2017b}%
  \BibitemOpen
  \bibfield  {author} {\bibinfo {author} {\bibfnamefont {K.}~\bibnamefont {Takasan}}, \bibinfo {author} {\bibfnamefont {A.}~\bibnamefont {Daido}}, \bibinfo {author} {\bibfnamefont {N.}~\bibnamefont {Kawakami}},\ and\ \bibinfo {author} {\bibfnamefont {Y.}~\bibnamefont {Yanase}},\ }\bibfield  {title} {\bibinfo {title} {{Laser-induced topological superconductivity in cuprate thin films}},\ }\href {https://doi.org/10.1103/PhysRevB.95.134508} {\bibfield  {journal} {\bibinfo  {journal} {Phys. Rev. B}\ }\textbf {\bibinfo {volume} {95}},\ \bibinfo {pages} {134508} (\bibinfo {year} {2017})}\BibitemShut {NoStop}%
\bibitem [{\citenamefont {Chono}\ \emph {et~al.}(2020)\citenamefont {Chono}, \citenamefont {Takasan},\ and\ \citenamefont {Yanase}}]{Chono2020a}%
  \BibitemOpen
  \bibfield  {author} {\bibinfo {author} {\bibfnamefont {H.}~\bibnamefont {Chono}}, \bibinfo {author} {\bibfnamefont {K.}~\bibnamefont {Takasan}},\ and\ \bibinfo {author} {\bibfnamefont {Y.}~\bibnamefont {Yanase}},\ }\bibfield  {title} {\bibinfo {title} {Laser-induced topological $s$-wave superconductivity in bilayer transition metal dichalcogenides},\ }\href {https://doi.org/10.1103/PhysRevB.102.174508} {\bibfield  {journal} {\bibinfo  {journal} {Phys. Rev. B}\ }\textbf {\bibinfo {volume} {102}},\ \bibinfo {pages} {174508} (\bibinfo {year} {2020})}\BibitemShut {NoStop}%
\bibitem [{\citenamefont {Wenk}\ \emph {et~al.}(2022)\citenamefont {Wenk}, \citenamefont {Grifoni},\ and\ \citenamefont {Schliemann}}]{Wenk2022}%
  \BibitemOpen
  \bibfield  {author} {\bibinfo {author} {\bibfnamefont {P.}~\bibnamefont {Wenk}}, \bibinfo {author} {\bibfnamefont {M.}~\bibnamefont {Grifoni}},\ and\ \bibinfo {author} {\bibfnamefont {J.}~\bibnamefont {Schliemann}},\ }\bibfield  {title} {\bibinfo {title} {{Topological transitions in two-dimensional Floquet superconductors}},\ }\href {https://doi.org/10.1103/PhysRevB.106.134508} {\bibfield  {journal} {\bibinfo  {journal} {Phys. Rev. B}\ }\textbf {\bibinfo {volume} {106}},\ \bibinfo {pages} {134508} (\bibinfo {year} {2022})}\BibitemShut {NoStop}%
\bibitem [{\citenamefont {Yanase}\ \emph {et~al.}(2022)\citenamefont {Yanase}, \citenamefont {Daido}, \citenamefont {Takasan},\ and\ \citenamefont {Yoshida}}]{Yanase2022}%
  \BibitemOpen
  \bibfield  {author} {\bibinfo {author} {\bibfnamefont {Y.}~\bibnamefont {Yanase}}, \bibinfo {author} {\bibfnamefont {A.}~\bibnamefont {Daido}}, \bibinfo {author} {\bibfnamefont {K.}~\bibnamefont {Takasan}},\ and\ \bibinfo {author} {\bibfnamefont {T.}~\bibnamefont {Yoshida}},\ }\bibfield  {title} {\bibinfo {title} {Topological $d$-wave superconductivity in two dimensions},\ }\href {https://doi.org/10.1016/j.physe.2022.115143} {\bibfield  {journal} {\bibinfo  {journal} {Phys. E: Low-Dimens. Syst. Nanostructures}\ }\textbf {\bibinfo {volume} {140}},\ \bibinfo {pages} {115143} (\bibinfo {year} {2022})}\BibitemShut {NoStop}%
\bibitem [{\citenamefont {Kitamura}\ and\ \citenamefont {Aoki}(2022)}]{Kitamura2022a}%
  \BibitemOpen
  \bibfield  {author} {\bibinfo {author} {\bibfnamefont {S.}~\bibnamefont {Kitamura}}\ and\ \bibinfo {author} {\bibfnamefont {H.}~\bibnamefont {Aoki}},\ }\bibfield  {title} {\bibinfo {title} {{Floquet topological superconductivity induced by chiral many-body interaction}},\ }\href {https://doi.org/10.1038/s42005-022-00936-w} {\bibfield  {journal} {\bibinfo  {journal} {Commun. Phys.}\ }\textbf {\bibinfo {volume} {5}},\ \bibinfo {pages} {174} (\bibinfo {year} {2022})}\BibitemShut {NoStop}%
\bibitem [{\citenamefont {Cayao}\ \emph {et~al.}(2021)\citenamefont {Cayao}, \citenamefont {Triola},\ and\ \citenamefont {Black-Schaffer}}]{Cayao2021}%
  \BibitemOpen
  \bibfield  {author} {\bibinfo {author} {\bibfnamefont {J.}~\bibnamefont {Cayao}}, \bibinfo {author} {\bibfnamefont {C.}~\bibnamefont {Triola}},\ and\ \bibinfo {author} {\bibfnamefont {A.~M.}\ \bibnamefont {Black-Schaffer}},\ }\bibfield  {title} {\bibinfo {title} {{Floquet engineering bulk odd-frequency superconducting pairs}},\ }\href {https://doi.org/10.1103/PhysRevB.103.104505} {\bibfield  {journal} {\bibinfo  {journal} {Phys. Rev. B}\ }\textbf {\bibinfo {volume} {103}},\ \bibinfo {pages} {104505} (\bibinfo {year} {2021})}\BibitemShut {NoStop}%
\bibitem [{\citenamefont {Cayao}\ and\ \citenamefont {Black-Schaffer}(2022)}]{Cayao2022}%
  \BibitemOpen
  \bibfield  {author} {\bibinfo {author} {\bibfnamefont {J.}~\bibnamefont {Cayao}}\ and\ \bibinfo {author} {\bibfnamefont {A.~M.}\ \bibnamefont {Black-Schaffer}},\ }\bibfield  {title} {\bibinfo {title} {{Exceptional odd-frequency pairing in non-Hermitian superconducting systems}},\ }\href {https://doi.org/10.1103/PhysRevB.105.094502} {\bibfield  {journal} {\bibinfo  {journal} {Phys. Rev. B}\ }\textbf {\bibinfo {volume} {105}},\ \bibinfo {pages} {094502} (\bibinfo {year} {2022})}\BibitemShut {NoStop}%
\bibitem [{\citenamefont {Kaneko}\ \emph {et~al.}(2019)\citenamefont {Kaneko}, \citenamefont {Shirakawa}, \citenamefont {Sorella},\ and\ \citenamefont {Yunoki}}]{Kaneko2019a}%
  \BibitemOpen
  \bibfield  {author} {\bibinfo {author} {\bibfnamefont {T.}~\bibnamefont {Kaneko}}, \bibinfo {author} {\bibfnamefont {T.}~\bibnamefont {Shirakawa}}, \bibinfo {author} {\bibfnamefont {S.}~\bibnamefont {Sorella}},\ and\ \bibinfo {author} {\bibfnamefont {S.}~\bibnamefont {Yunoki}},\ }\bibfield  {title} {\bibinfo {title} {{Photoinduced $\ensuremath{\eta}$ Pairing in the Hubbard Model}},\ }\href {https://doi.org/10.1103/PhysRevLett.122.077002} {\bibfield  {journal} {\bibinfo  {journal} {Phys. Rev. Lett.}\ }\textbf {\bibinfo {volume} {122}},\ \bibinfo {pages} {077002} (\bibinfo {year} {2019})}\BibitemShut {NoStop}%
\bibitem [{\citenamefont {Malakhov}\ and\ \citenamefont {Avdeev}(2021)}]{Malakhov2021a}%
  \BibitemOpen
  \bibfield  {author} {\bibinfo {author} {\bibfnamefont {M.}~\bibnamefont {Malakhov}}\ and\ \bibinfo {author} {\bibfnamefont {M.}~\bibnamefont {Avdeev}},\ }\bibfield  {title} {\bibinfo {title} {Non-equilibrium $d$-wave pair density wave order parameter in superconducting cuprates},\ }\href {https://doi.org/10.1016/j.physc.2021.1353820} {\bibfield  {journal} {\bibinfo  {journal} {Phys. C: Supercond. its Appl.}\ }\textbf {\bibinfo {volume} {581}},\ \bibinfo {pages} {1353820} (\bibinfo {year} {2021})}\BibitemShut {NoStop}%
\bibitem [{\citenamefont {Bernien}\ \emph {et~al.}(2017)\citenamefont {Bernien}, \citenamefont {Schwartz}, \citenamefont {Keesling}, \citenamefont {Levine}, \citenamefont {Omran}, \citenamefont {Pichler}, \citenamefont {Choi}, \citenamefont {Zibrov}, \citenamefont {Endres}, \citenamefont {Greiner}, \citenamefont {Vuleti{\'{c}}},\ and\ \citenamefont {Lukin}}]{Bernien2017}%
  \BibitemOpen
  \bibfield  {author} {\bibinfo {author} {\bibfnamefont {H.}~\bibnamefont {Bernien}}, \bibinfo {author} {\bibfnamefont {S.}~\bibnamefont {Schwartz}}, \bibinfo {author} {\bibfnamefont {A.}~\bibnamefont {Keesling}}, \bibinfo {author} {\bibfnamefont {H.}~\bibnamefont {Levine}}, \bibinfo {author} {\bibfnamefont {A.}~\bibnamefont {Omran}}, \bibinfo {author} {\bibfnamefont {H.}~\bibnamefont {Pichler}}, \bibinfo {author} {\bibfnamefont {S.}~\bibnamefont {Choi}}, \bibinfo {author} {\bibfnamefont {A.~S.}\ \bibnamefont {Zibrov}}, \bibinfo {author} {\bibfnamefont {M.}~\bibnamefont {Endres}}, \bibinfo {author} {\bibfnamefont {M.}~\bibnamefont {Greiner}}, \bibinfo {author} {\bibfnamefont {V.}~\bibnamefont {Vuleti{\'{c}}}},\ and\ \bibinfo {author} {\bibfnamefont {M.~D.}\ \bibnamefont {Lukin}},\ }\bibfield  {title} {\bibinfo {title} {{Probing many-body dynamics on a 51-atom quantum simulator}},\ }\href {https://doi.org/10.1038/nature24622} {\bibfield  {journal} {\bibinfo  {journal} {Nature}\ }\textbf {\bibinfo {volume}
  {551}},\ \bibinfo {pages} {579} (\bibinfo {year} {2017})}\BibitemShut {NoStop}%
\bibitem [{\citenamefont {Shiraishi}\ and\ \citenamefont {Mori}(2017)}]{Shiraishi2017}%
  \BibitemOpen
  \bibfield  {author} {\bibinfo {author} {\bibfnamefont {N.}~\bibnamefont {Shiraishi}}\ and\ \bibinfo {author} {\bibfnamefont {T.}~\bibnamefont {Mori}},\ }\bibfield  {title} {\bibinfo {title} {{Systematic Construction of Counterexamples to the Eigenstate Thermalization Hypothesis}},\ }\href {https://doi.org/10.1103/PhysRevLett.119.030601} {\bibfield  {journal} {\bibinfo  {journal} {Phys. Rev. Lett.}\ }\textbf {\bibinfo {volume} {119}},\ \bibinfo {pages} {030601} (\bibinfo {year} {2017})}\BibitemShut {NoStop}%
\bibitem [{\citenamefont {Turner}\ \emph {et~al.}(2018)\citenamefont {Turner}, \citenamefont {Michailidis}, \citenamefont {Abanin}, \citenamefont {Serbyn},\ and\ \citenamefont {Papi{\'{c}}}}]{Turner2018}%
  \BibitemOpen
  \bibfield  {author} {\bibinfo {author} {\bibfnamefont {C.~J.}\ \bibnamefont {Turner}}, \bibinfo {author} {\bibfnamefont {A.~A.}\ \bibnamefont {Michailidis}}, \bibinfo {author} {\bibfnamefont {D.~A.}\ \bibnamefont {Abanin}}, \bibinfo {author} {\bibfnamefont {M.}~\bibnamefont {Serbyn}},\ and\ \bibinfo {author} {\bibfnamefont {Z.}~\bibnamefont {Papi{\'{c}}}},\ }\bibfield  {title} {\bibinfo {title} {{Weak ergodicity breaking from quantum many-body scars}},\ }\href {https://doi.org/10.1038/s41567-018-0137-5} {\bibfield  {journal} {\bibinfo  {journal} {Nat. Phys.}\ }\textbf {\bibinfo {volume} {14}},\ \bibinfo {pages} {745} (\bibinfo {year} {2018})}\BibitemShut {NoStop}%
\bibitem [{\citenamefont {Deutsch}(1991)}]{Deutsch1991}%
  \BibitemOpen
  \bibfield  {author} {\bibinfo {author} {\bibfnamefont {J.~M.}\ \bibnamefont {Deutsch}},\ }\bibfield  {title} {\bibinfo {title} {{Quantum statistical mechanics in a closed system}},\ }\href {https://doi.org/10.1103/PhysRevA.43.2046} {\bibfield  {journal} {\bibinfo  {journal} {Phys. Rev. A}\ }\textbf {\bibinfo {volume} {43}},\ \bibinfo {pages} {2046} (\bibinfo {year} {1991})}\BibitemShut {NoStop}%
\bibitem [{\citenamefont {Srednicki}(1994)}]{Srednicki1994}%
  \BibitemOpen
  \bibfield  {author} {\bibinfo {author} {\bibfnamefont {M.}~\bibnamefont {Srednicki}},\ }\bibfield  {title} {\bibinfo {title} {{Chaos and quantum thermalization}},\ }\href {https://doi.org/10.1103/PhysRevE.50.888} {\bibfield  {journal} {\bibinfo  {journal} {Phys. Rev. E}\ }\textbf {\bibinfo {volume} {50}},\ \bibinfo {pages} {888} (\bibinfo {year} {1994})}\BibitemShut {NoStop}%
\bibitem [{\citenamefont {Rigol}\ \emph {et~al.}(2008)\citenamefont {Rigol}, \citenamefont {Dunjko},\ and\ \citenamefont {Olshanii}}]{Rigol2008a}%
  \BibitemOpen
  \bibfield  {author} {\bibinfo {author} {\bibfnamefont {M.}~\bibnamefont {Rigol}}, \bibinfo {author} {\bibfnamefont {V.}~\bibnamefont {Dunjko}},\ and\ \bibinfo {author} {\bibfnamefont {M.}~\bibnamefont {Olshanii}},\ }\bibfield  {title} {\bibinfo {title} {{Thermalization and its mechanism for generic isolated quantum systems}},\ }\href {https://doi.org/10.1038/nature06838} {\bibfield  {journal} {\bibinfo  {journal} {Nature}\ }\textbf {\bibinfo {volume} {452}},\ \bibinfo {pages} {854} (\bibinfo {year} {2008})}\BibitemShut {NoStop}%
\bibitem [{\citenamefont {Deutsch}(2018)}]{Deutsch2018a}%
  \BibitemOpen
  \bibfield  {author} {\bibinfo {author} {\bibfnamefont {J.~M.}\ \bibnamefont {Deutsch}},\ }\bibfield  {title} {\bibinfo {title} {{Eigenstate thermalization hypothesis}},\ }\href {https://doi.org/10.1088/1361-6633/aac9f1} {\bibfield  {journal} {\bibinfo  {journal} {Rep. Prog. Phys.}\ }\textbf {\bibinfo {volume} {81}},\ \bibinfo {pages} {082001} (\bibinfo {year} {2018})}\BibitemShut {NoStop}%
\bibitem [{\citenamefont {Serbyn}\ \emph {et~al.}(2021)\citenamefont {Serbyn}, \citenamefont {Abanin},\ and\ \citenamefont {Papi{\'{c}}}}]{Serbyn2021}%
  \BibitemOpen
  \bibfield  {author} {\bibinfo {author} {\bibfnamefont {M.}~\bibnamefont {Serbyn}}, \bibinfo {author} {\bibfnamefont {D.~A.}\ \bibnamefont {Abanin}},\ and\ \bibinfo {author} {\bibfnamefont {Z.}~\bibnamefont {Papi{\'{c}}}},\ }\bibfield  {title} {\bibinfo {title} {{Quantum many-body scars and weak breaking of ergodicity}},\ }\href {https://doi.org/10.1038/s41567-021-01230-2} {\bibfield  {journal} {\bibinfo  {journal} {Nat. Phys.}\ }\textbf {\bibinfo {volume} {17}},\ \bibinfo {pages} {675} (\bibinfo {year} {2021})}\BibitemShut {NoStop}%
\bibitem [{\citenamefont {Papi{\'{c}}}(2022)}]{Papic2022}%
  \BibitemOpen
  \bibfield  {author} {\bibinfo {author} {\bibfnamefont {Z.}~\bibnamefont {Papi{\'{c}}}},\ }\bibinfo {title} {{Weak Ergodicity Breaking Through the Lens of Quantum Entanglement}},\ in\ \href {https://doi.org/10.1007/978-3-031-03998-0_13} {\emph {\bibinfo {booktitle} {{Entanglement in Spin Chains: From Theory to Quantum Technology Applications}}}},\ \bibinfo {editor} {edited by\ \bibinfo {editor} {\bibfnamefont {A.}~\bibnamefont {Bayat}}, \bibinfo {editor} {\bibfnamefont {S.}~\bibnamefont {Bose}},\ and\ \bibinfo {editor} {\bibfnamefont {H.}~\bibnamefont {Johannesson}}}\ (\bibinfo  {publisher} {{Springer, Cham}},\ \bibinfo {year} {2022})\ pp.\ \bibinfo {pages} {341--395}\BibitemShut {NoStop}%
\bibitem [{\citenamefont {Moudgalya}\ \emph {et~al.}(2022)\citenamefont {Moudgalya}, \citenamefont {Bernevig},\ and\ \citenamefont {Regnault}}]{Moudgalya2022}%
  \BibitemOpen
  \bibfield  {author} {\bibinfo {author} {\bibfnamefont {S.}~\bibnamefont {Moudgalya}}, \bibinfo {author} {\bibfnamefont {B.~A.}\ \bibnamefont {Bernevig}},\ and\ \bibinfo {author} {\bibfnamefont {N.}~\bibnamefont {Regnault}},\ }\bibfield  {title} {\bibinfo {title} {{Quantum many-body scars and Hilbert space fragmentation: a review of exact results}},\ }\href {https://doi.org/10.1088/1361-6633/ac73a0} {\bibfield  {journal} {\bibinfo  {journal} {Rep. Prog. Phys.}\ }\textbf {\bibinfo {volume} {85}},\ \bibinfo {pages} {086501} (\bibinfo {year} {2022})}\BibitemShut {NoStop}%
\bibitem [{\citenamefont {Chandran}\ \emph {et~al.}(2023)\citenamefont {Chandran}, \citenamefont {Iadecola}, \citenamefont {Khemani},\ and\ \citenamefont {Moessner}}]{Chandran2023}%
  \BibitemOpen
  \bibfield  {author} {\bibinfo {author} {\bibfnamefont {A.}~\bibnamefont {Chandran}}, \bibinfo {author} {\bibfnamefont {T.}~\bibnamefont {Iadecola}}, \bibinfo {author} {\bibfnamefont {V.}~\bibnamefont {Khemani}},\ and\ \bibinfo {author} {\bibfnamefont {R.}~\bibnamefont {Moessner}},\ }\bibfield  {title} {\bibinfo {title} {{Quantum Many-Body Scars: A Quasiparticle Perspective}},\ }\href {https://doi.org/10.1146/annurev-conmatphys-031620-101617} {\bibfield  {journal} {\bibinfo  {journal} {Annu. Rev. Condens. Matter Phys.}\ }\textbf {\bibinfo {volume} {14}},\ \bibinfo {pages} {443} (\bibinfo {year} {2023})}\BibitemShut {NoStop}%
\bibitem [{\citenamefont {Moudgalya}\ \emph {et~al.}(2018)\citenamefont {Moudgalya}, \citenamefont {Regnault},\ and\ \citenamefont {Bernevig}}]{Moudgalya2018a}%
  \BibitemOpen
  \bibfield  {author} {\bibinfo {author} {\bibfnamefont {S.}~\bibnamefont {Moudgalya}}, \bibinfo {author} {\bibfnamefont {N.}~\bibnamefont {Regnault}},\ and\ \bibinfo {author} {\bibfnamefont {B.~A.}\ \bibnamefont {Bernevig}},\ }\bibfield  {title} {\bibinfo {title} {{Entanglement of exact excited states of Affleck-Kennedy-Lieb-Tasaki models: Exact results, many-body scars, and violation of the strong eigenstate thermalization hypothesis}},\ }\href {https://doi.org/10.1103/PhysRevB.98.235156} {\bibfield  {journal} {\bibinfo  {journal} {Phys. Rev. B}\ }\textbf {\bibinfo {volume} {98}},\ \bibinfo {pages} {235156} (\bibinfo {year} {2018})}\BibitemShut {NoStop}%
\bibitem [{\citenamefont {Schecter}\ and\ \citenamefont {Iadecola}(2019)}]{Schecter2019a}%
  \BibitemOpen
  \bibfield  {author} {\bibinfo {author} {\bibfnamefont {M.}~\bibnamefont {Schecter}}\ and\ \bibinfo {author} {\bibfnamefont {T.}~\bibnamefont {Iadecola}},\ }\bibfield  {title} {\bibinfo {title} {{Weak Ergodicity Breaking and Quantum Many-Body Scars in Spin-1 $XY$ Magnets}},\ }\href {https://doi.org/10.1103/PhysRevLett.123.147201} {\bibfield  {journal} {\bibinfo  {journal} {Phys. Rev. Lett.}\ }\textbf {\bibinfo {volume} {123}},\ \bibinfo {pages} {147201} (\bibinfo {year} {2019})}\BibitemShut {NoStop}%
\bibitem [{\citenamefont {Pai}\ and\ \citenamefont {Pretko}(2019)}]{Pai2019}%
  \BibitemOpen
  \bibfield  {author} {\bibinfo {author} {\bibfnamefont {S.}~\bibnamefont {Pai}}\ and\ \bibinfo {author} {\bibfnamefont {M.}~\bibnamefont {Pretko}},\ }\bibfield  {title} {\bibinfo {title} {{Dynamical Scar States in Driven Fracton Systems}},\ }\href {https://doi.org/10.1103/PhysRevLett.123.136401} {\bibfield  {journal} {\bibinfo  {journal} {Phys. Rev. Lett.}\ }\textbf {\bibinfo {volume} {123}},\ \bibinfo {pages} {136401} (\bibinfo {year} {2019})}\BibitemShut {NoStop}%
\bibitem [{\citenamefont {Ren}\ \emph {et~al.}(2021)\citenamefont {Ren}, \citenamefont {Liang},\ and\ \citenamefont {Fang}}]{Ren2021a}%
  \BibitemOpen
  \bibfield  {author} {\bibinfo {author} {\bibfnamefont {J.}~\bibnamefont {Ren}}, \bibinfo {author} {\bibfnamefont {C.}~\bibnamefont {Liang}},\ and\ \bibinfo {author} {\bibfnamefont {C.}~\bibnamefont {Fang}},\ }\bibfield  {title} {\bibinfo {title} {{Quasisymmetry Groups and Many-Body Scar Dynamics}},\ }\href {https://doi.org/10.1103/PhysRevLett.126.120604} {\bibfield  {journal} {\bibinfo  {journal} {Phys. Rev. Lett.}\ }\textbf {\bibinfo {volume} {126}},\ \bibinfo {pages} {120604} (\bibinfo {year} {2021})}\BibitemShut {NoStop}%
\bibitem [{\citenamefont {Yu}\ \emph {et~al.}(2018)\citenamefont {Yu}, \citenamefont {Luo},\ and\ \citenamefont {Clark}}]{Yu2018e}%
  \BibitemOpen
  \bibfield  {author} {\bibinfo {author} {\bibfnamefont {X.}~\bibnamefont {Yu}}, \bibinfo {author} {\bibfnamefont {D.}~\bibnamefont {Luo}},\ and\ \bibinfo {author} {\bibfnamefont {B.~K.}\ \bibnamefont {Clark}},\ }\bibfield  {title} {\bibinfo {title} {{Beyond many-body localized states in a spin-disordered Hubbard model}},\ }\href {https://doi.org/10.1103/PhysRevB.98.115106} {\bibfield  {journal} {\bibinfo  {journal} {Phys. Rev. B}\ }\textbf {\bibinfo {volume} {98}},\ \bibinfo {pages} {115106} (\bibinfo {year} {2018})}\BibitemShut {NoStop}%
\bibitem [{\citenamefont {Scherg}\ \emph {et~al.}(2021)\citenamefont {Scherg}, \citenamefont {Kohlert}, \citenamefont {Sala}, \citenamefont {Pollmann}, \citenamefont {{Hebbe Madhusudhana}}, \citenamefont {Bloch},\ and\ \citenamefont {Aidelsburger}}]{Scherg2021}%
  \BibitemOpen
  \bibfield  {author} {\bibinfo {author} {\bibfnamefont {S.}~\bibnamefont {Scherg}}, \bibinfo {author} {\bibfnamefont {T.}~\bibnamefont {Kohlert}}, \bibinfo {author} {\bibfnamefont {P.}~\bibnamefont {Sala}}, \bibinfo {author} {\bibfnamefont {F.}~\bibnamefont {Pollmann}}, \bibinfo {author} {\bibfnamefont {B.}~\bibnamefont {{Hebbe Madhusudhana}}}, \bibinfo {author} {\bibfnamefont {I.}~\bibnamefont {Bloch}},\ and\ \bibinfo {author} {\bibfnamefont {M.}~\bibnamefont {Aidelsburger}},\ }\bibfield  {title} {\bibinfo {title} {{Observing non-ergodicity due to kinetic constraints in tilted Fermi-Hubbard chains}},\ }\href {https://doi.org/10.1038/s41467-021-24726-0} {\bibfield  {journal} {\bibinfo  {journal} {Nat. Commun.}\ }\textbf {\bibinfo {volume} {12}},\ \bibinfo {pages} {4490} (\bibinfo {year} {2021})}\BibitemShut {NoStop}%
\bibitem [{\citenamefont {Chattopadhyay}\ \emph {et~al.}(2020)\citenamefont {Chattopadhyay}, \citenamefont {Pichler}, \citenamefont {Lukin},\ and\ \citenamefont {Ho}}]{Chattopadhyay2020a}%
  \BibitemOpen
  \bibfield  {author} {\bibinfo {author} {\bibfnamefont {S.}~\bibnamefont {Chattopadhyay}}, \bibinfo {author} {\bibfnamefont {H.}~\bibnamefont {Pichler}}, \bibinfo {author} {\bibfnamefont {M.~D.}\ \bibnamefont {Lukin}},\ and\ \bibinfo {author} {\bibfnamefont {W.~W.}\ \bibnamefont {Ho}},\ }\bibfield  {title} {\bibinfo {title} {{Quantum many-body scars from virtual entangled pairs}},\ }\href {https://doi.org/10.1103/PhysRevB.101.174308} {\bibfield  {journal} {\bibinfo  {journal} {Phys. Rev. B}\ }\textbf {\bibinfo {volume} {101}},\ \bibinfo {pages} {174308} (\bibinfo {year} {2020})}\BibitemShut {NoStop}%
\bibitem [{\citenamefont {Lin}\ \emph {et~al.}(2020)\citenamefont {Lin}, \citenamefont {Calvera},\ and\ \citenamefont {Hsieh}}]{Lin2020a}%
  \BibitemOpen
  \bibfield  {author} {\bibinfo {author} {\bibfnamefont {C.-J.}\ \bibnamefont {Lin}}, \bibinfo {author} {\bibfnamefont {V.}~\bibnamefont {Calvera}},\ and\ \bibinfo {author} {\bibfnamefont {T.~H.}\ \bibnamefont {Hsieh}},\ }\bibfield  {title} {\bibinfo {title} {{Quantum many-body scar states in two-dimensional Rydberg atom arrays}},\ }\href {https://doi.org/10.1103/PhysRevB.101.220304} {\bibfield  {journal} {\bibinfo  {journal} {Phys. Rev. B}\ }\textbf {\bibinfo {volume} {101}},\ \bibinfo {pages} {220304} (\bibinfo {year} {2020})}\BibitemShut {NoStop}%
\bibitem [{\citenamefont {Kuno}\ \emph {et~al.}(2020)\citenamefont {Kuno}, \citenamefont {Mizoguchi},\ and\ \citenamefont {Hatsugai}}]{Kuno2020}%
  \BibitemOpen
  \bibfield  {author} {\bibinfo {author} {\bibfnamefont {Y.}~\bibnamefont {Kuno}}, \bibinfo {author} {\bibfnamefont {T.}~\bibnamefont {Mizoguchi}},\ and\ \bibinfo {author} {\bibfnamefont {Y.}~\bibnamefont {Hatsugai}},\ }\bibfield  {title} {\bibinfo {title} {{Flat band quantum scar}},\ }\href {https://doi.org/10.1103/PhysRevB.102.241115} {\bibfield  {journal} {\bibinfo  {journal} {Phys. Rev. B}\ }\textbf {\bibinfo {volume} {102}},\ \bibinfo {pages} {241115} (\bibinfo {year} {2020})}\BibitemShut {NoStop}%
\bibitem [{\citenamefont {Sugiura}\ \emph {et~al.}(2021)\citenamefont {Sugiura}, \citenamefont {Kuwahara},\ and\ \citenamefont {Saito}}]{Sugiura2021}%
  \BibitemOpen
  \bibfield  {author} {\bibinfo {author} {\bibfnamefont {S.}~\bibnamefont {Sugiura}}, \bibinfo {author} {\bibfnamefont {T.}~\bibnamefont {Kuwahara}},\ and\ \bibinfo {author} {\bibfnamefont {K.}~\bibnamefont {Saito}},\ }\bibfield  {title} {\bibinfo {title} {{Many-body scar state intrinsic to periodically driven system}},\ }\href {https://doi.org/10.1103/PhysRevResearch.3.L012010} {\bibfield  {journal} {\bibinfo  {journal} {Phys. Rev. Res.}\ }\textbf {\bibinfo {volume} {3}},\ \bibinfo {pages} {L012010} (\bibinfo {year} {2021})}\BibitemShut {NoStop}%
\bibitem [{\citenamefont {Zhang}\ \emph {et~al.}(2023)\citenamefont {Zhang}, \citenamefont {Dong}, \citenamefont {Gao}, \citenamefont {Zhao}, \citenamefont {Hao}, \citenamefont {Desaules}, \citenamefont {Guo}, \citenamefont {Chen}, \citenamefont {Deng}, \citenamefont {Liu}, \citenamefont {Ren}, \citenamefont {Yao}, \citenamefont {Zhang}, \citenamefont {Xu}, \citenamefont {Wang}, \citenamefont {Jin}, \citenamefont {Zhu}, \citenamefont {Zhang}, \citenamefont {Li}, \citenamefont {Song}, \citenamefont {Wang}, \citenamefont {Liu}, \citenamefont {Papi{\'{c}}}, \citenamefont {Ying}, \citenamefont {Wang},\ and\ \citenamefont {Lai}}]{Zhang2023h}%
  \BibitemOpen
  \bibfield  {author} {\bibinfo {author} {\bibfnamefont {P.}~\bibnamefont {Zhang}}, \bibinfo {author} {\bibfnamefont {H.}~\bibnamefont {Dong}}, \bibinfo {author} {\bibfnamefont {Y.}~\bibnamefont {Gao}}, \bibinfo {author} {\bibfnamefont {L.}~\bibnamefont {Zhao}}, \bibinfo {author} {\bibfnamefont {J.}~\bibnamefont {Hao}}, \bibinfo {author} {\bibfnamefont {J.-Y.}\ \bibnamefont {Desaules}}, \bibinfo {author} {\bibfnamefont {Q.}~\bibnamefont {Guo}}, \bibinfo {author} {\bibfnamefont {J.}~\bibnamefont {Chen}}, \bibinfo {author} {\bibfnamefont {J.}~\bibnamefont {Deng}}, \bibinfo {author} {\bibfnamefont {B.}~\bibnamefont {Liu}}, \bibinfo {author} {\bibfnamefont {W.}~\bibnamefont {Ren}}, \bibinfo {author} {\bibfnamefont {Y.}~\bibnamefont {Yao}}, \bibinfo {author} {\bibfnamefont {X.}~\bibnamefont {Zhang}}, \bibinfo {author} {\bibfnamefont {S.}~\bibnamefont {Xu}}, \bibinfo {author} {\bibfnamefont {K.}~\bibnamefont {Wang}}, \bibinfo {author} {\bibfnamefont {F.}~\bibnamefont {Jin}}, \bibinfo {author} {\bibfnamefont
  {X.}~\bibnamefont {Zhu}}, \bibinfo {author} {\bibfnamefont {B.}~\bibnamefont {Zhang}}, \bibinfo {author} {\bibfnamefont {H.}~\bibnamefont {Li}}, \bibinfo {author} {\bibfnamefont {C.}~\bibnamefont {Song}}, \bibinfo {author} {\bibfnamefont {Z.}~\bibnamefont {Wang}}, \bibinfo {author} {\bibfnamefont {F.}~\bibnamefont {Liu}}, \bibinfo {author} {\bibfnamefont {Z.}~\bibnamefont {Papi{\'{c}}}}, \bibinfo {author} {\bibfnamefont {L.}~\bibnamefont {Ying}}, \bibinfo {author} {\bibfnamefont {H.}~\bibnamefont {Wang}},\ and\ \bibinfo {author} {\bibfnamefont {Y.-C.}\ \bibnamefont {Lai}},\ }\bibfield  {title} {\bibinfo {title} {{Many-body Hilbert space scarring on a superconducting processor}},\ }\href {https://doi.org/10.1038/s41567-022-01784-9} {\bibfield  {journal} {\bibinfo  {journal} {Nat. Phys.}\ }\textbf {\bibinfo {volume} {19}},\ \bibinfo {pages} {120} (\bibinfo {year} {2023})}\BibitemShut {NoStop}%
\bibitem [{\citenamefont {Su}\ \emph {et~al.}(2023)\citenamefont {Su}, \citenamefont {Sun}, \citenamefont {Hudomal}, \citenamefont {Desaules}, \citenamefont {Zhou}, \citenamefont {Yang}, \citenamefont {Halimeh}, \citenamefont {Yuan}, \citenamefont {Papi{\'{c}}},\ and\ \citenamefont {Pan}}]{Su2023}%
  \BibitemOpen
  \bibfield  {author} {\bibinfo {author} {\bibfnamefont {G.-X.}\ \bibnamefont {Su}}, \bibinfo {author} {\bibfnamefont {H.}~\bibnamefont {Sun}}, \bibinfo {author} {\bibfnamefont {A.}~\bibnamefont {Hudomal}}, \bibinfo {author} {\bibfnamefont {J.-Y.}\ \bibnamefont {Desaules}}, \bibinfo {author} {\bibfnamefont {Z.-Y.}\ \bibnamefont {Zhou}}, \bibinfo {author} {\bibfnamefont {B.}~\bibnamefont {Yang}}, \bibinfo {author} {\bibfnamefont {J.~C.}\ \bibnamefont {Halimeh}}, \bibinfo {author} {\bibfnamefont {Z.-S.}\ \bibnamefont {Yuan}}, \bibinfo {author} {\bibfnamefont {Z.}~\bibnamefont {Papi{\'{c}}}},\ and\ \bibinfo {author} {\bibfnamefont {J.-W.}\ \bibnamefont {Pan}},\ }\bibfield  {title} {\bibinfo {title} {{Observation of many-body scarring in a Bose-Hubbard quantum simulator}},\ }\href {https://doi.org/10.1103/PhysRevResearch.5.023010} {\bibfield  {journal} {\bibinfo  {journal} {Phys. Rev. Res.}\ }\textbf {\bibinfo {volume} {5}},\ \bibinfo {pages} {023010} (\bibinfo {year} {2023})}\BibitemShut {NoStop}%
\bibitem [{\citenamefont {Omiya}\ and\ \citenamefont {M{\"{u}}ller}(2023{\natexlab{a}})}]{Omiya2023}%
  \BibitemOpen
  \bibfield  {author} {\bibinfo {author} {\bibfnamefont {K.}~\bibnamefont {Omiya}}\ and\ \bibinfo {author} {\bibfnamefont {M.}~\bibnamefont {M{\"{u}}ller}},\ }\bibfield  {title} {\bibinfo {title} {{Quantum many-body scars in bipartite Rydberg arrays originating from hidden projector embedding}},\ }\href {https://doi.org/10.1103/PhysRevA.107.023318} {\bibfield  {journal} {\bibinfo  {journal} {Phys. Rev. A}\ }\textbf {\bibinfo {volume} {107}},\ \bibinfo {pages} {023318} (\bibinfo {year} {2023}{\natexlab{a}})}\BibitemShut {NoStop}%
\bibitem [{\citenamefont {Omiya}\ and\ \citenamefont {M{\"{u}}ller}(2023{\natexlab{b}})}]{Omiya2023a}%
  \BibitemOpen
  \bibfield  {author} {\bibinfo {author} {\bibfnamefont {K.}~\bibnamefont {Omiya}}\ and\ \bibinfo {author} {\bibfnamefont {M.}~\bibnamefont {M{\"{u}}ller}},\ }\bibfield  {title} {\bibinfo {title} {{Fractionalization paves the way to local projector embeddings of quantum many-body scars}},\ }\href {https://doi.org/10.1103/PhysRevB.108.054412} {\bibfield  {journal} {\bibinfo  {journal} {Phys. Rev. B}\ }\textbf {\bibinfo {volume} {108}},\ \bibinfo {pages} {054412} (\bibinfo {year} {2023}{\natexlab{b}})}\BibitemShut {NoStop}%
\bibitem [{\citenamefont {Desaules}\ \emph {et~al.}(2021)\citenamefont {Desaules}, \citenamefont {Hudomal}, \citenamefont {Turner},\ and\ \citenamefont {Papi{\'{c}}}}]{Desaules2021}%
  \BibitemOpen
  \bibfield  {author} {\bibinfo {author} {\bibfnamefont {J.-Y.}\ \bibnamefont {Desaules}}, \bibinfo {author} {\bibfnamefont {A.}~\bibnamefont {Hudomal}}, \bibinfo {author} {\bibfnamefont {C.~J.}\ \bibnamefont {Turner}},\ and\ \bibinfo {author} {\bibfnamefont {Z.}~\bibnamefont {Papi{\'{c}}}},\ }\bibfield  {title} {\bibinfo {title} {{Proposal for Realizing Quantum Scars in the Tilted 1D Fermi-Hubbard Model}},\ }\href {https://doi.org/10.1103/PhysRevLett.126.210601} {\bibfield  {journal} {\bibinfo  {journal} {Phys. Rev. Lett.}\ }\textbf {\bibinfo {volume} {126}},\ \bibinfo {pages} {210601} (\bibinfo {year} {2021})}\BibitemShut {NoStop}%
\bibitem [{\citenamefont {Kaneko}\ \emph {et~al.}(2024)\citenamefont {Kaneko}, \citenamefont {Kunimi},\ and\ \citenamefont {Danshita}}]{Kaneko2024}%
  \BibitemOpen
  \bibfield  {author} {\bibinfo {author} {\bibfnamefont {R.}~\bibnamefont {Kaneko}}, \bibinfo {author} {\bibfnamefont {M.}~\bibnamefont {Kunimi}},\ and\ \bibinfo {author} {\bibfnamefont {I.}~\bibnamefont {Danshita}},\ }\bibfield  {title} {\bibinfo {title} {{Quantum many-body scars in the Bose-Hubbard model with a three-body constraint}},\ }\href {https://doi.org/10.1103/PhysRevA.109.L011301} {\bibfield  {journal} {\bibinfo  {journal} {Phys. Rev. A}\ }\textbf {\bibinfo {volume} {109}},\ \bibinfo {pages} {L011301} (\bibinfo {year} {2024})}\BibitemShut {NoStop}%
\bibitem [{\citenamefont {Matsui}(2024)}]{Matsui2024}%
  \BibitemOpen
  \bibfield  {author} {\bibinfo {author} {\bibfnamefont {C.}~\bibnamefont {Matsui}},\ }\bibfield  {title} {\bibinfo {title} {{Exactly solvable subspaces of nonintegrable spin chains with boundaries and quasiparticle interactions}},\ }\href {https://doi.org/10.1103/PhysRevB.109.104307} {\bibfield  {journal} {\bibinfo  {journal} {Phys. Rev. B}\ }\textbf {\bibinfo {volume} {109}},\ \bibinfo {pages} {104307} (\bibinfo {year} {2024})}\BibitemShut {NoStop}%
\bibitem [{\citenamefont {Pakrouski}\ \emph {et~al.}(2020)\citenamefont {Pakrouski}, \citenamefont {Pallegar}, \citenamefont {Popov},\ and\ \citenamefont {Klebanov}}]{Pakrouski2020a}%
  \BibitemOpen
  \bibfield  {author} {\bibinfo {author} {\bibfnamefont {K.}~\bibnamefont {Pakrouski}}, \bibinfo {author} {\bibfnamefont {P.~N.}\ \bibnamefont {Pallegar}}, \bibinfo {author} {\bibfnamefont {F.~K.}\ \bibnamefont {Popov}},\ and\ \bibinfo {author} {\bibfnamefont {I.~R.}\ \bibnamefont {Klebanov}},\ }\bibfield  {title} {\bibinfo {title} {{Many-Body Scars as a Group Invariant Sector of Hilbert Space}},\ }\href {https://doi.org/10.1103/PhysRevLett.125.230602} {\bibfield  {journal} {\bibinfo  {journal} {Phys. Rev. Lett.}\ }\textbf {\bibinfo {volume} {125}},\ \bibinfo {pages} {230602} (\bibinfo {year} {2020})}\BibitemShut {NoStop}%
\bibitem [{\citenamefont {Yang}(1989)}]{Yang1989}%
  \BibitemOpen
  \bibfield  {author} {\bibinfo {author} {\bibfnamefont {C.~N.}\ \bibnamefont {Yang}},\ }\bibfield  {title} {\bibinfo {title} {{$\eta$ pairing and off-diagonal long-range order in a Hubbard model}},\ }\href {https://doi.org/10.1103/PhysRevLett.63.2144} {\bibfield  {journal} {\bibinfo  {journal} {Phys. Rev. Lett.}\ }\textbf {\bibinfo {volume} {63}},\ \bibinfo {pages} {2144} (\bibinfo {year} {1989})}\BibitemShut {NoStop}%
\bibitem [{\citenamefont {Yang}\ and\ \citenamefont {Zhang}(1990)}]{Yang1990}%
  \BibitemOpen
  \bibfield  {author} {\bibinfo {author} {\bibfnamefont {C.~N.}\ \bibnamefont {Yang}}\ and\ \bibinfo {author} {\bibfnamefont {S.}~\bibnamefont {Zhang}},\ }\bibfield  {title} {\bibinfo {title} {{${\mathrm{SO}}_4$ SYMMETRY IN A HUBBARD MODEL}},\ }\href {https://doi.org/10.1142/S0217984990000933} {\bibfield  {journal} {\bibinfo  {journal} {Mod. Phys. Lett. B}\ }\textbf {\bibinfo {volume} {04}},\ \bibinfo {pages} {759} (\bibinfo {year} {1990})}\BibitemShut {NoStop}%
\bibitem [{\citenamefont {Vafek}\ \emph {et~al.}(2017)\citenamefont {Vafek}, \citenamefont {Regnault},\ and\ \citenamefont {Bernevig}}]{Vafek2017}%
  \BibitemOpen
  \bibfield  {author} {\bibinfo {author} {\bibfnamefont {O.}~\bibnamefont {Vafek}}, \bibinfo {author} {\bibfnamefont {N.}~\bibnamefont {Regnault}},\ and\ \bibinfo {author} {\bibfnamefont {B.~A.}\ \bibnamefont {Bernevig}},\ }\bibfield  {title} {\bibinfo {title} {{Entanglement of exact excited eigenstates of the Hubbard model in arbitrary dimension}},\ }\href {https://doi.org/10.21468/SciPostPhys.3.6.043} {\bibfield  {journal} {\bibinfo  {journal} {SciPost Phys.}\ }\textbf {\bibinfo {volume} {3}},\ \bibinfo {pages} {043} (\bibinfo {year} {2017})}\BibitemShut {NoStop}%
\bibitem [{\citenamefont {Li}(2020)}]{Li2020l}%
  \BibitemOpen
  \bibfield  {author} {\bibinfo {author} {\bibfnamefont {K.}~\bibnamefont {Li}},\ }\bibfield  {title} {\bibinfo {title} {$\eta$-pairing in correlated fermion models with spin-orbit coupling},\ }\href {https://doi.org/10.1103/PhysRevB.102.165150} {\bibfield  {journal} {\bibinfo  {journal} {Phys. Rev. B}\ }\textbf {\bibinfo {volume} {102}},\ \bibinfo {pages} {165150} (\bibinfo {year} {2020})}\BibitemShut {NoStop}%
\bibitem [{\citenamefont {Mark}\ and\ \citenamefont {Motrunich}(2020)}]{Mark2020}%
  \BibitemOpen
  \bibfield  {author} {\bibinfo {author} {\bibfnamefont {D.~K.}\ \bibnamefont {Mark}}\ and\ \bibinfo {author} {\bibfnamefont {O.~I.}\ \bibnamefont {Motrunich}},\ }\bibfield  {title} {\bibinfo {title} {{$\eta$-pairing states as true scars in an extended Hubbard model}},\ }\href {https://doi.org/10.1103/PhysRevB.102.075132} {\bibfield  {journal} {\bibinfo  {journal} {Phys. Rev. B}\ }\textbf {\bibinfo {volume} {102}},\ \bibinfo {pages} {075132} (\bibinfo {year} {2020})}\BibitemShut {NoStop}%
\bibitem [{\citenamefont {Moudgalya}\ \emph {et~al.}(2020)\citenamefont {Moudgalya}, \citenamefont {Regnault},\ and\ \citenamefont {Bernevig}}]{Moudgalya2020}%
  \BibitemOpen
  \bibfield  {author} {\bibinfo {author} {\bibfnamefont {S.}~\bibnamefont {Moudgalya}}, \bibinfo {author} {\bibfnamefont {N.}~\bibnamefont {Regnault}},\ and\ \bibinfo {author} {\bibfnamefont {B.~A.}\ \bibnamefont {Bernevig}},\ }\bibfield  {title} {\bibinfo {title} {{$\eta$-pairing in Hubbard models: From spectrum generating algebras to quantum many-body scars}},\ }\href {https://doi.org/10.1103/PhysRevB.102.085140} {\bibfield  {journal} {\bibinfo  {journal} {Phys. Rev. B}\ }\textbf {\bibinfo {volume} {102}},\ \bibinfo {pages} {085140} (\bibinfo {year} {2020})}\BibitemShut {NoStop}%
\bibitem [{\citenamefont {Pakrouski}\ \emph {et~al.}(2021)\citenamefont {Pakrouski}, \citenamefont {Pallegar}, \citenamefont {Popov},\ and\ \citenamefont {Klebanov}}]{Pakrouski2021}%
  \BibitemOpen
  \bibfield  {author} {\bibinfo {author} {\bibfnamefont {K.}~\bibnamefont {Pakrouski}}, \bibinfo {author} {\bibfnamefont {P.~N.}\ \bibnamefont {Pallegar}}, \bibinfo {author} {\bibfnamefont {F.~K.}\ \bibnamefont {Popov}},\ and\ \bibinfo {author} {\bibfnamefont {I.~R.}\ \bibnamefont {Klebanov}},\ }\bibfield  {title} {\bibinfo {title} {{Group theoretic approach to many-body scar states in fermionic lattice models}},\ }\href {https://doi.org/10.1103/PhysRevResearch.3.043156} {\bibfield  {journal} {\bibinfo  {journal} {Phys. Rev. Res.}\ }\textbf {\bibinfo {volume} {3}},\ \bibinfo {pages} {043156} (\bibinfo {year} {2021})}\BibitemShut {NoStop}%
\bibitem [{\citenamefont {Wildeboer}\ \emph {et~al.}(2022)\citenamefont {Wildeboer}, \citenamefont {Langlett}, \citenamefont {Yang}, \citenamefont {Gorshkov}, \citenamefont {Iadecola},\ and\ \citenamefont {Xu}}]{Wildeboer2022a}%
  \BibitemOpen
  \bibfield  {author} {\bibinfo {author} {\bibfnamefont {J.}~\bibnamefont {Wildeboer}}, \bibinfo {author} {\bibfnamefont {C.~M.}\ \bibnamefont {Langlett}}, \bibinfo {author} {\bibfnamefont {Z.-C.}\ \bibnamefont {Yang}}, \bibinfo {author} {\bibfnamefont {A.~V.}\ \bibnamefont {Gorshkov}}, \bibinfo {author} {\bibfnamefont {T.}~\bibnamefont {Iadecola}},\ and\ \bibinfo {author} {\bibfnamefont {S.}~\bibnamefont {Xu}},\ }\bibfield  {title} {\bibinfo {title} {{Quantum many-body scars from Einstein-Podolsky-Rosen states in bilayer systems}},\ }\href {https://doi.org/10.1103/PhysRevB.106.205142} {\bibfield  {journal} {\bibinfo  {journal} {Phys. Rev. B}\ }\textbf {\bibinfo {volume} {106}},\ \bibinfo {pages} {205142} (\bibinfo {year} {2022})}\BibitemShut {NoStop}%
\bibitem [{\citenamefont {Sun}\ \emph {et~al.}(2023)\citenamefont {Sun}, \citenamefont {Popov}, \citenamefont {Klebanov},\ and\ \citenamefont {Pakrouski}}]{Sun2023b}%
  \BibitemOpen
  \bibfield  {author} {\bibinfo {author} {\bibfnamefont {Z.}~\bibnamefont {Sun}}, \bibinfo {author} {\bibfnamefont {F.~K.}\ \bibnamefont {Popov}}, \bibinfo {author} {\bibfnamefont {I.~R.}\ \bibnamefont {Klebanov}},\ and\ \bibinfo {author} {\bibfnamefont {K.}~\bibnamefont {Pakrouski}},\ }\bibfield  {title} {\bibinfo {title} {{Majorana scars as group singlets}},\ }\href {https://doi.org/10.1103/PhysRevResearch.5.043208} {\bibfield  {journal} {\bibinfo  {journal} {Phys. Rev. Res.}\ }\textbf {\bibinfo {volume} {5}},\ \bibinfo {pages} {043208} (\bibinfo {year} {2023})}\BibitemShut {NoStop}%
\bibitem [{\citenamefont {Kolb}\ and\ \citenamefont {Pakrouski}(2023)}]{Kolb2023a}%
  \BibitemOpen
  \bibfield  {author} {\bibinfo {author} {\bibfnamefont {P.}~\bibnamefont {Kolb}}\ and\ \bibinfo {author} {\bibfnamefont {K.}~\bibnamefont {Pakrouski}},\ }\bibfield  {title} {\bibinfo {title} {{Stability of the Many-Body Scars in Fermionic Spin-1/2 Models}},\ }\href {https://doi.org/10.1103/PRXQuantum.4.040348} {\bibfield  {journal} {\bibinfo  {journal} {PRX Quantum}\ }\textbf {\bibinfo {volume} {4}},\ \bibinfo {pages} {040348} (\bibinfo {year} {2023})}\BibitemShut {NoStop}%
\bibitem [{\citenamefont {Hoshino}(2014)}]{Hoshino2014}%
  \BibitemOpen
  \bibfield  {author} {\bibinfo {author} {\bibfnamefont {S.}~\bibnamefont {Hoshino}},\ }\bibfield  {title} {\bibinfo {title} {{Mean-field description of odd-frequency superconductivity with staggered ordering vector}},\ }\href {https://doi.org/10.1103/PhysRevB.90.115154} {\bibfield  {journal} {\bibinfo  {journal} {Phys. Rev. B}\ }\textbf {\bibinfo {volume} {90}},\ \bibinfo {pages} {115154} (\bibinfo {year} {2014})}\BibitemShut {NoStop}%
\bibitem [{\citenamefont {Tsuji}\ \emph {et~al.}()\citenamefont {Tsuji}, \citenamefont {Nakagawa},\ and\ \citenamefont {Ueda}}]{Tsuji2021}%
  \BibitemOpen
  \bibfield  {author} {\bibinfo {author} {\bibfnamefont {N.}~\bibnamefont {Tsuji}}, \bibinfo {author} {\bibfnamefont {M.}~\bibnamefont {Nakagawa}},\ and\ \bibinfo {author} {\bibfnamefont {M.}~\bibnamefont {Ueda}},\ }\bibfield  {title} {\bibinfo {title} {{Tachyonic and Plasma Instabilities of $\eta$-Pairing States Coupled to Electromagnetic Fields}},\ }\Eprint {https://arxiv.org/abs/2103.01547} {arXiv:2103.01547} \BibitemShut {NoStop}%
\bibitem [{\citenamefont {Shibata}\ \emph {et~al.}(2020)\citenamefont {Shibata}, \citenamefont {Yoshioka},\ and\ \citenamefont {Katsura}}]{Shibata2020}%
  \BibitemOpen
  \bibfield  {author} {\bibinfo {author} {\bibfnamefont {N.}~\bibnamefont {Shibata}}, \bibinfo {author} {\bibfnamefont {N.}~\bibnamefont {Yoshioka}},\ and\ \bibinfo {author} {\bibfnamefont {H.}~\bibnamefont {Katsura}},\ }\bibfield  {title} {\bibinfo {title} {{Onsager's Scars in Disordered Spin Chains}},\ }\href {https://doi.org/10.1103/PhysRevLett.124.180604} {\bibfield  {journal} {\bibinfo  {journal} {Phys. Rev. Lett.}\ }\textbf {\bibinfo {volume} {124}},\ \bibinfo {pages} {180604} (\bibinfo {year} {2020})}\BibitemShut {NoStop}%
\bibitem [{\citenamefont {Tamura}\ and\ \citenamefont {Katsura}(2022)}]{Tamura2022}%
  \BibitemOpen
  \bibfield  {author} {\bibinfo {author} {\bibfnamefont {K.}~\bibnamefont {Tamura}}\ and\ \bibinfo {author} {\bibfnamefont {H.}~\bibnamefont {Katsura}},\ }\bibfield  {title} {\bibinfo {title} {{Quantum many-body scars of spinless fermions with density-assisted hopping in higher dimensions}},\ }\href {https://doi.org/10.1103/PhysRevB.106.144306} {\bibfield  {journal} {\bibinfo  {journal} {Phys. Rev. B}\ }\textbf {\bibinfo {volume} {106}},\ \bibinfo {pages} {144306} (\bibinfo {year} {2022})}\BibitemShut {NoStop}%
\bibitem [{\citenamefont {Gotta}\ \emph {et~al.}(2022)\citenamefont {Gotta}, \citenamefont {Mazza}, \citenamefont {Simon},\ and\ \citenamefont {Roux}}]{Gotta2022}%
  \BibitemOpen
  \bibfield  {author} {\bibinfo {author} {\bibfnamefont {L.}~\bibnamefont {Gotta}}, \bibinfo {author} {\bibfnamefont {L.}~\bibnamefont {Mazza}}, \bibinfo {author} {\bibfnamefont {P.}~\bibnamefont {Simon}},\ and\ \bibinfo {author} {\bibfnamefont {G.}~\bibnamefont {Roux}},\ }\bibfield  {title} {\bibinfo {title} {{Exact many-body scars based on pairs or multimers in a chain of spinless fermions}},\ }\href {https://doi.org/10.1103/PhysRevB.106.235147} {\bibfield  {journal} {\bibinfo  {journal} {Phys. Rev. B}\ }\textbf {\bibinfo {volume} {106}},\ \bibinfo {pages} {235147} (\bibinfo {year} {2022})}\BibitemShut {NoStop}%
\bibitem [{\citenamefont {Zhai}(2005)}]{Zhai2005}%
  \BibitemOpen
  \bibfield  {author} {\bibinfo {author} {\bibfnamefont {H.}~\bibnamefont {Zhai}},\ }\bibfield  {title} {\bibinfo {title} {{Two generalizations of $\eta$ pairing in extended Hubbard models}},\ }\href {https://doi.org/10.1103/PhysRevB.71.012512} {\bibfield  {journal} {\bibinfo  {journal} {Phys. Rev. B}\ }\textbf {\bibinfo {volume} {71}},\ \bibinfo {pages} {012512} (\bibinfo {year} {2005})}\BibitemShut {NoStop}%
\bibitem [{\citenamefont {Nakagawa}\ \emph {et~al.}(2024)\citenamefont {Nakagawa}, \citenamefont {Katsura},\ and\ \citenamefont {Ueda}}]{Nakagawa2022a}%
  \BibitemOpen
  \bibfield  {author} {\bibinfo {author} {\bibfnamefont {M.}~\bibnamefont {Nakagawa}}, \bibinfo {author} {\bibfnamefont {H.}~\bibnamefont {Katsura}},\ and\ \bibinfo {author} {\bibfnamefont {M.}~\bibnamefont {Ueda}},\ }\bibfield  {title} {\bibinfo {title} {{Exact eigenstates of multicomponent Hubbard models: SU($N$) magnetic $\ensuremath{\eta}$ pairing, weak ergodicity breaking, and partial integrability}},\ }\href {https://doi.org/10.1103/PhysRevResearch.6.043259} {\bibfield  {journal} {\bibinfo  {journal} {Phys. Rev. Res.}\ }\textbf {\bibinfo {volume} {6}},\ \bibinfo {pages} {043259} (\bibinfo {year} {2024})}\BibitemShut {NoStop}%
\bibitem [{\citenamefont {Yoshida}\ and\ \citenamefont {Katsura}(2022)}]{Yoshida2022a}%
  \BibitemOpen
  \bibfield  {author} {\bibinfo {author} {\bibfnamefont {H.}~\bibnamefont {Yoshida}}\ and\ \bibinfo {author} {\bibfnamefont {H.}~\bibnamefont {Katsura}},\ }\bibfield  {title} {\bibinfo {title} {{Exact eigenstates of extended $\mathrm{SU}(N)$ Hubbard models: Generalization of $\ensuremath{\eta}$-pairing states with $N$-particle off-diagonal long-range order}},\ }\href {https://doi.org/10.1103/PhysRevB.105.024520} {\bibfield  {journal} {\bibinfo  {journal} {Phys. Rev. B}\ }\textbf {\bibinfo {volume} {105}},\ \bibinfo {pages} {024520} (\bibinfo {year} {2022})}\BibitemShut {NoStop}%
\bibitem [{\citenamefont {Ray}\ \emph {et~al.}(2023)\citenamefont {Ray}, \citenamefont {Murakami},\ and\ \citenamefont {Werner}}]{Ray2023a}%
  \BibitemOpen
  \bibfield  {author} {\bibinfo {author} {\bibfnamefont {S.}~\bibnamefont {Ray}}, \bibinfo {author} {\bibfnamefont {Y.}~\bibnamefont {Murakami}},\ and\ \bibinfo {author} {\bibfnamefont {P.}~\bibnamefont {Werner}},\ }\bibfield  {title} {\bibinfo {title} {{Nonthermal superconductivity in photodoped multiorbital Hubbard systems}},\ }\href {https://doi.org/10.1103/PhysRevB.108.174515} {\bibfield  {journal} {\bibinfo  {journal} {Phys. Rev. B}\ }\textbf {\bibinfo {volume} {108}},\ \bibinfo {pages} {174515} (\bibinfo {year} {2023})}\BibitemShut {NoStop}%
\bibitem [{\citenamefont {Linder}\ and\ \citenamefont {Balatsky}(2019)}]{Linder2019}%
  \BibitemOpen
  \bibfield  {author} {\bibinfo {author} {\bibfnamefont {J.}~\bibnamefont {Linder}}\ and\ \bibinfo {author} {\bibfnamefont {A.~V.}\ \bibnamefont {Balatsky}},\ }\bibfield  {title} {\bibinfo {title} {{Odd-frequency superconductivity}},\ }\href {https://doi.org/10.1103/RevModPhys.91.045005} {\bibfield  {journal} {\bibinfo  {journal} {Rev. Mod. Phys.}\ }\textbf {\bibinfo {volume} {91}},\ \bibinfo {pages} {045005} (\bibinfo {year} {2019})}\BibitemShut {NoStop}%
\bibitem [{\citenamefont {Lu}\ \emph {et~al.}(2014)\citenamefont {Lu}, \citenamefont {Xiang},\ and\ \citenamefont {Lee}}]{Lu2014b}%
  \BibitemOpen
  \bibfield  {author} {\bibinfo {author} {\bibfnamefont {Y.-M.}\ \bibnamefont {Lu}}, \bibinfo {author} {\bibfnamefont {T.}~\bibnamefont {Xiang}},\ and\ \bibinfo {author} {\bibfnamefont {D.-H.}\ \bibnamefont {Lee}},\ }\bibfield  {title} {\bibinfo {title} {{Underdoped superconducting cuprates as topological superconductors}},\ }\href {https://doi.org/10.1038/nphys3021} {\bibfield  {journal} {\bibinfo  {journal} {Nat. Phys.}\ }\textbf {\bibinfo {volume} {10}},\ \bibinfo {pages} {634} (\bibinfo {year} {2014})}\BibitemShut {NoStop}%
\bibitem [{\citenamefont {Chen}\ \emph {et~al.}(2020)\citenamefont {Chen}, \citenamefont {Wang}, \citenamefont {Zhou},\ and\ \citenamefont {Wang}}]{Chen2020d}%
  \BibitemOpen
  \bibfield  {author} {\bibinfo {author} {\bibfnamefont {L.-H.}\ \bibnamefont {Chen}}, \bibinfo {author} {\bibfnamefont {D.}~\bibnamefont {Wang}}, \bibinfo {author} {\bibfnamefont {Y.}~\bibnamefont {Zhou}},\ and\ \bibinfo {author} {\bibfnamefont {Q.-H.}\ \bibnamefont {Wang}},\ }\bibfield  {title} {\bibinfo {title} {{Superconductivity, Pair Density Wave, and N{\'{e}}el Order in Cuprates*}},\ }\href {https://doi.org/10.1088/0256-307X/37/1/017403} {\bibfield  {journal} {\bibinfo  {journal} {Chinese Phys. Lett.}\ }\textbf {\bibinfo {volume} {37}},\ \bibinfo {pages} {017403} (\bibinfo {year} {2020})}\BibitemShut {NoStop}%
\bibitem [{\citenamefont {Georgiou}\ and\ \citenamefont {Varelogiannis}(2020)}]{Georgiou2020}%
  \BibitemOpen
  \bibfield  {author} {\bibinfo {author} {\bibfnamefont {M.}~\bibnamefont {Georgiou}}\ and\ \bibinfo {author} {\bibfnamefont {G.}~\bibnamefont {Varelogiannis}},\ }\bibfield  {title} {\bibinfo {title} {{Pair density waves in spinless media}},\ }\href {https://doi.org/10.1103/PhysRevB.102.094514} {\bibfield  {journal} {\bibinfo  {journal} {Phys. Rev. B}\ }\textbf {\bibinfo {volume} {102}},\ \bibinfo {pages} {094514} (\bibinfo {year} {2020})}\BibitemShut {NoStop}%
\bibitem [{\citenamefont {Zhu}\ \emph {et~al.}()\citenamefont {Zhu}, \citenamefont {Sun}, \citenamefont {Gong}, \citenamefont {Huang}, \citenamefont {Feng}, \citenamefont {Scalettar},\ and\ \citenamefont {Guo}}]{Zhu2024}%
  \BibitemOpen
  \bibfield  {author} {\bibinfo {author} {\bibfnamefont {X.}~\bibnamefont {Zhu}}, \bibinfo {author} {\bibfnamefont {J.}~\bibnamefont {Sun}}, \bibinfo {author} {\bibfnamefont {S.-S.}\ \bibnamefont {Gong}}, \bibinfo {author} {\bibfnamefont {W.}~\bibnamefont {Huang}}, \bibinfo {author} {\bibfnamefont {S.}~\bibnamefont {Feng}}, \bibinfo {author} {\bibfnamefont {R.~T.}\ \bibnamefont {Scalettar}},\ and\ \bibinfo {author} {\bibfnamefont {H.}~\bibnamefont {Guo}},\ }\bibfield  {title} {\bibinfo {title} {{Exact Demonstration of pair-density-wave superconductivity in the $\sigma_z$-Hubbard model}},\ }\Eprint {https://arxiv.org/abs/2404.11043} {arXiv:2404.11043} \BibitemShut {NoStop}%
\bibitem [{\citenamefont {Kitaev}(2001)}]{Kitaev2001}%
  \BibitemOpen
  \bibfield  {author} {\bibinfo {author} {\bibfnamefont {A.~Y.}\ \bibnamefont {Kitaev}},\ }\bibfield  {title} {\bibinfo {title} {{Unpaired Majorana fermions in quantum wires}},\ }\href {https://doi.org/10.1070/1063-7869/44/10S/S29} {\bibfield  {journal} {\bibinfo  {journal} {Phys.-Usp.}\ }\textbf {\bibinfo {volume} {44}},\ \bibinfo {pages} {131} (\bibinfo {year} {2001})}\BibitemShut {NoStop}%
\bibitem [{\citenamefont {Xu}\ and\ \citenamefont {Wu}(2022)}]{Xu2022b}%
  \BibitemOpen
  \bibfield  {author} {\bibinfo {author} {\bibfnamefont {S.}~\bibnamefont {Xu}}\ and\ \bibinfo {author} {\bibfnamefont {C.}~\bibnamefont {Wu}},\ }\bibfield  {title} {\bibinfo {title} {{Orbital-active Dirac materials from the symmetry principle}},\ }\href {https://doi.org/10.1007/s44214-022-00025-7} {\bibfield  {journal} {\bibinfo  {journal} {Quantum Front.}\ }\textbf {\bibinfo {volume} {1}},\ \bibinfo {pages} {24} (\bibinfo {year} {2022})}\BibitemShut {NoStop}%
\bibitem [{\citenamefont {Bianchi}\ \emph {et~al.}(2022)\citenamefont {Bianchi}, \citenamefont {Hackl}, \citenamefont {Kieburg}, \citenamefont {Rigol},\ and\ \citenamefont {Vidmar}}]{Bianchi2022}%
  \BibitemOpen
  \bibfield  {author} {\bibinfo {author} {\bibfnamefont {E.}~\bibnamefont {Bianchi}}, \bibinfo {author} {\bibfnamefont {L.}~\bibnamefont {Hackl}}, \bibinfo {author} {\bibfnamefont {M.}~\bibnamefont {Kieburg}}, \bibinfo {author} {\bibfnamefont {M.}~\bibnamefont {Rigol}},\ and\ \bibinfo {author} {\bibfnamefont {L.}~\bibnamefont {Vidmar}},\ }\bibfield  {title} {\bibinfo {title} {{Volume-Law Entanglement Entropy of Typical Pure Quantum States}},\ }\href {https://doi.org/10.1103/PRXQuantum.3.030201} {\bibfield  {journal} {\bibinfo  {journal} {PRX Quantum}\ }\textbf {\bibinfo {volume} {3}},\ \bibinfo {pages} {030201} (\bibinfo {year} {2022})}\BibitemShut {NoStop}%
\bibitem [{\citenamefont {Page}(1993)}]{Page1993}%
  \BibitemOpen
  \bibfield  {author} {\bibinfo {author} {\bibfnamefont {D.~N.}\ \bibnamefont {Page}},\ }\bibfield  {title} {\bibinfo {title} {{Average entropy of a subsystem}},\ }\href {https://doi.org/10.1103/PhysRevLett.71.1291} {\bibfield  {journal} {\bibinfo  {journal} {Phys. Rev. Lett.}\ }\textbf {\bibinfo {volume} {71}},\ \bibinfo {pages} {1291} (\bibinfo {year} {1993})}\BibitemShut {NoStop}%
\bibitem [{\citenamefont {Bruus}\ and\ \citenamefont {Angl`es~d'Auriac}(1997)}]{Bruus1997}%
  \BibitemOpen
  \bibfield  {author} {\bibinfo {author} {\bibfnamefont {H.}~\bibnamefont {Bruus}}\ and\ \bibinfo {author} {\bibfnamefont {J.-C.}\ \bibnamefont {Angl`es~d'Auriac}},\ }\bibfield  {title} {\bibinfo {title} {{Energy level statistics of the two-dimensional Hubbard model at low filling}},\ }\href {https://doi.org/10.1103/PhysRevB.55.9142} {\bibfield  {journal} {\bibinfo  {journal} {Phys. Rev. B}\ }\textbf {\bibinfo {volume} {55}},\ \bibinfo {pages} {9142} (\bibinfo {year} {1997})}\BibitemShut {NoStop}%
\bibitem [{\citenamefont {G{\'{o}}mez}\ \emph {et~al.}(2002)\citenamefont {G{\'{o}}mez}, \citenamefont {Molina}, \citenamefont {Rela{\~{n}}o},\ and\ \citenamefont {Retamosa}}]{Gomez2002}%
  \BibitemOpen
  \bibfield  {author} {\bibinfo {author} {\bibfnamefont {J.~M.~G.}\ \bibnamefont {G{\'{o}}mez}}, \bibinfo {author} {\bibfnamefont {R.~A.}\ \bibnamefont {Molina}}, \bibinfo {author} {\bibfnamefont {A.}~\bibnamefont {Rela{\~{n}}o}},\ and\ \bibinfo {author} {\bibfnamefont {J.}~\bibnamefont {Retamosa}},\ }\bibfield  {title} {\bibinfo {title} {{Misleading signatures of quantum chaos}},\ }\href {https://doi.org/10.1103/PhysRevE.66.036209} {\bibfield  {journal} {\bibinfo  {journal} {Phys. Rev. E}\ }\textbf {\bibinfo {volume} {66}},\ \bibinfo {pages} {036209} (\bibinfo {year} {2002})}\BibitemShut {NoStop}%
\bibitem [{\citenamefont {Tekur}\ and\ \citenamefont {Santhanam}(2020)}]{Tekur2020}%
  \BibitemOpen
  \bibfield  {author} {\bibinfo {author} {\bibfnamefont {S.~H.}\ \bibnamefont {Tekur}}\ and\ \bibinfo {author} {\bibfnamefont {M.~S.}\ \bibnamefont {Santhanam}},\ }\bibfield  {title} {\bibinfo {title} {{Symmetry deduction from spectral fluctuations in complex quantum systems}},\ }\href {https://doi.org/10.1103/PhysRevResearch.2.032063} {\bibfield  {journal} {\bibinfo  {journal} {Phys. Rev. Res.}\ }\textbf {\bibinfo {volume} {2}},\ \bibinfo {pages} {032063} (\bibinfo {year} {2020})}\BibitemShut {NoStop}%
\bibitem [{\citenamefont {Bhosale}(2021)}]{Bhosale2021}%
  \BibitemOpen
  \bibfield  {author} {\bibinfo {author} {\bibfnamefont {U.~T.}\ \bibnamefont {Bhosale}},\ }\bibfield  {title} {\bibinfo {title} {{Superposition and higher-order spacing ratios in random matrix theory with application to complex systems}},\ }\href {https://doi.org/10.1103/PhysRevB.104.054204} {\bibfield  {journal} {\bibinfo  {journal} {Phys. Rev. B}\ }\textbf {\bibinfo {volume} {104}},\ \bibinfo {pages} {054204} (\bibinfo {year} {2021})}\BibitemShut {NoStop}%
\bibitem [{\citenamefont {Giraud}\ \emph {et~al.}(2022)\citenamefont {Giraud}, \citenamefont {Mac{\'{e}}}, \citenamefont {Vernier},\ and\ \citenamefont {Alet}}]{Giraud2022}%
  \BibitemOpen
  \bibfield  {author} {\bibinfo {author} {\bibfnamefont {O.}~\bibnamefont {Giraud}}, \bibinfo {author} {\bibfnamefont {N.}~\bibnamefont {Mac{\'{e}}}}, \bibinfo {author} {\bibfnamefont {{\'{E}}.}~\bibnamefont {Vernier}},\ and\ \bibinfo {author} {\bibfnamefont {F.}~\bibnamefont {Alet}},\ }\bibfield  {title} {\bibinfo {title} {{Probing Symmetries of Quantum Many-Body Systems through Gap Ratio Statistics}},\ }\href {https://doi.org/10.1103/PhysRevX.12.011006} {\bibfield  {journal} {\bibinfo  {journal} {Phys. Rev. X}\ }\textbf {\bibinfo {volume} {12}},\ \bibinfo {pages} {011006} (\bibinfo {year} {2022})}\BibitemShut {NoStop}%
\bibitem [{\citenamefont {Atas}\ \emph {et~al.}(2013)\citenamefont {Atas}, \citenamefont {Bogomolny}, \citenamefont {Giraud},\ and\ \citenamefont {Roux}}]{Atas2013}%
  \BibitemOpen
  \bibfield  {author} {\bibinfo {author} {\bibfnamefont {Y.~Y.}\ \bibnamefont {Atas}}, \bibinfo {author} {\bibfnamefont {E.}~\bibnamefont {Bogomolny}}, \bibinfo {author} {\bibfnamefont {O.}~\bibnamefont {Giraud}},\ and\ \bibinfo {author} {\bibfnamefont {G.}~\bibnamefont {Roux}},\ }\bibfield  {title} {\bibinfo {title} {{Distribution of the Ratio of Consecutive Level Spacings in Random Matrix Ensembles}},\ }\href {https://doi.org/10.1103/PhysRevLett.110.084101} {\bibfield  {journal} {\bibinfo  {journal} {Phys. Rev. Lett.}\ }\textbf {\bibinfo {volume} {110}},\ \bibinfo {pages} {084101} (\bibinfo {year} {2013})}\BibitemShut {NoStop}%
\bibitem [{\citenamefont {Chertkov}\ and\ \citenamefont {Clark}(2018)}]{Chertkov2018}%
  \BibitemOpen
  \bibfield  {author} {\bibinfo {author} {\bibfnamefont {E.}~\bibnamefont {Chertkov}}\ and\ \bibinfo {author} {\bibfnamefont {B.~K.}\ \bibnamefont {Clark}},\ }\bibfield  {title} {\bibinfo {title} {{Computational Inverse Method for Constructing Spaces of Quantum Models from Wave Functions}},\ }\href {https://doi.org/10.1103/PhysRevX.8.031029} {\bibfield  {journal} {\bibinfo  {journal} {Phys. Rev. X}\ }\textbf {\bibinfo {volume} {8}},\ \bibinfo {pages} {031029} (\bibinfo {year} {2018})}\BibitemShut {NoStop}%
\bibitem [{\citenamefont {Qi}\ and\ \citenamefont {Ranard}(2019)}]{Qi2019}%
  \BibitemOpen
  \bibfield  {author} {\bibinfo {author} {\bibfnamefont {X.-L.}\ \bibnamefont {Qi}}\ and\ \bibinfo {author} {\bibfnamefont {D.}~\bibnamefont {Ranard}},\ }\bibfield  {title} {\bibinfo {title} {{Determining a local Hamiltonian from a single eigenstate}},\ }\href {https://doi.org/10.22331/q-2019-07-08-159} {\bibfield  {journal} {\bibinfo  {journal} {Quantum}\ }\textbf {\bibinfo {volume} {3}},\ \bibinfo {pages} {159} (\bibinfo {year} {2019})}\BibitemShut {NoStop}%
\bibitem [{Note1()}]{Note1}%
  \BibitemOpen
  \bibinfo {note} {In the one-dimensional case, the spin-singlet and parity-odd $\eta $-pairing operator is mentioned in Ref.~\cite {Mark2020}. Specifically, it is shown that the state obtained by acting with the $\eta _{p_x}^+$ operator [Eq.~\protect \eqref {eq:pxeta_operator}] only once on Yang’s $\eta $-pairing state [Eq.~\protect \eqref {eq:eta_state}] becomes an eigenstate of the Hirsch model. To extend this result to states where the $\eta _{p_x}^+$ operator is applied multiple times or to higher-dimensional systems, the multibody interactions proposed in this study are required.}\BibitemShut {Stop}%
\bibitem [{\citenamefont {Kitamura}\ and\ \citenamefont {Aoki}(2016)}]{Kitamura2016}%
  \BibitemOpen
  \bibfield  {author} {\bibinfo {author} {\bibfnamefont {S.}~\bibnamefont {Kitamura}}\ and\ \bibinfo {author} {\bibfnamefont {H.}~\bibnamefont {Aoki}},\ }\bibfield  {title} {\bibinfo {title} {{$\eta$-pairing superfluid in periodically-driven fermionic Hubbard model with strong attraction}},\ }\href {https://doi.org/10.1103/PhysRevB.94.174503} {\bibfield  {journal} {\bibinfo  {journal} {Phys. Rev. B}\ }\textbf {\bibinfo {volume} {94}},\ \bibinfo {pages} {174503} (\bibinfo {year} {2016})}\BibitemShut {NoStop}%
\bibitem [{\citenamefont {Peronaci}\ \emph {et~al.}(2020)\citenamefont {Peronaci}, \citenamefont {Parcollet},\ and\ \citenamefont {Schir{\'{o}}}}]{Peronaci2020}%
  \BibitemOpen
  \bibfield  {author} {\bibinfo {author} {\bibfnamefont {F.}~\bibnamefont {Peronaci}}, \bibinfo {author} {\bibfnamefont {O.}~\bibnamefont {Parcollet}},\ and\ \bibinfo {author} {\bibfnamefont {M.}~\bibnamefont {Schir{\'{o}}}},\ }\bibfield  {title} {\bibinfo {title} {{Enhancement of local pairing correlations in periodically driven Mott insulators}},\ }\href {https://doi.org/10.1103/PhysRevB.101.161101} {\bibfield  {journal} {\bibinfo  {journal} {Phys. Rev. B}\ }\textbf {\bibinfo {volume} {101}},\ \bibinfo {pages} {161101} (\bibinfo {year} {2020})}\BibitemShut {NoStop}%
\bibitem [{\citenamefont {Cook}\ and\ \citenamefont {Clark}(2020)}]{Cook2020}%
  \BibitemOpen
  \bibfield  {author} {\bibinfo {author} {\bibfnamefont {M.~W.}\ \bibnamefont {Cook}}\ and\ \bibinfo {author} {\bibfnamefont {S.~R.}\ \bibnamefont {Clark}},\ }\bibfield  {title} {\bibinfo {title} {{Controllable finite-momenta dynamical quasicondensation in the periodically driven one-dimensional Fermi-Hubbard model}},\ }\href {https://doi.org/10.1103/PhysRevA.101.033604} {\bibfield  {journal} {\bibinfo  {journal} {Phys. Rev. A}\ }\textbf {\bibinfo {volume} {101}},\ \bibinfo {pages} {33604} (\bibinfo {year} {2020})}\BibitemShut {NoStop}%
\bibitem [{\citenamefont {Tindall}\ \emph {et~al.}(2021)\citenamefont {Tindall}, \citenamefont {Schlawin}, \citenamefont {Sentef},\ and\ \citenamefont {Jaksch}}]{Tindall2021}%
  \BibitemOpen
  \bibfield  {author} {\bibinfo {author} {\bibfnamefont {J.}~\bibnamefont {Tindall}}, \bibinfo {author} {\bibfnamefont {F.}~\bibnamefont {Schlawin}}, \bibinfo {author} {\bibfnamefont {M.~A.}\ \bibnamefont {Sentef}},\ and\ \bibinfo {author} {\bibfnamefont {D.}~\bibnamefont {Jaksch}},\ }\bibfield  {title} {\bibinfo {title} {{Analytical solution for the steady states of the driven Hubbard model}},\ }\href {https://doi.org/10.1103/PhysRevB.103.035146} {\bibfield  {journal} {\bibinfo  {journal} {Phys. Rev. B}\ }\textbf {\bibinfo {volume} {103}},\ \bibinfo {pages} {035146} (\bibinfo {year} {2021})}\BibitemShut {NoStop}%
\bibitem [{\citenamefont {Werner}\ \emph {et~al.}(2018)\citenamefont {Werner}, \citenamefont {Strand}, \citenamefont {Hoshino}, \citenamefont {Murakami},\ and\ \citenamefont {Eckstein}}]{Werner2018a}%
  \BibitemOpen
  \bibfield  {author} {\bibinfo {author} {\bibfnamefont {P.}~\bibnamefont {Werner}}, \bibinfo {author} {\bibfnamefont {H.~U.~R.}\ \bibnamefont {Strand}}, \bibinfo {author} {\bibfnamefont {S.}~\bibnamefont {Hoshino}}, \bibinfo {author} {\bibfnamefont {Y.}~\bibnamefont {Murakami}},\ and\ \bibinfo {author} {\bibfnamefont {M.}~\bibnamefont {Eckstein}},\ }\bibfield  {title} {\bibinfo {title} {{Enhanced pairing susceptibility in a photodoped two-orbital Hubbard model}},\ }\href {https://doi.org/10.1103/PhysRevB.97.165119} {\bibfield  {journal} {\bibinfo  {journal} {Phys. Rev. B}\ }\textbf {\bibinfo {volume} {97}},\ \bibinfo {pages} {165119} (\bibinfo {year} {2018})}\BibitemShut {NoStop}%
\bibitem [{\citenamefont {Werner}\ \emph {et~al.}(2019)\citenamefont {Werner}, \citenamefont {Li}, \citenamefont {Gole{\v{z}}},\ and\ \citenamefont {Eckstein}}]{Werner2019}%
  \BibitemOpen
  \bibfield  {author} {\bibinfo {author} {\bibfnamefont {P.}~\bibnamefont {Werner}}, \bibinfo {author} {\bibfnamefont {J.}~\bibnamefont {Li}}, \bibinfo {author} {\bibfnamefont {D.}~\bibnamefont {Gole{\v{z}}}},\ and\ \bibinfo {author} {\bibfnamefont {M.}~\bibnamefont {Eckstein}},\ }\bibfield  {title} {\bibinfo {title} {{Entropy-cooled nonequilibrium states of the Hubbard model}},\ }\href {https://doi.org/10.1103/PhysRevB.100.155130} {\bibfield  {journal} {\bibinfo  {journal} {Phys. Rev. B}\ }\textbf {\bibinfo {volume} {100}},\ \bibinfo {pages} {155130} (\bibinfo {year} {2019})}\BibitemShut {NoStop}%
\bibitem [{\citenamefont {Li}\ \emph {et~al.}(2020)\citenamefont {Li}, \citenamefont {Golez}, \citenamefont {Werner},\ and\ \citenamefont {Eckstein}}]{Li2020j}%
  \BibitemOpen
  \bibfield  {author} {\bibinfo {author} {\bibfnamefont {J.}~\bibnamefont {Li}}, \bibinfo {author} {\bibfnamefont {D.}~\bibnamefont {Golez}}, \bibinfo {author} {\bibfnamefont {P.}~\bibnamefont {Werner}},\ and\ \bibinfo {author} {\bibfnamefont {M.}~\bibnamefont {Eckstein}},\ }\bibfield  {title} {\bibinfo {title} {{$\eta$-paired superconducting hidden phase in photodoped Mott insulators}},\ }\href {https://doi.org/10.1103/PhysRevB.102.165136} {\bibfield  {journal} {\bibinfo  {journal} {Phys. Rev. B}\ }\textbf {\bibinfo {volume} {102}},\ \bibinfo {pages} {165136} (\bibinfo {year} {2020})}\BibitemShut {NoStop}%
\bibitem [{\citenamefont {Murakami}\ \emph {et~al.}(2022)\citenamefont {Murakami}, \citenamefont {Takayoshi}, \citenamefont {Kaneko}, \citenamefont {Sun}, \citenamefont {Gole{\v{z}}}, \citenamefont {Millis},\ and\ \citenamefont {Werner}}]{Murakami2022f}%
  \BibitemOpen
  \bibfield  {author} {\bibinfo {author} {\bibfnamefont {Y.}~\bibnamefont {Murakami}}, \bibinfo {author} {\bibfnamefont {S.}~\bibnamefont {Takayoshi}}, \bibinfo {author} {\bibfnamefont {T.}~\bibnamefont {Kaneko}}, \bibinfo {author} {\bibfnamefont {Z.}~\bibnamefont {Sun}}, \bibinfo {author} {\bibfnamefont {D.}~\bibnamefont {Gole{\v{z}}}}, \bibinfo {author} {\bibfnamefont {A.~J.}\ \bibnamefont {Millis}},\ and\ \bibinfo {author} {\bibfnamefont {P.}~\bibnamefont {Werner}},\ }\bibfield  {title} {\bibinfo {title} {{Exploring nonequilibrium phases of photo-doped Mott insulators with generalized Gibbs ensembles}},\ }\href {https://doi.org/10.1038/s42005-021-00799-7} {\bibfield  {journal} {\bibinfo  {journal} {Commun. Phys.}\ }\textbf {\bibinfo {volume} {5}},\ \bibinfo {pages} {23} (\bibinfo {year} {2022})}\BibitemShut {NoStop}%
\bibitem [{\citenamefont {Murakami}\ \emph {et~al.}(2023)\citenamefont {Murakami}, \citenamefont {Takayoshi}, \citenamefont {Kaneko}, \citenamefont {L{\"{a}}uchli},\ and\ \citenamefont {Werner}}]{Murakami2023b}%
  \BibitemOpen
  \bibfield  {author} {\bibinfo {author} {\bibfnamefont {Y.}~\bibnamefont {Murakami}}, \bibinfo {author} {\bibfnamefont {S.}~\bibnamefont {Takayoshi}}, \bibinfo {author} {\bibfnamefont {T.}~\bibnamefont {Kaneko}}, \bibinfo {author} {\bibfnamefont {A.~M.}\ \bibnamefont {L{\"{a}}uchli}},\ and\ \bibinfo {author} {\bibfnamefont {P.}~\bibnamefont {Werner}},\ }\bibfield  {title} {\bibinfo {title} {{Spin, Charge, and $\eta$-Spin Separation in One-Dimensional Photodoped Mott Insulators}},\ }\href {https://doi.org/10.1103/PhysRevLett.130.106501} {\bibfield  {journal} {\bibinfo  {journal} {Phys. Rev. Lett.}\ }\textbf {\bibinfo {volume} {130}},\ \bibinfo {pages} {106501} (\bibinfo {year} {2023})}\BibitemShut {NoStop}%
\bibitem [{\citenamefont {Diehl}\ \emph {et~al.}(2008)\citenamefont {Diehl}, \citenamefont {Micheli}, \citenamefont {Kantian}, \citenamefont {Kraus}, \citenamefont {B{\"{u}}chler},\ and\ \citenamefont {Zoller}}]{Diehl2008}%
  \BibitemOpen
  \bibfield  {author} {\bibinfo {author} {\bibfnamefont {S.}~\bibnamefont {Diehl}}, \bibinfo {author} {\bibfnamefont {A.}~\bibnamefont {Micheli}}, \bibinfo {author} {\bibfnamefont {A.}~\bibnamefont {Kantian}}, \bibinfo {author} {\bibfnamefont {B.}~\bibnamefont {Kraus}}, \bibinfo {author} {\bibfnamefont {H.~P.}\ \bibnamefont {B{\"{u}}chler}},\ and\ \bibinfo {author} {\bibfnamefont {P.}~\bibnamefont {Zoller}},\ }\bibfield  {title} {\bibinfo {title} {{Quantum states and phases in driven open quantum systems with cold atoms}},\ }\href {https://doi.org/10.1038/nphys1073} {\bibfield  {journal} {\bibinfo  {journal} {Nat. Phys.}\ }\textbf {\bibinfo {volume} {4}},\ \bibinfo {pages} {878} (\bibinfo {year} {2008})}\BibitemShut {NoStop}%
\bibitem [{\citenamefont {Kraus}\ \emph {et~al.}(2008)\citenamefont {Kraus}, \citenamefont {B{\"{u}}chler}, \citenamefont {Diehl}, \citenamefont {Kantian}, \citenamefont {Micheli},\ and\ \citenamefont {Zoller}}]{Kraus2008}%
  \BibitemOpen
  \bibfield  {author} {\bibinfo {author} {\bibfnamefont {B.}~\bibnamefont {Kraus}}, \bibinfo {author} {\bibfnamefont {H.~P.}\ \bibnamefont {B{\"{u}}chler}}, \bibinfo {author} {\bibfnamefont {S.}~\bibnamefont {Diehl}}, \bibinfo {author} {\bibfnamefont {A.}~\bibnamefont {Kantian}}, \bibinfo {author} {\bibfnamefont {A.}~\bibnamefont {Micheli}},\ and\ \bibinfo {author} {\bibfnamefont {P.}~\bibnamefont {Zoller}},\ }\bibfield  {title} {\bibinfo {title} {{Preparation of entangled states by quantum Markov processes}},\ }\href {https://doi.org/10.1103/PhysRevA.78.042307} {\bibfield  {journal} {\bibinfo  {journal} {Phys. Rev. A}\ }\textbf {\bibinfo {volume} {78}},\ \bibinfo {pages} {042307} (\bibinfo {year} {2008})}\BibitemShut {NoStop}%
\bibitem [{\citenamefont {Nakagawa}\ \emph {et~al.}()\citenamefont {Nakagawa}, \citenamefont {Tsuji}, \citenamefont {Kawakami},\ and\ \citenamefont {Ueda}}]{Nakagawa2021}%
  \BibitemOpen
  \bibfield  {author} {\bibinfo {author} {\bibfnamefont {M.}~\bibnamefont {Nakagawa}}, \bibinfo {author} {\bibfnamefont {N.}~\bibnamefont {Tsuji}}, \bibinfo {author} {\bibfnamefont {N.}~\bibnamefont {Kawakami}},\ and\ \bibinfo {author} {\bibfnamefont {M.}~\bibnamefont {Ueda}},\ }\bibfield  {title} {\bibinfo {title} {{$\eta$ Pairing of Light-Emitting Fermions: Nonequilibrium Pairing Mechanism at High Temperatures}},\ }\Eprint {https://arxiv.org/abs/2103.13624} {arXiv:2103.13624} \BibitemShut {NoStop}%
\bibitem [{\citenamefont {Yang}\ and\ \citenamefont {Song}(2022)}]{Yang2022e}%
  \BibitemOpen
  \bibfield  {author} {\bibinfo {author} {\bibfnamefont {X.~M.}\ \bibnamefont {Yang}}\ and\ \bibinfo {author} {\bibfnamefont {Z.}~\bibnamefont {Song}},\ }\bibfield  {title} {\bibinfo {title} {{Dynamic transition from insulating state to $\eta$-pairing state in a composite non-Hermitian system}},\ }\href {https://doi.org/10.1103/PhysRevB.105.195132} {\bibfield  {journal} {\bibinfo  {journal} {Phys. Rev. B}\ }\textbf {\bibinfo {volume} {105}},\ \bibinfo {pages} {195132} (\bibinfo {year} {2022})}\BibitemShut {NoStop}%
\bibitem [{\citenamefont {Kantian}\ \emph {et~al.}(2010)\citenamefont {Kantian}, \citenamefont {Daley},\ and\ \citenamefont {Zoller}}]{Kantian2010}%
  \BibitemOpen
  \bibfield  {author} {\bibinfo {author} {\bibfnamefont {A.}~\bibnamefont {Kantian}}, \bibinfo {author} {\bibfnamefont {A.~J.}\ \bibnamefont {Daley}},\ and\ \bibinfo {author} {\bibfnamefont {P.}~\bibnamefont {Zoller}},\ }\bibfield  {title} {\bibinfo {title} {{$\eta$ Condensate of Fermionic Atom Pairs via Adiabatic State Preparation}},\ }\href {https://doi.org/10.1103/PhysRevLett.104.240406} {\bibfield  {journal} {\bibinfo  {journal} {Phys. Rev. Lett.}\ }\textbf {\bibinfo {volume} {104}},\ \bibinfo {pages} {240406} (\bibinfo {year} {2010})}\BibitemShut {NoStop}%
\bibitem [{\citenamefont {Gotta}\ \emph {et~al.}(2023)\citenamefont {Gotta}, \citenamefont {Moudgalya},\ and\ \citenamefont {Mazza}}]{Gotta2023a}%
  \BibitemOpen
  \bibfield  {author} {\bibinfo {author} {\bibfnamefont {L.}~\bibnamefont {Gotta}}, \bibinfo {author} {\bibfnamefont {S.}~\bibnamefont {Moudgalya}},\ and\ \bibinfo {author} {\bibfnamefont {L.}~\bibnamefont {Mazza}},\ }\bibfield  {title} {\bibinfo {title} {{Asymptotic Quantum Many-Body Scars}},\ }\href {https://doi.org/10.1103/PhysRevLett.131.190401} {\bibfield  {journal} {\bibinfo  {journal} {Phys. Rev. Lett.}\ }\textbf {\bibinfo {volume} {131}},\ \bibinfo {pages} {190401} (\bibinfo {year} {2023})}\BibitemShut {NoStop}%
\bibitem [{\citenamefont {Nandkishore}\ and\ \citenamefont {Hermele}(2019)}]{Nandkishore2019}%
  \BibitemOpen
  \bibfield  {author} {\bibinfo {author} {\bibfnamefont {R.~M.}\ \bibnamefont {Nandkishore}}\ and\ \bibinfo {author} {\bibfnamefont {M.}~\bibnamefont {Hermele}},\ }\bibfield  {title} {\bibinfo {title} {{Fractons}},\ }\href {https://doi.org/10.1146/annurev-conmatphys-031218-013604} {\bibfield  {journal} {\bibinfo  {journal} {Annu. Rev. Condens. Matter Phys.}\ }\textbf {\bibinfo {volume} {10}},\ \bibinfo {pages} {295} (\bibinfo {year} {2019})}\BibitemShut {NoStop}%
\bibitem [{\citenamefont {Pretko}\ \emph {et~al.}(2020)\citenamefont {Pretko}, \citenamefont {Chen},\ and\ \citenamefont {You}}]{Pretko2020a}%
  \BibitemOpen
  \bibfield  {author} {\bibinfo {author} {\bibfnamefont {M.}~\bibnamefont {Pretko}}, \bibinfo {author} {\bibfnamefont {X.}~\bibnamefont {Chen}},\ and\ \bibinfo {author} {\bibfnamefont {Y.}~\bibnamefont {You}},\ }\bibfield  {title} {\bibinfo {title} {{Fracton phases of matter}},\ }\href {https://doi.org/10.1142/S0217751X20300033} {\bibfield  {journal} {\bibinfo  {journal} {Int. J. Mod. Phys. A}\ }\textbf {\bibinfo {volume} {35}},\ \bibinfo {pages} {2030003} (\bibinfo {year} {2020})}\BibitemShut {NoStop}%
\bibitem [{\citenamefont {Xavier}\ and\ \citenamefont {Pereira}(2021)}]{Xavier2021}%
  \BibitemOpen
  \bibfield  {author} {\bibinfo {author} {\bibfnamefont {H.~B.}\ \bibnamefont {Xavier}}\ and\ \bibinfo {author} {\bibfnamefont {R.~G.}\ \bibnamefont {Pereira}},\ }\bibfield  {title} {\bibinfo {title} {{Fractons from a liquid of singlet pairs}},\ }\href {https://doi.org/10.1103/PhysRevB.103.085101} {\bibfield  {journal} {\bibinfo  {journal} {Phys. Rev. B}\ }\textbf {\bibinfo {volume} {103}},\ \bibinfo {pages} {085101} (\bibinfo {year} {2021})}\BibitemShut {NoStop}%
\bibitem [{\citenamefont {B{\"{u}}chler}\ \emph {et~al.}(2007)\citenamefont {B{\"{u}}chler}, \citenamefont {Micheli},\ and\ \citenamefont {Zoller}}]{Buchler2007}%
  \BibitemOpen
  \bibfield  {author} {\bibinfo {author} {\bibfnamefont {H.~P.}\ \bibnamefont {B{\"{u}}chler}}, \bibinfo {author} {\bibfnamefont {A.}~\bibnamefont {Micheli}},\ and\ \bibinfo {author} {\bibfnamefont {P.}~\bibnamefont {Zoller}},\ }\bibfield  {title} {\bibinfo {title} {{Three-body interactions with cold polar molecules}},\ }\href {https://doi.org/10.1038/nphys678} {\bibfield  {journal} {\bibinfo  {journal} {Nat. Phys.}\ }\textbf {\bibinfo {volume} {3}},\ \bibinfo {pages} {726} (\bibinfo {year} {2007})}\BibitemShut {NoStop}%
\bibitem [{\citenamefont {Han}(2010)}]{Han2010}%
  \BibitemOpen
  \bibfield  {author} {\bibinfo {author} {\bibfnamefont {J.}~\bibnamefont {Han}},\ }\bibfield  {title} {\bibinfo {title} {{Direct evidence of three-body interactions in a cold $^{85}\mathrm{Rb}$ Rydberg gas}},\ }\href {https://doi.org/10.1103/PhysRevA.82.052501} {\bibfield  {journal} {\bibinfo  {journal} {Phys. Rev. A}\ }\textbf {\bibinfo {volume} {82}},\ \bibinfo {pages} {052501} (\bibinfo {year} {2010})}\BibitemShut {NoStop}%
\bibitem [{\citenamefont {Will}\ \emph {et~al.}(2010)\citenamefont {Will}, \citenamefont {Best}, \citenamefont {Schneider}, \citenamefont {Hackerm{\"{u}}ller}, \citenamefont {L{\"{u}}hmann},\ and\ \citenamefont {Bloch}}]{Will2010}%
  \BibitemOpen
  \bibfield  {author} {\bibinfo {author} {\bibfnamefont {S.}~\bibnamefont {Will}}, \bibinfo {author} {\bibfnamefont {T.}~\bibnamefont {Best}}, \bibinfo {author} {\bibfnamefont {U.}~\bibnamefont {Schneider}}, \bibinfo {author} {\bibfnamefont {L.}~\bibnamefont {Hackerm{\"{u}}ller}}, \bibinfo {author} {\bibfnamefont {D.-S.}\ \bibnamefont {L{\"{u}}hmann}},\ and\ \bibinfo {author} {\bibfnamefont {I.}~\bibnamefont {Bloch}},\ }\bibfield  {title} {\bibinfo {title} {{Time-resolved observation of coherent multi-body interactions in quantum phase revivals}},\ }\href {https://doi.org/10.1038/nature09036} {\bibfield  {journal} {\bibinfo  {journal} {Nature}\ }\textbf {\bibinfo {volume} {465}},\ \bibinfo {pages} {197} (\bibinfo {year} {2010})}\BibitemShut {NoStop}%
\bibitem [{\citenamefont {Hammer}\ \emph {et~al.}(2013)\citenamefont {Hammer}, \citenamefont {Nogga},\ and\ \citenamefont {Schwenk}}]{Hammer2013}%
  \BibitemOpen
  \bibfield  {author} {\bibinfo {author} {\bibfnamefont {H.-W.}\ \bibnamefont {Hammer}}, \bibinfo {author} {\bibfnamefont {A.}~\bibnamefont {Nogga}},\ and\ \bibinfo {author} {\bibfnamefont {A.}~\bibnamefont {Schwenk}},\ }\bibfield  {title} {\bibinfo {title} {{Colloquium : Three-body forces: From cold atoms to nuclei}},\ }\href {https://doi.org/10.1103/RevModPhys.85.197} {\bibfield  {journal} {\bibinfo  {journal} {Rev. Mod. Phys.}\ }\textbf {\bibinfo {volume} {85}},\ \bibinfo {pages} {197} (\bibinfo {year} {2013})}\BibitemShut {NoStop}%
\bibitem [{\citenamefont {Ren}\ \emph {et~al.}(2015)\citenamefont {Ren}, \citenamefont {Wu},\ and\ \citenamefont {Xu}}]{Ren2015}%
  \BibitemOpen
  \bibfield  {author} {\bibinfo {author} {\bibfnamefont {J.}~\bibnamefont {Ren}}, \bibinfo {author} {\bibfnamefont {Y.-Z.}\ \bibnamefont {Wu}},\ and\ \bibinfo {author} {\bibfnamefont {X.-F.}\ \bibnamefont {Xu}},\ }\bibfield  {title} {\bibinfo {title} {{Expansion dynamics in a one-dimensional hard-core boson model with three-body interactions}},\ }\href {https://doi.org/10.1038/srep14743} {\bibfield  {journal} {\bibinfo  {journal} {Sci. Rep.}\ }\textbf {\bibinfo {volume} {5}},\ \bibinfo {pages} {14743} (\bibinfo {year} {2015})}\BibitemShut {NoStop}%
\bibitem [{\citenamefont {Valiente}(2019)}]{Valiente2019a}%
  \BibitemOpen
  \bibfield  {author} {\bibinfo {author} {\bibfnamefont {M.}~\bibnamefont {Valiente}},\ }\bibfield  {title} {\bibinfo {title} {{Three-body repulsive forces among identical bosons in one dimension}},\ }\href {https://doi.org/10.1103/PhysRevA.100.013614} {\bibfield  {journal} {\bibinfo  {journal} {Phys. Rev. A}\ }\textbf {\bibinfo {volume} {100}},\ \bibinfo {pages} {013614} (\bibinfo {year} {2019})}\BibitemShut {NoStop}%
\bibitem [{\citenamefont {Gurian}\ \emph {et~al.}(2012)\citenamefont {Gurian}, \citenamefont {Cheinet}, \citenamefont {Huillery}, \citenamefont {Fioretti}, \citenamefont {Zhao}, \citenamefont {Gould}, \citenamefont {Comparat},\ and\ \citenamefont {Pillet}}]{Gurian2012}%
  \BibitemOpen
  \bibfield  {author} {\bibinfo {author} {\bibfnamefont {J.~H.}\ \bibnamefont {Gurian}}, \bibinfo {author} {\bibfnamefont {P.}~\bibnamefont {Cheinet}}, \bibinfo {author} {\bibfnamefont {P.}~\bibnamefont {Huillery}}, \bibinfo {author} {\bibfnamefont {A.}~\bibnamefont {Fioretti}}, \bibinfo {author} {\bibfnamefont {J.}~\bibnamefont {Zhao}}, \bibinfo {author} {\bibfnamefont {P.~L.}\ \bibnamefont {Gould}}, \bibinfo {author} {\bibfnamefont {D.}~\bibnamefont {Comparat}},\ and\ \bibinfo {author} {\bibfnamefont {P.}~\bibnamefont {Pillet}},\ }\bibfield  {title} {\bibinfo {title} {{Observation of a Resonant Four-Body Interaction in Cold Cesium Rydberg Atoms}},\ }\href {https://doi.org/10.1103/PhysRevLett.108.023005} {\bibfield  {journal} {\bibinfo  {journal} {Phys. Rev. Lett.}\ }\textbf {\bibinfo {volume} {108}},\ \bibinfo {pages} {023005} (\bibinfo {year} {2012})}\BibitemShut {NoStop}%
\bibitem [{\citenamefont {Honda}\ \emph {et~al.}()\citenamefont {Honda}, \citenamefont {Takasu}, \citenamefont {Haruna}, \citenamefont {Nishida},\ and\ \citenamefont {Takahashi}}]{Honda2024}%
  \BibitemOpen
  \bibfield  {author} {\bibinfo {author} {\bibfnamefont {K.}~\bibnamefont {Honda}}, \bibinfo {author} {\bibfnamefont {Y.}~\bibnamefont {Takasu}}, \bibinfo {author} {\bibfnamefont {Y.}~\bibnamefont {Haruna}}, \bibinfo {author} {\bibfnamefont {Y.}~\bibnamefont {Nishida}},\ and\ \bibinfo {author} {\bibfnamefont {Y.}~\bibnamefont {Takahashi}},\ }\bibfield  {title} {\bibinfo {title} {{Evidence of a Four-Body Force in an Interaction-Tunable Trapped Cold-Atom System}},\ }\Eprint {https://arxiv.org/abs/2402.16254} {arXiv:2402.16254} \BibitemShut {NoStop}%
\end{thebibliography}%
\end{document}